\documentclass[a4paper,12pt]{article}

\usepackage{amsmath,amsfonts,amssymb,epsfig}
\usepackage{slashed}
\numberwithin{equation}{section}
\usepackage{dsfont}
\usepackage{cancel}
\usepackage{cite}
\usepackage{color}

\def\nn{\nonumber}
\def\vo{\mathcal{V}}
\def\mc{\mathcal}
\def\rhoo{a}
\def\bra{\langle}
\def\ket{\rangle}

\def\tr{\mathrm{tr}}

\def\beq{\begin{equation}}
\def\eeq{\end{equation}}
\def\be{\begin{equation}}
\def\ee{\end{equation}}
\def\bea{\begin{eqnarray}}
\def\eea{\end{eqnarray}}
\def\bal{\begin{align}}
\def\eal{\end{align}}
\def\aC[#1,#2]{C_{#1 #2}}
\def\mC[#1,#2]{C_{#1 #2}}
\def\CC{\mathcal{C}}

\def\2b2[#1,#2][#3,#4]{\left( \begin{array}{cc} #1 & #2 \\ #3 & #4 \end{array}
\right)}
\def\3b3[#1,#2,#3][#4,#5,#6][#7,#8,#9]{\left( \begin{array}{ccc} #1 & #2 #3 \\
#4 & #5 & #6\\#7&#8&#9\end{array} \right)}

\def\ov{\overline}

\newcommand{\exclude}[1]{}

\author{Michele Cicoli$^\clubsuit $\footnote{mcicoli@ictp.it}, \,
Mark~D.~Goodsell$^\diamondsuit $\footnote{mark.goodsell@cern.ch}\,  and  Andreas Ringwald$^\spadesuit $\footnote{andreas.ringwald@desy.de} }
\date{}
\title{\vspace{-3cm}\small{
\hfill{CERN-PH-TH/2012-153}\\
\hfill{DESY 12-058}}\\[2cm]
\huge{The type IIB string axiverse and its low-energy phenomenology}}
\begin{document}
\maketitle
\vspace{-1cm}
\begin{center}
\emph{$^\clubsuit $Abdus Salam ICTP, Strada Costiera 11, Trieste 34014, Italy\\
$^\clubsuit $INFN, Sezione di Trieste, Italy\\
$^\diamondsuit$CERN, Theory Division, CH-1211 Geneva 23, Switzerland\\
$^\spadesuit$Deutsches Elektronen-Synchrotron, DESY, Notkestra\ss e 85, 22607  Hamburg,
Germany}
\end{center}
\abstract{We study closed string axions
in type IIB orientifold compactifications. We show that
for natural values of the background fluxes
the moduli stabilisation mechanism of the LARGE Volume Scenario (LVS) gives
rise to an axiverse characterised by the presence of a QCD axion
plus many light axion-like particles whose masses are logarithmically hierarchical.
We study the phenomenological features of the LVS axiverse, deriving the masses of the axions
and their couplings to matter and gauge fields.
We also determine when closed string axions can solve the strong CP problem,
and analyse the first explicit examples of semi-realistic models with stable moduli
and a QCD axion candidate which is not eaten by an anomalous Abelian gauge boson.
We discuss the impact of the choice of inflationary scenario on the LVS axiverse,
and summarise the astrophysical, cosmological and experimental constraints upon it.
Moreover, we show how models can be constructed with additional light axion-like particles
that could explain some intriguing astrophysical anomalies,
and could be searched for in the next generation of axion helioscopes
and light-shining-through-a-wall experiments.}

\newpage

\tableofcontents

\section{Introduction}
\label{intro}

The QCD axion `$a$' is predicted~\cite{Weinberg:1977ma,Wilczek:1977pj} in the most plausible
explanation of the strong CP problem, that is the non-observation of a $\theta$-angle term. To accomplish this,
it must act as a field-dependent $\theta$-angle term, i.e. it must have a coupling to
the topological charge density in QCD:
\begin{align}
\label{axion}
\mc{L} \supset
\frac{1}{2}\, \partial_\mu a\, \partial^\mu a -  \frac{g_{3}^2}{32\pi^2} \frac{a\, \aC[a,3] }{f_a} \,F_{3,\mu\nu}^b \tilde{F}_{3}^{b,\mu\nu},
\end{align}
where $F_{3,\mu\nu}^b$ is the $b$th component of the gluon field strength tensor, $\tilde{F}_3^{b,\mu\nu} = \frac{1}{2} \epsilon^{\mu\nu\rho \sigma}F_{3,\rho\sigma}^b $ its dual, $g_3$ is the strong coupling, and $f_a$ is the axion decay constant. We take conventions such that the axion is periodic under $a \rightarrow a + 2\pi f_a$; $\aC[a,3]$ is then a constant describing the coupling of the axion to QCD, which for a field theory model is equal to the anomaly of the Peccei-Quinn field under QCD. It is usually absorbed into the definition of $f_a$ but it shall be more convenient for string models to retain it.

Non-perturbative QCD effects at low energies then generate a potential for the QCD axion plus the $\theta$ angle, stabilising the physical value at zero, thereby solving the strong CP problem~\cite{Peccei:1977hh}, and giving the axion a parametrically small mass $m_a\sim m_\pi f_\pi/f_a$,
where $m_\pi$ and $f_\pi$ are the pion mass and decay constant, respectively.

\begin{center}
\begin{figure}[!h]
\epsfig{file=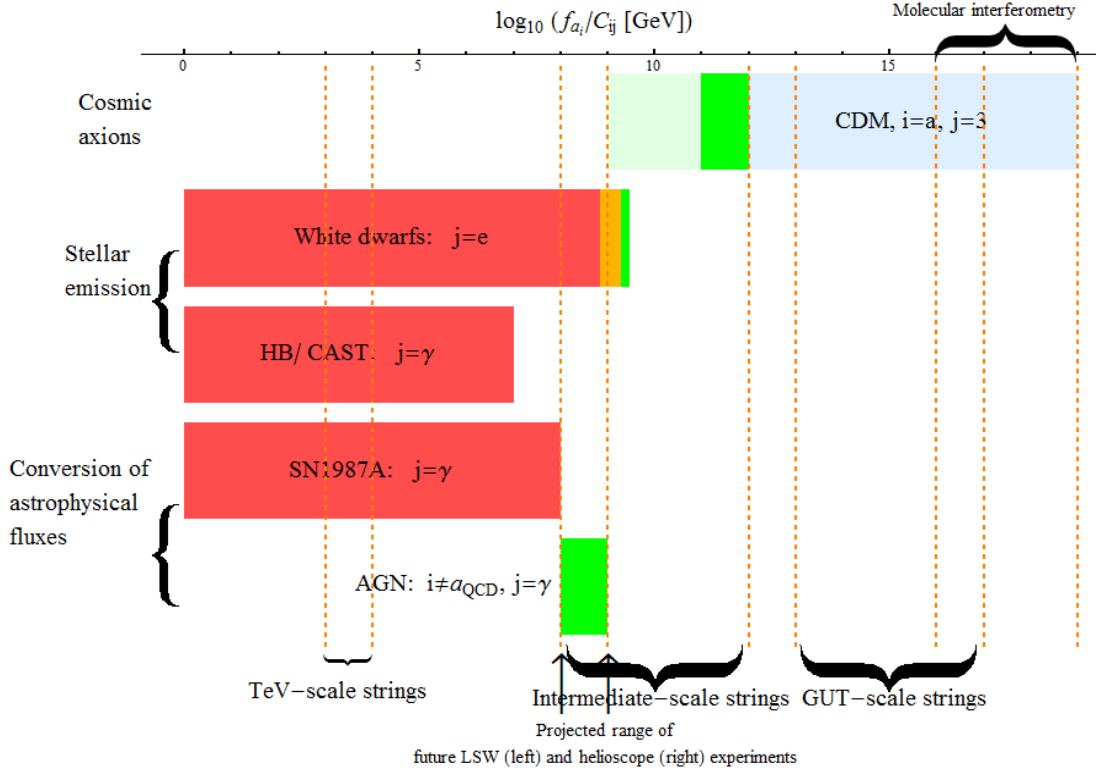,width=0.9\textwidth}
\caption{\footnotesize A summary of constraints on and hints for the couplings $f_{a_i}/C_{ij}$ of axion-like particles $a_i$ to fields $j$.
The green regions from top to bottom correspond respectively to the classic `axion dark matter window', hints of an axion from
white dwarf cooling and transparency of the Universe to very high energy gamma rays. Red regions are excluded,
and the orange region would be excluded by red giants but is compatible with the hints from white dwarfs. The blue region would be excluded by dark matter overproduction in the absence of a dilution mechanism or tuning of the misalignment angle.
Legend: `Molecular interferometry' refers to reach of future experiments discussed in \cite{Graham:2011qk},
see appendix \ref{APP:COSMIC}; `CDM' is constraints on the QCD axion coupling from dark matter overproduction in the standard cosmology,
see section \ref{APP:COSMIC}; constraints from `White dwarfs' are discussed in section \ref{APP:STELLAR};
`HB/CAST' are constraints from horizontal branch stars and the CAST experiment respectively, see section 
\ref{APP:STELLAR};
`SN1987A' are the bounds from observation of high energy gamma rays from that supernova, see section \ref{APP:FLUXES};
AGN are the hints from observation of gamma rays from active galactic nuclei, also in section \ref{APP:FLUXES}.}
\label{FIG:Constraints}\end{figure}\end{center}

There are beam dump, astrophysical and experimental constraints on the decay constant of the QCD axion
and axion-like particles (ALPs) $a_i$ via their couplings to photons and electrons:
\be
\mc{L} \supset -  \frac{e^2 }{32\pi^2} \frac{\aC[i,\gamma] }{f_{a_i}} \,a_i \,F_{\mu\nu} \tilde{F}^{\mu\nu} + \frac{\mC[i,e]}{2 f_{a_i}} \,\bar{e} \gamma^\mu\gamma_5 e \partial_\mu a_i\,,
\label{EQ:DEFCAGG}
\ee
that are summarised in appendix \ref{APP:CONSTRAINTS}. There are also searches underway in
light-shining-through-a-wall (LSW) experiments \cite{Redondo:2010dp},
in searches for solar axions \cite{Irastorza:2011gs} and in direct haloscope
searches for axion cold dark matter (CDM) \cite{Arias:2012mb}.
Direct searches for the QCD axion in beam dumps and the non-observation of its effect on the cooling of stars
constrain the QCD axion decay constant to be $f_a/\aC[a,3]\gtrsim 10^9$ GeV.
Furthermore, there are also several tantalising astrophysical hints, which together point to a light ALP with couplings and mass \cite{DeAngelis:2007dy,Simet:2007sa,SanchezConde:2009wu,Dominguez:2011xy,Tavecchio:2012um,Isern:2012ef}:
\be
\frac{f_{a_i}}{\mC[i,e]}\simeq (0.7\div 2.6)\times 10^9\ {\rm GeV}\,,
\quad \frac{f_{a_i}}{\aC[i,\gamma]} \sim 10^8\ {\rm GeV},
\quad m_{a_i} \lesssim 10^{-9} \div 10^{-10}\ {\rm eV}\,, \nn
\ee
hinting at $\aC[i,\gamma]/\mC[i,e] \sim 10$, which we shall review in appendix \ref{APP:CONSTRAINTS}.
A summary of constraints and hints is given in figure \ref{FIG:Constraints}.

As a result of these constraints and hints, the QCD axion is necessarily associated with a very high energy scale,
and so it is natural to search for it in ultra-violet completions of the Standard Model (SM) such as string theory.
Indeed, it has long been known that the low-energy effective field theory (EFT) of string compactifications predicts
natural candidates for the QCD axion~\cite{Witten:1984dg,Joe,Svrcek:2006yi,Conlon:2006ur,Choi:2006qj},
often even an `axiverse'~\cite{Arvanitaki:2009fg} containing
many additional light ALPs whose masses are logarithmically hierarchical.

Despite all of these promising considerations, it has been very hard to construct explicit
string theoretic examples with a successful QCD axion candidate, let alone additional light
ALPs. In order to understand the origins of these difficulties, we focus on type IIB flux compactifications
since this is the theory where moduli stabilisation is best understood.
Let us summarise the main features and problems encountered in the study of closed string axions from type IIB compactifications:
\begin{enumerate}
\item The low-energy spectrum below the compactification scale generically contains many axion-like particles
which are the Kaluza-Klein zero modes of antisymmetric form fields belonging to the massless spectrum of the bosonic type IIB string.
Their number is related to the topology of the internal manifold, namely to the number of harmonic forms
which can be defined on a Calabi-Yau. Given that this number is generically of the order of hundreds,
we expect the low-energy theory to be populated by many axion-like particles.
These are all closed string axions which live in the bulk. On top of them there can also be open string
axions living on a brane but their presence is more model-dependent, and so
we shall focus just on closed string axions.

\item Type IIB string theory compactified on a Calabi-Yau three-fold
gives rise to an $\mc{N}=2$ 4D EFT which is not chiral.
An $\mc{N}=1$ EFT can be obtained via an orientifold projection which projects out many degrees of freedom of the low-energy theory.
The kind of axions which are projected out depends on the particular orientifold projection considered,
but certainly several axions are removed from the low-energy spectrum.

\item Each axion comes with a real scalar field, the corresponding `saxion', forming a
complex field which is the lowest component of an $\mc{N}=1$ chiral superfield. The real part of this field is the
saxion whereas the axion is the imaginary part. The saxions are moduli which parameterise the size
of the extra dimensions and their VEV determines crucial features of the EFT like the gauge couplings and the mass spectrum.
It is then of primary importance to develop a potential for these moduli also because, if massless, they would mediate long-range
unobserved fifth-forces. Moreover the moduli may suffer from the famous cosmological moduli problem (CMP)
either decaying after Big-Bang Nucleosynthesis (BBN) has begun or, if stable, overclosing the Universe.
Typically this is solved if they have masses larger than $\mc{O}(10)$ TeV.

\item The mechanism which stabilises the moduli, giving them a mass of the order $m_{\rm mod}\gtrsim 10$ TeV,
might also develop a potential  for the axions rendering them too heavy.
In fact, the axions enjoy a shift symmetry which forbids the presence of axion-dependent operators
but only at the perturbative level since this symmetry is broken by non-perturbative effects.
Therefore if the saxions are fixed perturbatively, the axions do not develop any potential,
whereas if the moduli are fixed by non-perturbative effects, the axions might
obtain a mass of the same order of magnitude. This is indeed the case for KKLT \cite{KKLT} scenarios
where the K\"ahler moduli are fixed by assuming that the superpotential has non-perturbative contributions
for each K\"ahler modulus. All the corresponding axions would then have the same mass as the K\"ahler moduli,
which is generically of the order the gravitino mass, rendering all of them too heavy.

\item In type IIB string compactifications, the visible sector (the MSSM or GUT-like generalisations thereof)
is built via space-time filling D7-branes wrapping internal four-cycles whose volumes are measured by the saxions
(the so-called K\"ahler moduli). Each stack of $N$ D7-branes supports a $U(N)=SU(N)\times U(1)$ field theory
where the $U(1)$ factor is generically anomalous. This anomaly is cancelled via the Green-Schwarz mechanism
which induces a mass term for the $U(1)$ gauge bosons.
The gauge boson masses generated in this way may be as small as an meV \cite{HiddenPhotons} for a low string scale,
but will often be much heavier than observable physics.
This is a standard St\"uckelberg mass term since the Abelian gauge boson is made massive by eating the axion
which is the imaginary part of the K\"ahler modulus parameterising the size of the wrapped four-cycle.
Thus all the axions which are eaten disappear from the low-energy EFT.

\item Even if some axions are not projected out by the orientifold projection, do not
become too heavy due to non-perturbative moduli stabilisation and are not eaten
by anomalous $U(1)$s, it is very hard to obtain axion decay constants
$f_a \lesssim 10^{12}$ GeV, where this bound comes from
the requirement in standard thermal cosmology of avoiding the overclosure of the Universe by axion CDM
produced by the misalignment mechanism (see appendix \ref{APP:COSMIC} for a brief review).
In fact, in many string compactifications the QCD axion and other ALPs generically
have a larger decay constant of the order
the GUT scale $M_{\rm GUT}\sim 10^{16}$ GeV.\footnote{This is true not just for type IIB strings but also for
heterotic orbifolds~\cite{Choi:2006qj} or M-theory compactifications~\cite{Acharya:2010zx}.}
This is not necessarily a phenomenological disaster if the axions are diluted by the late out-of-equilibrium decay
of some heavy moduli, or have their initial misalignment angles tuned small. Nonetheless only ALPs with an intermediate scale decay constant allow
a possible observation of axion related effects in astrophysics, cosmology and even in the laboratory.
\end{enumerate}

At present, the best realisation of the axiverse from type IIB string compactifications
is based on the generation of a single non-perturbative correction to the superpotential
which fixes all the $h^{1,1}$ K\"ahler moduli plus one axion combination, leaving
all the other $h^{1,1}-1$ axions massless at leading order \cite{Acharya:2010zx}.
These axions then develop a potential at subleading order via higher order
instanton effects. In order to fix the moduli within the regime of validity
of the EFT, the tree-level superpotential $W_0$ has to be fine-tuned very small, $W_0\ll 1$,
of the order the leading non-perturbative correction. This effect
can be generated by a Euclidean D3-brane instanton or gaugino condensation
on D7-branes wrapping an `ample divisor' which is given by a combination
 of all the K\"ahler moduli with positive coefficients.
A microscopic realisation of this interesting scenario based on a Calabi-Yau example with explicit brane set-up and fluxes
has not been found yet due to the difficulty to find a rigid ample divisor
and to avoid chiral intersections between the instanton and the visible sector.

In this paper we shall present a different realisation of the axiverse
from type IIB strings which is valid in the natural regime $W_0 \sim \mc{O}(1)$.
It is based on a moduli stabilisation mechanism, the so-called LARGE Volume Scenario (LVS),
which exploits both perturbative and non-perturbative effects
and is characterised by an exponentially large volume of the extra dimensions \cite{LVS,K3Stab}.

Explicit LVS examples of globally consistent Calabi-Yau constructions with magnetised D-branes and chirality
have been recently derived in \cite{arXiv:1110.3333}. The same paper presents a general
discussion of various tensions between brane fluxes and moduli stabilisation
(chirality \emph{vs} non-vanishing non-perturbative effects, cancellation of Freed-Witten
anomalies \emph{vs} generation of more than one non-perturbative contribution, D-term induced
shrinking of the cycles supporting the visible sector) showing how
the LARGE Volume Scenario provides a very attractive solution to these problems.
Then ref. \cite{arXiv:1110.3333} outlined a general strategy to fix the moduli
within the LVS. We shall now show how this moduli fixing mechanism gives rise to an axiverse:
\begin{itemize}
\item In the LVS, usually all the $h_{1,1}^-$ moduli are projected out by
the orientifold involution so that $h_{1,1}^+=h_{1,1}$.

\item The only condition to realise the LVS is the presence
of a del Pezzo four-cycle $\tau_{\rm dP}$ supporting a single non-perturbative effect whose interplay
with the leading order $\alpha'$ correction to the EFT naturally yields
an exponentially large internal volume $\vo \sim W_0 \,e^{2\pi \tau_{\rm dP}/n}$. Notice that the rigidity
of a del Pezzo divisor guarantees the generation of this non-perturbative effect,
and the fact that the LVS does not require more than one non-perturbative contribution implies that
there is no problem with the cancellation of Freed-Witten anomalies.
Moreover, a del Pezzo divisor is intrinsically `local', in the sense that
a basis can always be found where this four-cycle enters the volume in a
completely diagonal way. Therefore it is easy to decouple this modulus
from the one supporting the visible sector, avoiding problems arising from chiral intersections.
The final upshot is that this blow-up mode does not appear in the D-terms
and the corresponding axion is very heavy, of the order of the gravitino mass.

\item The D-terms are the leading order effects in an expansion in inverse powers of
the huge internal volume and they might in principle generate a potential for $h^{1,1} - 1$ moduli
via the flux induced generation of moduli-dependent Fayet-Iliopoulos (FI) terms.
However, if this is the case, the only solution to $D=0$ in the absence of
SM singlets which can have non-zero VEVs, is the one where all the cycles shrink to zero size.
This can be avoided if the FI-terms depend only on $d$ moduli with $d< h^{1,1} - 1$,
which can be obtained by an appropriate choice of brane set-up and fluxes.
Thus exactly $d$ axions are eaten up, corresponding to the $d$ combinations of K\"ahler moduli fixed
by the D-terms.

\item All the remaining $n_{\rm ax} = h^{1,1} -1 - d\,$ flat directions
have to be fixed perturbatively via $\alpha'$ or $g_s$ effects.
As we have already pointed out, the volume
is fixed by the leading order $\alpha'$ effects whereas all the other cycles
are stabilised by string loops.

\item The number of massless axions is $n_{\rm ax}$ which
for an arbitrary Calabi-Yau with $h^{1,1}\sim\mc{O}(100)$ is likely to be very large.
Subleading higher-order non-perturbative effects then slightly lift the axion directions giving
rise to the `LVS axiverse'.
\end{itemize}
As explained in \cite{arXiv:1110.3333}, the LARGE Volume Scenario requires $n_{\rm ax}\geq 2$,
implying that the simplest model involves at least two light axions. If $n_{\rm ax}=1$, this
remaining flat direction would parameterise the overall volume and
be constrained to be fixed at small size by the requirement of
obtaining a visible sector gauge coupling $g_{\rm vs}$ of the correct order of magnitude,
given that the D-term relations force the volume to scale as the visible sector cycle:
$\vo^{2/3}\sim \tau_{\rm vs}\sim g_{\rm vs}^{-2}$.
Hence the volume would be rather small in string units and the tree-level
superpotential $W_0$ fine-tuned small otherwise the relation $\vo \sim W_0 \,e^{2\pi \tau_{\rm dP}/n}$
would yield an exponentially small gauge coupling.
However, if $n_{\rm ax}\geq 2$, only one of these flat directions is forced to be small
to obtain the correct value of $g_{\rm vs}$ while the others can be exponentially large allowing a
standard realisation of the LVS for $W_0\sim\mc{O}(1)$.

Clearly, providing a string realisation of the axiverse is not enough if none of
these light axions can behave as the QCD axion. This depends mainly on
the strength of the corresponding axion coupling to QCD which, in turn,
depends on the topology of the Calabi-Yau three-fold and the choice of brane set-up
and fluxes. In fact, if the visible sector wraps a small rigid divisor which
does not intersect the cycles controlling the size of the large overall volume,
the decay constant of the corresponding axion is of the order the string
scale $f_a\sim M_s/\sqrt{4\pi}$ due to the local nature of this interaction. In all the other
cases, this interaction is non-local resulting in axion couplings
suppressed by the Planck scale.
We shall discuss the consequences of this in section \ref{SEC:MIXINGCPDM}
where we shall find that, in the cases when non-local axions
can play the r\^ole of the QCD axion, their decay constants are
of the order $f_a \sim M_{\rm GUT} \sim 10^{16}$ GeV.

Given that by dimensional reduction the string scale can be expressed as $M_s = M_P / \sqrt{4\pi\vo}$,
exponentially large values of the volume can naturally yield a very low string scale.
The gravitino mass is instead 
$m_{3/2} = \sqrt{g_s/2} W_0 M_P/\vo$
, and so TeV-scale soft-terms $M_{\rm soft}\sim m_{3/2} \sim 1$ TeV
can be obtained for $\vo \sim 10^{14}$ if $W_0\sim\mc{O}(1)$ and $g_s\simeq 0.1$.
Consequently the string scale turns out to be intermediate, $M_s \sim 5\cdot 10^{10}$ GeV, or a few orders of magnitude larger as we moderately tune $W_0$ smaller,
providing an axion decay constant in the classic phenomenological window
for the case of a local interaction \cite{Joe}.  In this paper we shall describe how
to obtain a QCD axion candidate plus additional ALPs
which are not eaten by any anomalous $U(1)$s and exhibit these interesting phenomenological features.

We stress that LVS models with an intermediate string scale involve light moduli which may suffer from the CMP.
For example, the mass of the volume mode $\vo$ is $m_{\vo}= m_{3/2}/\vo^{1/2}\sim 0.1$ MeV.
In order to solve the CMP, these light moduli could be diluted either by the entropy released by the decay
of some heavy moduli \cite{Acharya:2008bk} or by a late time inflationary period caused by thermal effects \cite{K3norm}.
If a non-local axion with a GUT-scale decay constant plays the r\^ole of the QCD axion,
this dilution mechanism would also dilute this axion,
potentially allowing it to evade the present bounds from dark matter overproduction.
Consequently, non-local axions might form dark matter and be detectable via future molecular interferometry experiments (see appendix \ref{APP:COSMIC}).
On the other hand, if the QCD axion is realised as a local axion with
an intermediate scale decay constant at the lower end discussed above, $f_a\sim M_s/\sqrt{4\pi} \sim 10^{10}$ GeV, this opens up
the intriguing possibility to detect it directly in the laboratory in
the next generation of LSW experiments. Furthermore,
these local axions can provide a nice explanation to several astrophysical puzzles,
such as the anomalous transparency of the Universe for TeV photons
and the anomalous cooling of white dwarfs; all of these hints and constraints are summarised in appendix \ref{APP:CONSTRAINTS}.
Moreover, it is worth stressing that in this case axions do not form a sizeable contribution to dark matter, since 
the latter requires a decay constant in the upper end discussed above, $f_a\sim M_s/\sqrt{4\pi} \sim 10^{12}$ GeV, 
cf. appendix \ref{APP:COSMIC}.

We shall illustrate our general claims with four Calabi-Yau examples with stabilised moduli
and an explicit choice of brane set-up and fluxes. The main features of each example are:
\begin{enumerate}
\item GUT-like model with $n_{\rm ax}=1$ non-local axion with $f_a \sim M^{\rm 10D}_{\rm KK}/(4\pi) \simeq 5\cdot 10^{14}$ GeV that behaves as
the QCD axion in the presence of a dilution mechanism.
\item MSSM-like model with $n_{\rm ax}=2$ non-local axions, $a_f$ with $f_{a_f} \simeq 10^{16}$ GeV,
and $a_b$ with $f_{a_b}\sim M^{\rm 6D}_{\rm KK}/(4\pi) \simeq 5$ TeV. The axion $a_f$ can be the QCD axion
in the presence of a dilution mechanism, whereas $a_b$ is a very light and almost decoupled ALP ($C_{bj} \ll 1$).
\item Chiral model with $n_{\rm ax}=3$ light axions: one local axion $a_s$ with $f_{a_s} \sim M_s/\sqrt{4\pi} \simeq 10^{10}$ GeV,
and two non-local axions, $a_f$ and $a_b$, with $f_{a_f} \sim f_{a_b} \sim M^{\rm 10D}_{\rm KK}/(4\pi) \simeq 10^8$ GeV.
The local axion $a_s$ provides the first explicit example of a closed string QCD axion which is not eaten
by any anomalous $U(1)$ and has an intermediate scale decay constant. The two non-local axions are very light and almost decoupled ALPs ($C_{fj}, C_{bj} \ll 1$).
\item Chiral model with more than one light local axion with $f_a \sim M_s/\sqrt{4\pi}\simeq 10^{10}$ GeV.
\end{enumerate}
We stress again that in this analysis we focused only on axions
which come from closed strings since only this case is model-independent.
In fact, axion-like particles can also be realised as open strings which live on D-branes
but this realisation is more model-dependent,
since it depends crucially on the particular D-brane construction considered.

This paper is organised as follows.
Section \ref{IIBaxions} presents a simple review of the low-energy 4D EFT of type IIB string theory
compactified on Calabi-Yau orientifolds paying particular attention to axions and their
interplay with massive anomalous $U(1)$s. In section \ref{LVSaxions} we describe
in detail how the LVS moduli stabilisation gives rise to an axiverse
giving formulae for the sizes of decay constants, couplings to gauge and matter fields for Swiss-cheese and fibred Calabi-Yau compactifications.
The explicit LVS examples mentioned above are analysed in section \ref{LVSexamples},
whereas we present our conclusions in section \ref{conclusions}.
Due to this success to derive the axiverse from type IIB strings, in appendix \ref{sec:pheno},
we work out the low-energy effective action of many ALPs
computing in particular their couplings to photons and electrons which are particularly relevant for phenomenology.
In appendix \ref{APP:CONSTRAINTS}
we summarise the results of axion searches in astrophysics, cosmology, and laboratory
arguing that various astrophysical hints point to an axion decay constant at the intermediate scale.
Finally, in appendix \ref{SEC:LOOPS} we compute the axionic couplings to matter at one-loop,
and in appendix \ref{GenfibCY} we discuss the axion decay constants for general fibred Calabi-Yau manifolds.

\section{Closed string axions from type IIB strings}
\label{IIBaxions}

In this section we will describe how axion-like fields arise in type IIB string theory models
and how to determine which of the axion-like fields are eaten by massive $U(1)$ gauge bosons.
The low-energy couplings, decay constants, and masses can be obtained by comparing the
EFT obtained in this section from dimensional reduction of the IIB string from 10 to 4 dimensions
with the general low-energy effective Lagrangian of axion-like particles~\eqref{leff} which
we shall derive in appendix \ref{sec:pheno}.

\subsection{Effective axionic action}

In type IIB flux compactifications on Calabi-Yau orientifolds,
the Standard Model or a suitable extension like a Grand Unified Theory (GUT) is localised on a stack of space-time filling D3 or D7 branes.
Pseudo-scalar axion-like fields $c^a$ and $c_\alpha$ arise as Kaluza-Klein zero modes of the Ramond-Ramond (RR) antisymmetric tensor fields $C_2$ and $C_4$~\cite{Grimm:2004uq,Jockers:2004yj} (for a short review, see the appendices
in~\cite{Goodsell:2009xc,HiddenPhotons}), that is as coefficients of the two- and four-form term, respectively,
in their expansion in harmonics of the Calabi-Yau orientifold:
\be
C_2 = \,c^a (x) \omega_a,\quad a = 1,...,h_-^{1,1} \qquad\text{and}\qquad
C_4 = \,c_\alpha (x) \tilde\omega^\alpha +...,\quad \alpha = 1,...,h_+^{1,1}. \nn
\ee
The axionic couplings to gauge field strengths $\propto F\wedge F$ arise then from the Kaluza-Klein reduction of
the Chern-Simons term in the action of the D-branes.

The effective Lagrangian including $U(1)$ gauge bosons $A_i$ and couplings to axions $c^a$ and $c_\alpha$ can be written as:
\begin{align}
\mc{L}&\supset - \frac{ e^{\Phi}}{4\vo^2}\left( dc^a + \frac{e^{-\Phi}}{2\pi} M_P A_i r^{ia} \right)
\mc{K}^{ab} \wedge \star \left( dc^b + \frac{e^{-\Phi}}{2\pi} M_P A_j r^{jb} \right)  \\
& + \frac{2\pi M_P^2}{\vo^2 (2\pi)^3} \,e^{-\Phi} A_i A_j r^{ia} r^{jb} \mc{K}^{ab}
-  \left( d c_\alpha + \frac{M_P}{\pi} A_i q_{i\alpha}\right) \frac{\mc{K}_{\alpha\beta}}{8}
\wedge \star \left( d c_\beta + \frac{M_P}{\pi} A_j q_{j\beta}\right) \nn\\
&+ \frac{M_P^2}{2(2\pi)^2}  A_i A_j q_{i\alpha}  \mc{K}_{\alpha\beta} q_{j\beta}
+\frac{1}{4\pi M_P}\left( r^{i\alpha} c_\alpha  + q_{ia} c^a \right) \tr(F\wedge F)  - \frac{r^{i\alpha}\tau_\alpha}{4 \pi M_P} \tr(F_i \wedge \star F_i) .\nn
\label{EQ:BIGAXIONCOUPLINGS}
\end{align}
The quantities in the above expression are defined as follows. $M_P$ is the reduced Planck mass
$(8\pi G_N)^{-1/2} \simeq 2.4 \cdot 10^{18}$ GeV, $\Phi$ is the axio-dilaton,
$\vo$ is the Einstein-frame volume of the internal space in string units, while $q_{iA}$ and $r^{iA}$ are the
couplings of the two-forms $\omega_A$ and four-forms $\tilde{\omega}^A$ to branes, ignoring any globally-trivial forms,
and are given by:
\begin{align}
q_{iA} \equiv& \,\ell_s^{-2} \int_{D_i} \omega_A \wedge \frac{\mc{F}}{2\pi}  = \ell_s^{-4} \int_{D_i} \omega_A \wedge \ell_s^2\mc{F}, \nn\\
r^{iA} \equiv& \,\ell_s^{-4} \int_{D_i} \tilde{\omega}^A =  \ell_s^{-4} \int \hat{D}_i \wedge \tilde{\omega}^A,\quad A=1,...,h^{1,1}=h^{1,1}_-+h^{1,1}_+,
\end{align}
where $\mc{F}$ is the gauge flux on $D_i$ which is a generic divisor whose Poincar\'e dual two-form can be expressed
in terms of the basis forms as $\hat{D}_i = r^{iA} \omega_A$ with $\ell_s^{-4} \int \omega_B \wedge \tilde{\omega}^A = \delta_B^A$.
Writing the K\"ahler form as $J = t^A \omega_A$, we have:
\begin{eqnarray}
&&K = -2\ln\vo, \qquad \tau_A \equiv \frac{1}{2}\,k_{ABC} t^A t^C, \qquad
\mc{T}_{AB} \equiv k_{ABC} t^C,
\label{EQ:METRICS} \\
&&\mc{K}_{AB} \equiv \frac{\partial^2 K}{\partial\tau_A\partial\tau_B} = \frac{t_A t_B}{2\vo^2}
- \frac{(\mc{T}^{-1})_{AB}}{\vo}, \qquad
\mc{K}^{AB} = \tau_A \tau_B - \mc{T}_{AB}\vo, \nn
\end{eqnarray}
where $k_{ABC} = \ell_s^{-6} \int \omega_A \wedge \omega_B \wedge \omega_C$ are the triple intersection numbers.

The above expression makes clear that some axions are eaten by the $U(1)$ gauge symmetries in the model, while the axionic couplings to other gauge groups are inversely proportional to the Planck mass. However, the physical axion decay constants will be determined by the kinetic terms, which - particularly for large volumes - can differ from unity by many orders of magnitude. In this work, we are interested in ALPs whose decay constants may be equal to or less than the string scale, so that their couplings may be equal to or enhanced relative to the QCD axion; an examination of the above action and the expressions for the metrics (\ref{EQ:METRICS}) will make clear that the orientifold-odd fields $c^a$ will have decay constants equal to or parametrically greater than the string scale (similar to the case for $U(1)$ masses~\cite{Goodsell:2009xc}) and so we will neglect them, or focus on models with $h^{1,1}_- =0$. In this case
the K\"ahler moduli are defined as:
\be
T_\alpha = \tau_\alpha + {\rm i} \,c_\alpha \qquad\text{with}\qquad \alpha=1,...,h^{1,1}_+=h^{1,1}\,.
\ee
However, it should be stated that the analysis of the odd two-form fields $c^a$ would proceed almost identically to that of the even fields.

\subsection{Canonical normalisation}

From (\ref{EQ:BIGAXIONCOUPLINGS}) it can be seen via topological charge quantisation that the axions have (unconventionally for axions, but conventionally for K\"ahler moduli) periods equal to \emph{integer} multiples of $M_P$; this derives from the quantisation of the Ramond-Ramond forms from which they descend. We diagonalise the axions in two steps: first we make an orthogonal transformation, after which the new fields are still periodic under integer shifts by $M_P$. This leads to kinetic terms:
\be
\mc{L} \supset - \frac{1}{8} \lambda_\alpha \partial_\mu c_\alpha^\prime \partial^\mu c_\alpha^\prime\,, \nn
\ee
where $\lambda_\alpha$ are the eigenvalues of $\mc{K}_{\alpha \beta} $. We then canonically normalise by defining $c_\alpha^\prime \equiv 2 \lambda_\alpha^{-1/2} a_\alpha$; the new fields are then periodic under shifts by $\frac{1}{2} \lambda_\alpha^{1/2} M_P$ - this allows us to define the decay constants:
\begin{align}
f_\alpha \equiv \frac{\sqrt{\lambda_\alpha} M_P}{4\pi}
\end{align}
so that the fields have the conventional periodicity $a_\alpha = a_\alpha + 2 \pi f_\alpha$.

Some of these fields will be eaten by a $U(1)$ gauge boson and thus cannot enter into a term in the superpotential of the form $W \supset A \,e^{-aT}$ where $A$ is a constant (not a function of matter fields) due to gauge invariance - the shift symmetry is gauged. Hence these axions are in some sense orthogonal to their light ALP brothers, and we must first remove them from our consideration.
To find the massless axions coupling to a given gauge group, we must find the null vectors to eliminate the $d c \wedge \star A$ couplings. Hence the full transformation, leaving only the ungauged axions, is:
\begin{eqnarray}
&&c_\alpha = 2\, a_{\gamma} V_{\gamma \beta} \CC_{\beta \alpha}, \qquad
\CC_{\gamma' \alpha} \mc{K}_{\alpha \beta} \CC_{\beta \delta'}^T = \delta_{\gamma' \delta'}, \qquad
\CC_{\gamma' \alpha}  \CC_{\alpha \delta'}^T = \lambda_{\gamma'}^{-1} \delta_{\gamma' \delta'}, \nn\\
&& V_{\gamma \gamma^\prime } \CC_{\gamma^\prime \beta} \mc{K}_{\beta \alpha } q^T_{\alpha i} = 0, \qquad
(VV^T)_{\alpha \beta} = \left\{ \begin{array}{ll} \delta_{\alpha \beta} & a\ {\rm ungauged}
\\ 0 &  a\ {\rm gauged}\end{array} \right.,
\end{eqnarray}
i.e. the matrix $V$ is the set of null vectors of the mass couplings.
When both axions and gauge fields are canonically normalised we have a term $\mc{L} \supset \frac{g_i^2}{2\pi M_P} \, \tr(F_i \wedge F_i) \,r^{i\alpha} \CC^T_{\alpha \beta} V^T_{\beta \gamma} a_{\gamma}$, so in the form of couplings (\ref{EQ:DEFCAGG}) this gives:
\begin{align}
\frac{f_{a_j}}{\aC[j,i]} =&  \frac{1}{8\pi } \frac{M_P}{r^{j\alpha} \CC^T_{\alpha \beta} V^T_{\beta i}}  \times
\left\{ \begin{array}{ll} 1/2 & U(1) \\ 1 & SU(N)\end{array} \right. .
\end{align}
Note that for $U(1) \supset U(N)$ we normalise the generators so that $\tr (T_{U(1)} T_{U(1)}) = 1$, which means that $T_{U(1)} = \mathds{1}/\sqrt{N}$ and $\tr(T_{U(1)}) = \sqrt{N}$. There is an alternative notation much used in the literature, involving physical couplings $g_{ji}$:
\begin{align}
\mc{L} \supset& - \frac{1}{4} a_j g_{ji} F_{i,\mu\nu}^b \tilde{F}_{i}^{b,\mu\nu} \nn\\
\Rightarrow g_{ji} =& -\frac{2 g_i^2}{\pi M_P} \, r^{i\alpha} \CC^T_{\alpha \beta} V^T_{\beta j} \times
\left\{ \begin{array}{ll} 1 & U(1) \\ \frac{1}{2} & SU(N)\end{array} \right.
= -\frac{4}{ M_P} \frac{1}{\tau_i} \,r^{i\alpha} \CC^T_{\alpha \beta} V^T_{\beta j}\,,
\label{EQ:STRINGAXIONCOUPLINGS}
\end{align}
where $\tau_i$ is the volume of the four-cycle wrapped by the brane in string units.

It is often convenient to make a redefinition of the axions choosing a basis where $r^{i\alpha} a_\alpha \equiv \{\hat{a}_i, 0\}$ depending on whether the cycle is wrapped by a brane or not. Under this transformation, we must not forget to transform the charges appropriately.

With these couplings, we can work out which axions will obtain a mass through any non-perturbative effects - such as coupling to a gaugino condensate or D-brane instanton - since these axions cannot be gauged (for masses that arise through superpotential terms of the form $W \supset A\,e^{-bT}$ where $A$ is independent of matter fields). In determining the light axion and ALPs we must therefore make a further redefinition of the fields to eliminate the heavy massive ones.

\subsection{Volume scaling of axion decay constants and couplings to gauge bosons}
\label{AxionDC}

Being the imaginary component of the moduli, whose interaction with matter is gravitational,
we expect the physical couplings of axions to matter fields to be suppressed by the Planck scale.
This applies for moduli which are able to propagate throughout the Calabi-Yau volume.
However, for axion fields that only propagate over so-called `local' cycles (such as small rigid divisors),
i.e. four-cycles which do not intersect any divisor which controls the overall volume,
the natural coupling strength is the string scale.
Using conventional definitions, the physical coupling is inversely proportional to the axion decay constant $f_a$
and proportional to the square of the gauge coupling $g^2\simeq \tau_a^{-1}$. However, this argument only holds for the scaling with respect to the overall volume, rather than the volumes of small cycles - we expect:
\begin{align}
f_a\simeq \left\{ \begin{array}{ll} M_P/\tau_a & \quad{\rm non-local\ axion}
\\ M_P/\sqrt{\vo} \simeq M_s & \quad{\rm local\ axion} \end{array} \right.
\label{EQ:LOCALITY}
\end{align}

As pointed out in section \ref{intro}, local interactions open up the possibility to obtain axion decay
constants which are much lower than the Planck or GUT scale, since their size is controlled by $M_s$ instead of $M_P$ \cite{Joe}.
Hence if $M_s \sim 10^{10}$ GeV, $f_a$ turns out to be intermediate in a regime which is particularly interesting
for phenomenology. Notice that this value of $M_s$ can be naturally obtained
for exponentially large values of the overall volume which characterise LVS scenarios. A range of values of $f_a$ are then possible depending upon the exact numerical coefficients, the value of the gravitino mass and the tuning of $W_0$;
$M_s \simeq \sqrt{\frac{M_P m_{3/2}}{W_0}}$ so values of decay constants spanning the classic axion window ($10^9 \div 10^{12}$ GeV) are possible without fine tuning.

On the other hand, the decay constants of non-local axions may be even smaller - since $\tau_a$ in this case can scale as $\vo^{2/3}$ or even $\vo$ for anisotropic compactifications.\footnote{For anisotropic compactifications we show in appendix \ref{GenfibCY} that there is only one axion that has a Planck-sized decay constant - that corresponding to the fibre - with the others being much smaller.} This has important consequences for the amount of dark matter that they can produce, which we shall discuss in section \ref{SEC:COSMO}.

We shall now illustrate these general claims and derive the physical couplings to gauge bosons, $C_{ij}$, for explicit Swiss-cheese compactifications
and in section \ref{fibCY} for fibred Calabi-Yau examples.

\subsubsection{Swiss-cheese Calabi-Yau manifolds}
\label{ScCY}

In the simplest Swiss-cheese case, the Calabi-Yau volume takes the form \cite{SwissCheese}:
\be
\vo = \frac{1}{9\sqrt{2}} \left( \tau_b^{3/2} - \tau_s^{3/2}\right),
\label{volSwiss}
\ee
where $\tau_b$ is a large cycle controlling the overall size of the internal manifold whereas the small divisor $\tau_s$ is a local blow-up mode.
Writing $\epsilon^2 \equiv \tau_s/\tau_b\ll 1$, and working at leading order in a large volume expansion, we find:
\begin{align}
\mc{K}_0 =& \,\frac{3}{2 \tau_b^{3/2}\sqrt{\tau_s}} \left(\begin{array}{cc}
 2\,\epsilon & -3\, \epsilon^2   \\
 -3\, \epsilon^2   &  1
\end{array} \right).
\end{align}
The eigenvalues of this are $3\,\tau_b^{-2}, \frac{3}{2} \,\tau_b^{-3/2} \tau_s^{-1/2}$,
and so the decay constants of the relevant axions will be:
\begin{align}
f_{a_b} = \frac{\sqrt{3}}{4\pi} \frac{M_P}{\tau_b}\simeq \frac{M_P}{4\pi\vo^{2/3}} \simeq \frac{M^{\rm 10D}_{\rm KK}}{4\pi}\,, \qquad
f_{a_s} = \frac{1}{\sqrt{6}\left(2\tau_s\right)^{1/4}}  \frac{M_P}{4\pi\sqrt{\vo}}\simeq \frac{M_s}{\sqrt{4\pi}\tau_s^{1/4}}\,,
\label{fasSC}
\end{align}
in agreement with (\ref{EQ:LOCALITY}). In (\ref{fasSC}), $M^{\rm 10D}_{\rm KK}$ denotes the 10D Kaluza-Klein
scale:
\be
M^{\rm 10D}_{\rm KK}\simeq \frac{M_s}{\vo^{1/6}} \simeq \frac{M_P}{\vo^{2/3}}\,.
\ee

The diagonalising matrix is given by:
\begin{align}
\CC^T =&  \,\tau_b \left(
\begin{array}{cc}
 -\frac{1}{\sqrt{6}}\left(\frac{1-9\epsilon^3}{\sqrt{1- \frac 92 \epsilon^3}}\right) &
\sqrt{3} \,\epsilon^{3/2} \\
 -\epsilon^2{\sqrt{\frac{3}{2}\left(1-\frac 92 \epsilon^3\right)}} &
 \frac{1}{\sqrt{3}}\, \epsilon^{1/2}
\end{array}
\right),
\end{align}
and the matrix $r^{i\alpha}$ is the identity. Hence the original axion fields $c_b$ and $c_s$ can be
written in terms of the canonically normalised fields $a_b$ and $a_s$ as:
\begin{eqnarray}
\frac{c_b}{\tau_b} &\simeq & - \frac{1}{\sqrt{6}}\,a_b +\sqrt{3}\,\epsilon^{3/2} a_s \simeq
\mc{O}\left(1\right)\,a_b +\mc{O}\left(\tau_s^{3/4}\,\vo^{-1/2}\right) \,a_s\,, \\
\frac{c_s}{\tau_s} &\simeq& -\sqrt{\frac{3}{2}}\,a_b
+\frac{1}{\sqrt{3}\,\epsilon^{3/2}} \,a_s \simeq \mc{O}\left(1\right)\,a_b
+\mc{O}\left(\tau_s^{-3/4}\,\vo^{1/2}\right) a_s\,,
\end{eqnarray}
where we have used $\vo\simeq \tau_b^{2/3}$ and we have neglected subleading corrections proportional to $\epsilon$.
Thus the axionic couplings to gauge bosons scale as:
\begin{align}
\mc{L} \supset&  \frac{c_b }{M_P}\, g_b^2\,\tr(F_b \wedge F_b)  +  \frac{c_s}{M_P} \,g_s^2\,\tr(F_s \wedge F_s)  \nn\\
\simeq & \left[\mc{O}\left(\frac{1}{M_P}\right)\,a_b +\mc{O}\left(\frac{\tau_s^{3/4}}{\vo^{1/2} M_P}\right) \,a_s \right] \tr(F_b \wedge F_b) \nn \\
+ & \left[\mc{O}\left(\frac{1}{M_P}\right)\,a_b +\mc{O}\left(\frac{1}{\tau_s^{3/4}\,M_s}\right) \,a_s \right] \tr(F_s \wedge F_s).
\end{align}
The volume scaling of the leading terms of these couplings are exactly in accordance with (\ref{EQ:LOCALITY}).
The constants describing the coupling of the axions to gauge bosons instead become:
\bea
\aC[b,b] &\simeq & g_b^{-2} \frac{f_{a_b}}{M_P} \simeq \mc{O}\left(1\right), \qquad
 \aC[s,b]\simeq g_b^{-2} \frac{f_{a_s}\tau_s^{3/4}}{\vo^{1/2} M_P} \simeq \mc{O}\left(\epsilon\right)
\simeq \mc{O}\left(\vo^{-1/3}\right), \nn \\
\aC[b,s] &\simeq & g_s^{-2} \frac{f_{a_b}}{M_P} \simeq \mc{O}\left(\epsilon^2\right)
\simeq \mc{O}\left(\vo^{-2/3}\right), \qquad
\aC[s,s]\simeq g_s^{-2}\frac{f_{a_s}}{\tau_s^{3/4}\,M_s}\simeq \mc{O}\left(1\right)\,. \nn
\eea

\subsubsection{Fibred Calabi-Yau manifolds}
\label{fibCY}

Now let us consider fibred Calabi-Yau manifolds with a del Pezzo divisor.
The volume is given by \cite{FibCY}:
\be
\vo = t_b t_f^2 + t_s^3 = t_b \tau_f + t_s^3\,,
\label{volFib}
\ee
where $t_b$ is the size of the $\mathbb{P}^1$ base, $\tau_f=t_f^2$ the volume of the K3 or $T^4$ fibre and $\tau_s=3 t_s^2$
parameterises the volume of a small del Pezzo four-cycle. The volume of the divisor including the base of the fibration is instead
given by $\tau_b= 2 t_b t_f = 2 t_b \sqrt{\tau_f}$. We treat more general fibred examples in appendix \ref{GenfibCY}.

In the  limit $t_s\tau_f\gg \tau_s^{3/2}$, the eigenvalues of the mixing matrix are to leading order
$\tau_f^{-2}, 2\,\tau_b^{-2}, \frac{1}{2\sqrt{3}}\,\vo \sqrt{\tau_s}$, and so the decay constants are:
\be
f_{a_f} = \frac{1}{4\pi} \frac{M_P}{\tau_f}\,,
\qquad f_{a_b} = \frac{\sqrt{2}}{4\pi} \frac{M_P}{\tau_b}\,,
\qquad f_{a_s} = \frac{1}{\sqrt{2}\left(3 \tau_s\right)^{1/4}}  \frac{M_P}{4\pi\sqrt{\vo}}\simeq \frac{M_s}{\sqrt{4\pi}\tau_s^{1/4}}\,, \nn
\ee
again in agreement with (\ref{EQ:LOCALITY}): the first two axions are non-local. Note that the fibre axion has a potentially large decay constant in the anisotropic limit when $\tau_f \sim \tau_s \ll \tau_b$; we show in appendix \ref{GenfibCY} that
this is the only axion that can have a decay constant parametrically above the string scale.

We should, however, distinguish the decay constants from the physical couplings.
To determine these we can use the results of \cite{K3norm}, by which the original axion fields $c_b$, $c_f$ and $c_s$
can be written in terms of the canonically normalised fields $a_b$, $a_f$ and $a_s$ as:
\bea
\frac{c_f}{\tau_f} &\simeq & \mc{O}\left(1\right) a_f + \mc{O}\left(\tau_s^{3/2} \vo^{-1}\right) a_b
+\mc{O}\left(\tau_s^{3/4}\vo^{-1/2}\right) \,a_s\,,
\label{cf} \\
\frac{c_b}{\tau_b} &\simeq & \mc{O}\left(\tau_s^{3/2}\vo^{-1}\right)a_f + \mc{O}\left(1\right) a_b
+\mc{O}\left(\tau_s^{3/4}\vo^{-1/2}\right) \,a_s\,,
\label{cb} \\
\frac{c_s}{\tau_s} &\simeq& \mc{O}\left(1\right) a_f + \mc{O}\left(1\right)\,a_b +\mc{O}\left(\tau_s^{-3/4} \vo^{1/2}\right) a_s\,.
\label{cs}
\eea
The axionic couplings to gauge bosons following from (\ref{cf})-(\ref{cs}) scale as:
\begin{align}
\mc{L} \supset&  \frac{c_f }{M_P}\, g_f^2\,\tr(F_f \wedge F_f) + \frac{c_b }{M_P}\, g_b^2\,\tr(F_b \wedge F_b)
+ \frac{c_s}{M_P} \,g_s^2\,\tr(F_s \wedge F_s)  \nn\\
\simeq & \left[\mc{O}\left(\frac{1}{M_P}\right)\,a_f + \mc{O}\left(\frac{\tau_s^{3/2}}{\vo M_P}\right)\,a_b
+\mc{O}\left(\frac{\tau_s^{3/4}}{\vo^{1/2} M_P}\right) \,a_s \right] \tr(F_f \wedge F_f) \nn \\
+ & \left[\mc{O}\left(\frac{\tau_s^{3/2}}{\vo M_P}\right)\,a_f + \mc{O}\left(\frac{1}{M_P}\right)\,a_b
+\mc{O}\left(\frac{\tau_s^{3/4}}{\vo^{1/2} M_P}\right) \,a_s \right] \tr(F_b \wedge F_b) \nn \\
+ & \left[\mc{O}\left(\frac{1}{M_P}\right)\,a_f+\mc{O}\left(\frac{1}{M_P}\right)\,a_b
+\mc{O}\left(\frac{1}{\tau_s^{3/4} M_s}\right) \,a_s \right] \tr(F_s \wedge F_s).
\end{align}
Notice that volume scaling of the leading terms of these couplings are exactly in accordance with (\ref{EQ:LOCALITY}).
The constants describing the coupling of the axions to gauge bosons instead become:
\bea
\aC[f,f] &\simeq & \mc{O}\left(1\right), \qquad
\aC[b,f] \simeq \mc{O}\left(\frac{(\tau_s\tau_f)^{3/2}}{\vo^2}\right),
\qquad \aC[s,f]\simeq \mc{O}\left(\frac{\tau_f\sqrt{\tau_s}}{\vo}\right) \nn \\
\aC[f,b] &\simeq & \mc{O}\left(\left(\frac{\tau_s}{\tau_f}\right)^{3/2}\right), \qquad
\aC[b,b]\simeq \mc{O}\left(1\right),\qquad
\aC[s,b]\simeq \mc{O}\left(\sqrt{\frac{\tau_s}{\tau_f}}\right), \nn \\
\aC[f,s] &\simeq& \mc{O}\left(\frac{\tau_s}{\tau_f}\right),\quad
\aC[b,s]\simeq \mc{O}\left(\frac{\tau_s\sqrt{\tau_f}}{\vo}\right),
\quad \aC[s,s]\simeq \mc{O}\left(1\right)\,.
\label{Cs}
\eea
Let us now see how these quantities behave in the two different cases when the Calabi-Yau has an isotropic
or  anisotropic shape.
\begin{itemize}
\item {\bf Isotropic case:} in the isotropic limit $t_b \sim \sqrt{\tau_f} \sim \vo^{1/3}$ the decay constants have the same volume scaling as the Swiss-cheese example:
\be
f_{a_f}\simeq f_{a_b}\simeq \frac{M_P}{4\pi\vo^{2/3}} \simeq \frac{M^{\rm 10D}_{\rm KK}}{4\pi}\,,
\qquad f_{a_s}\simeq \frac{M_s}{\sqrt{4\pi}\tau_s^{1/4}}\,,
\ee
and the anomaly coefficients (\ref{Cs}) reduce to:
\bea
\aC[f,f] &\simeq & \aC[b,b] \simeq \mc{O}\left(1\right), \quad
\aC[b,f] \simeq \aC[f,b] \simeq \mc{O}\left(\vo^{-1}\right),
\quad \aC[s,f]\simeq \aC[s,b]\simeq \mc{O}\left(\vo^{-1/3}\right) \nn \\
\aC[f,s] &\simeq& \aC[b,s]\simeq \mc{O}\left(\vo^{-2/3}\right),
\quad \aC[s,s]\simeq \mc{O}\left(1\right)\,.
\label{Csiso}
\eea
\item {\bf Anisotropic case:} in the anisotropic limit $t_b \sim \vo \gg \sqrt{\tau_f} \sim \sqrt{\tau_s}$ considered in \cite{HiddenPhotons,ADDstrings},
the decay constants become:
\be
f_{a_f}\simeq \frac{M_P}{4\pi\tau_s}\,,\qquad f_{a_b}\simeq \frac{M_P}{4\pi\tau_b} \simeq \frac{M^{\rm 6D}_{\rm KK}}{4\pi}\,,
\qquad f_{a_s}\simeq \frac{M_s}{\sqrt{4\pi}\tau_s^{1/4}}\,,
\ee
where $f_{a_b}$ is now scaling as the Kaluza-Klein scale of the 6D EFT given by:
\be
M^{\rm 6D}_{\rm KK} \simeq \frac{M_s}{\sqrt{t_b}} \simeq \frac{M_P}{\sqrt{\vo t_b}} \simeq \frac{M_P}{t_b t_f} \simeq \frac{M_P}{\tau_b}\,.
\ee
Moreover, the anomaly coefficients (\ref{Cs}) take the simplified form:
\bea
\aC[f,f] &\simeq & \aC[f,b] \simeq \aC[b,b]\simeq \aC[s,b]\simeq \aC[f,s] \simeq \aC[s,s]\simeq \mc{O}\left(1\right), \nn \\
\aC[s,f] &\simeq& \aC[b,s] \simeq \mc{O}\left(\vo^{-1}\right), \qquad
\aC[b,f] \simeq \mc{O}\left(\vo^{-2}\right).
\label{Csaniso}
\eea
\end{itemize}

\subsection{Axionic couplings to matter fields}
\label{AxionicCouplings}

After having computed the axionic couplings to gauge fields, let us now move to
the computation of the axionic couplings to matter fields. Written in terms of two-component Weyl spinors, these couplings take the form
\be
\mc{L} \supset \frac{\hat{X}^i_{\psi}}{f_{a_i}} (\psi \sigma^\mu \ov{\psi}) \partial_\mu a_i .
\ee
We discuss the low energy phenomenology of these couplings in appendix \ref{sec:pheno}, where we also explain
that we shall neglect `Yukawa' couplings of the form $a_i \psi \tilde{\psi}$ - since they arise from non-generic and model dependent non-perturbative terms.

In string theory, closed string axions couple generically to matter fields from higher-order K\"ahler potential terms, and can thus be derived from the effective supergravity action. In a globally supersymmetric theory, the axion couplings arise from the $i\theta \sigma^\mu \ov{\theta} \partial_\mu T_\alpha$ component of the superfield $T_\alpha = \tau_\alpha + {\rm i}\, c_\alpha + ...$:
\be
\int d^4 \theta  \, \Phi \ov{\Phi} \left(T_\alpha+\ov{T}_\alpha\right)  \supset  (\psi \sigma^\mu \ov{\psi}) \partial_\mu c_\alpha\,.
\ee
If we write for simplicity a diagonal matter K\"ahler metric as $\partial_{\Psi \ov{\Psi}} K \equiv \hat{K}_\Psi $ (the generalisation to non-diagonal cases being obvious) then
the moduli dependence of this gives us the couplings:
\be
\mc{L} \supset \int d^4 \theta  \, \hat{K}_i \Phi \ov{\Phi}
\supset \hat{K}_\Phi|_{\theta = \bar{\theta} = 0} i \psi \sigma^\mu \partial_\mu \ov{\psi}
+ \frac{\partial \hat{K}_\Phi}{\partial T_\alpha}\bigg|_{\theta = \bar{\theta} = 0} (\psi \sigma^\mu \ov{\psi}) \partial_\mu c_\alpha .
\ee
So in the physical basis (when we canonically normalise the matter fields) the coupling of axions to fermions is
$\partial_{T_\alpha} \left(\ln \hat{K}_\Phi\right)$.
Given that the above expression has to be gauge invariant, the axions can only appear
in the combination $\partial_\mu c_\alpha + \frac{M_P}{\pi} \,q_{\alpha} A_\mu$, and so we have additional anomalous gauge couplings.
However, in a \emph{local} supergravity theory, there is an additional contribution:
\be
\mc{L} \supset \hat{K}_\Phi i\psi \sigma^\mu \partial_\mu \ov{\psi}
+  \psi \sigma^\mu \ov{\psi} \partial_\mu c_\alpha \left( \hat{K}_{\Phi\,T} - \frac{1}{2} K_{0, T} \hat{K}_\Phi \right),
\ee
and so when we canonically normalise the matter fields we obtain physical couplings:
\be
\mc{L} \supset \partial_{T_\alpha}\left[ \ln \left(e^{- \frac{K_0}{2}}\hat{K}_\Phi\right)\right]
(\psi \sigma^\mu \ov{\psi}) \partial_\mu c_\alpha\,.
\ee
Note that these are different to the coupling to the real parts of the moduli, which couple to the fields via the kinetic term only.

The best conjecture for the K\"ahler metric of chiral matter supported on curves at the intersection of cycles $a$ and $b$ is
\cite{Conlon:2006tj,sequester}:
\be
\hat{K}_\Phi = e^{K_0/3} t_{ab}^{1/3} \quad\text{where}\quad
t_{ab} = \int [D_a] \wedge [D_b] \wedge J= r^{ai} r^{bj} k_{ijk} t^k.
\ee
Thus:
\be
\frac{\partial \left[ \ln \left(e^{- \frac{K_0}{2}}\hat{K}_\Phi\right)\right]}{\partial T_\alpha}=
\frac{1}{6} \left[- \frac{1}{2} \frac{\partial K_0}{\partial \tau_\alpha}
+ \frac{1}{t_{ab}} \frac{\partial t_{ab}}{\partial \tau_\alpha} \right]
= \frac{1}{6} \left[ \frac{1}{2} \frac{t_\alpha}{\vo} + \frac{r^{ai} r^{bj}}{t_{ab}} \,k_{ijk} \mc{T}^{k\alpha}\right], \nn
\ee
For matter supported all on one cycle we have $\hat{K}_\Phi = \,e^{K_0/3} \tau_{a}^{1/3}$ where $\tau_a \equiv r^{ai} \tau_i$, so:
\be
\frac{\partial \left[ \ln \left(e^{- \frac{K_0}{2}}\hat{K}_\Phi\right)\right]}{\partial T_\alpha} =
\,\frac 16 \left[ - \frac 12 \frac{\partial K_0}{\partial \tau_\alpha} + \frac{r^{a\alpha}}{\tau_a} \right]
=\frac 16 \left[ \frac 12 \frac{t_\alpha}{\vo} + \frac{g^2_a}{4\pi}  \right].
\ee
We then translate these into ALP-matter couplings (to axions $\rhoo_i$):
\begin{align}
\frac{\hat{X}_{\psi}^i}{f_{a_i}} =& \frac{1}{3M_P} \,V_{i\beta} \CC_{\beta \alpha}
\left\{ \begin{array}{cl} \frac{t_\alpha}{2\vo} + \frac{r^{ai} r^{bj}}{t_{ab}}\, k_{ijk} \mc{T}^{k\alpha} & \quad {\rm matter\ on\ curve}\,t_{ab}\\
\frac{t_\alpha}{2\vo} + \frac{r^{a\alpha}}{\tau_a} & \quad {\rm matter\ on\ cycle}\, a \\
\frac{t_\alpha}{2\vo} &  \quad{\rm matter\ at\ a\ singularity}
\end{array} \right.
\label{EQ:ALPMATTERCOUPLINGS}
\end{align}
It is thus necessary to consider the canonical normalisation of the axions.
A local axion will thus couple typically as $\frac{\hat{X}_{\psi}^i}{f_{a_i}}
\sim \frac{\sqrt{\vo}}{3M_P}\, \mc{O}(g^{2})$ in the first two cases,
and non-local ones will couple much weaker than Planck strength (or not at all).  Note that we expect these couplings to \emph{violate parity}
-- the parity violating couplings are proportional to $\hat{X}_{\psi_L}^i - \hat{X}_{\psi_R}^i$,
which are generically non-zero since the left- and right-handed fields will be localised on different curves or branes
- but as discussed earlier these should not have severe consequences.
We can parameterise the coupling as:
\begin{align}
\mc{L} \supset& \frac{C_{i\psi}^A}{2 f_{a_i}} \,\bar{\psi} \gamma^\mu \gamma_5 \psi \partial_\mu a_i
+  \frac{C_{i\psi}^V}{2 f_{a_i}} \,\bar{\psi} \gamma^\mu \psi \partial_\mu a_i,
\end{align}
where $C_{i\psi}^{A,V} = \frac{1}{2}( \hat{X}_{\psi_R}^j \pm \hat{X}_{\psi_L}^j)$.

An important special case is the couplings of the `large' cycle axion.
When the SM is realised on a local cycle, $\CC_{bi} \simeq \mc{O}(1)$ and only the universal term contributes;
the couplings are then highly suppressed, with $\frac{\hat{X}_{\psi}^i}{f_{a_i}}
\sim \frac{1}{3M_P \vo^{2/3}} $ for a Swiss-cheese compactification. The coupling to matter fields will then be dominated by the effect generated at one loop via the photon coupling; most important among these for phenomenology is the coupling to electrons. We shall give a full discussion of the one-loop effects in appendix \ref{SEC:LOOPS}, but here present a useful approximation to the one-loop corrections to the above couplings:
\begin{align}
\Delta_{i \gamma\gamma}[ C_{ie}^{A}] \approx& \,\frac{3 \alpha^2}{4\pi^2}  \aC[i,\gamma] \ln \left(\frac{M_{\rm soft}}{m_e}\right) +  \frac{2\alpha^2}{4\pi^2}  \aC[i,\gamma] \ln \left(\frac{\Lambda}{M_{\rm soft}}\right),  \nn\\
\Delta_{i \gamma\gamma}[ C_{ie}^{V}] \approx& \,\frac{2\alpha^2}{4\pi^2}  \aC[i,\gamma] \ln \left(\frac{\Lambda}{M_{\rm soft}}\right),
\end{align}
while there is an additional correction from pion loops to the QCD axion to electron coupling, unchanged in SUSY theories, equal to $\Delta_{\rm QCD} [C_{ae}^A] = \frac{3 \alpha^2}{4\pi} \Delta \aC[a,\gamma] \ln (\Lambda_{\rm QCD}/m_a)$.

\section{The axiverse in the LARGE Volume Scenario}
\label{LVSaxions}

In this section we will focus on the type IIB LARGE Volume Scenario and
show how this moduli stabilisation mechanism gives rise to an
axiverse with many light axions. We shall then discuss the phenomenology and cosmology
of local and non-local axions in different scenarios where the Calabi-Yau has a
isotropic or anisotropic shape and the SM is supported on a local or non-local four-cycle.

\subsection{The axiverse and moduli stabilisation}
\label{AxionsMS}

As we have already pointed out in section \ref{intro}, there are two regimes for the type IIB axiverse
depending on the value of the VEV of the tree-level superpotential $W_0$.

\subsubsection*{Case $W_0 \ll 1$: the axiverse from an ample divisor}

In the case where $W_0 \ll 1$, the axiverse can be realised by an E3-instanton or
gaugino condensation on D7-branes wrapping an ample divisor $D_{\rm am}$ whose volume is given by $\tau_{\rm am}$ \cite{Acharya:2010zx}.
If $D_i$ is a basis of $H_4(X,\mathbb{Z})$ where $X$ is a generic Calabi-Yau three-fold, then:
\be
D_{\rm am}=\sum_{i=1}^{h^{1,1}} \lambda_i D_i,
\ee
is `ample' if and only if $\lambda_i >0$ $\forall\,i=1,...,h^{1,1}$.
The non-perturbative effects supported by this divisor yield a superpotential which looks like \cite{Bobkov:2010rf}:
\be
W=W_0 + A\,e^{-\,a\, T_{\rm am}}=W_0 + A\,e^{-\,a \sum_{i=1}^{h^{1,1}} \lambda_i T_i},
\ee
where $a=2\pi/n$ with $n=1$ for an E3-instanton and $n=N$ for gaugino condensation on $N$ D7-branes.
By fine-tuning $W_0 \sim A\,e^{-\,a\, T_{\rm am}} \ll 1$,
this single non-perturbative effect can generate a minimum for all the K\"ahler moduli within the regime of validity
of the EFT. However it can lift only one axion corresponding to the imaginary part of the ample divisor modulus: $c_{\rm am}={\rm Im}(T_{\rm am})$.
All the remaining $h^{1,1}-1$ axions are massless at leading order
and develop a potential only via tiny higher order instanton effects of the form \cite{Acharya:2010zx}:
\be
W=W_0 + A\,e^{-\,a \,T_{\rm am}} + \sum_{i=1}^{h^{1,1}-1} A_i\,e^{-\, n_i a_i T_i},
\label{higherNP}
\ee
where $T_i$ is a combination of moduli orthogonal to $T_{\rm am}$ $\forall\,i=1,...,h^{1,1}-1$.
Two main concerns regarding the microscopic realisation of this scenario
are related to the difficulty to find an ample divisor which is rigid (and so definitely receives non-perturbative effects),
and the possibility to choose gauge fluxes that avoid chiral intersections between the instanton and the visible sector.

\subsubsection*{Case $W_0 \sim \mc{O}(1)$: the LVS axiverse}

In the natural regime where $W_0 \sim \mc{O}(1)$, the LVS gives rise to an axiverse
since only one K\"ahler modulus, a del Pezzo divisor $T_{\rm dP}$, is fixed by non-perturbative effects
whereas all the other moduli (if not fixed by D-terms) are stabilised perturbatively by $\alpha'$ or $g_s$ effects.
Hence only one axion, $c_{\rm dP}={\rm Im}(T_{\rm dP})$, becomes heavy whereas all the other ones
(except those eaten up by anomalous $U(1)$s) remain light and develop a potential via subleading higher order instanton effects.

Before providing the details of this mechanism,
let us summarise three issues that have to be addressed
when combining moduli stabilisation with the presence of brane fluxes
within the same compactification \cite{arXiv:1110.3333,FibCY}:
\begin{enumerate}
\item \emph{Tension between chirality and non-vanishing non-perturbative effects}

Chiral modes at the intersection of D-branes are induced by a non-zero world-volume flux $\mc{F} \neq 0$.
In the presence of chiral intersections between the visible sector and a cycle $T_{\rm np}$ supporting
non-perturbative effects of the form:
\be
W_{\rm np} = A \,e^{-\,a\, T_{\rm np}}\,,
\label{NPchiral}
\ee
the prefactor $A$ depends on visible sector modes $\phi_i$: $A=A(\phi_i)$.
However, in order to prevent the breaking of the visible sector gauge group at a high scale,
$\langle\phi_i\rangle = 0$, and in turn $A = 0$, implying the vanishing of $W_{\rm np}$.

This observation sets the following constraint on the flux choice:
there has to be no chiral intersection between the visible sector and the cycle supporting
non-perturbative effects. A straightforward solution to this problem is obtained by
placing the non-perturbative effects on a del Pezzo divisor $D_{\rm dP}$ since,
due to the local nature of this blow-up mode, it is always possible
to find a basis of $H_4(X,\mathbb{Z})$ such that $D_{\rm dP}$ is an element of the
basis that intersects no other element. Moreover, due to the rigidity of this four-cycle,
the generation of the non-perturbative effect is definitely under control.

\item \emph{Tension between the cancellation of Freed-Witten (FW) anomalies
and the generation of more than one non-perturbative effect}

The cancellation of worldsheet FW anomalies requires to turn on a half-integer flux
on any divisor $D_{\rm ns}$ which is `non-spin', i.e. its first Chern class $c_1(D_{\rm ns})=-\hat{D}_{\rm ns}$ is odd:
\be
F_{\rm FW}= - \frac 12 \,c_1(D_{\rm ns})=\frac 12 \,\hat{D}_{\rm ns}\,.
\ee
In order to obtain a non-zero contribution to the superpotential,
the total flux $\mc{F} = F - B$ on an E3-instanton or gaugino condensation stack
wrapping an invariant four-cycle $D_a$, has to vanish: $\mc{F}_a=0$.
However, in the case of a non-spin divisor, $F_a = \frac 12 \,\hat{D}_a \neq 0$,
forcing to choose $B= F_a$ in order to obtain $\mc{F}_a=0$.
Notice, however, that once $B$ is fixed in this way, generically
it cannot be used anymore to cancel the FW flux on a second non-spin divisor $D_b$.
Thus the total flux on $D_b$ is non-zero, $\mc{F}_a = \frac 12 \left(\hat{D}_b-\hat{D}_a\right) \neq 0$,
preventing the generation of a second non-perturbative contribution to $W$.

This tension is clearly absent if the K\"ahler moduli are frozen by using only one
non-perturbative effect. If this is supported by a del Pezzo divisor, then there is also
no problem with chirality and the rigidity of this divisor guarantees the generation of $W_{\rm np}$.

\item \emph{D-term induced shrinking of the cycles supporting the visible sector}

The world-volume flux turned on to produce chirality generates also moduli-dependent Fayet-Iliopoulos (FI) terms:
\be
\xi_a = \frac{1}{4\pi\vo}\int_{D_a} J \wedge \mc{F}_a = \frac{1}{4\pi\vo} \,q_{aj} t^j
\quad\text{where}\quad q_{aj}\equiv k_{ajk} \mc{F}_a^k\,,
\label{FIterm}
\ee
resulting in a total D-term potential which looks like:
\be
V_D=\sum_a \frac{g_a^2}{2}\left(\sum_j \tilde{q}_{aj} |\phi_j|^2-\xi_a\right)^2.
\label{VD}
\ee
For vanishing VEVs of charged open string fields, $\langle\phi_j\rangle= 0$, the
supersymmetric D-term conditions give rise to a system of $a$ homogeneous
linear equations $\xi_a = 0$ $\Leftrightarrow$ $q_{aj} t^j = 0$ which might in general depend on
all the $h^{1,1}$ K\"ahler moduli.

Notice that in order to avoid unwanted matter in the adjoint representation,
the visible sector is generically built by wrapping D7-branes around rigid divisors.
If this divisor is a del Pezzo $T_{\rm dP}$, then one equation reduces just to $t_{\rm dP}=0$,
forcing the shrinking of the four-cycle supporting the visible sector. Hence the EFT is
driven to a regime where the supergravity approximation is not under good control.
For this reason, the visible sector should be built on rigid but not del Pezzo divisors.
On the other hand, as we have stressed above, a del Pezzo divisor should support the non-perturbative
effects. Thus this four-cycle does not enter in the D-term equations which turn out to
depend in general on $h^{1,1}-1$ unknowns. In order to prevent the shrinking of some four-cycles,
the brane configuration and the fluxes have to be chosen such that
the rank $d$ of the matrix of the system is $d < h^{1,1}-1$, implying that
the D-term conditions stabilise $d$ combinations of K\"ahler moduli leaving $n_{\rm ax}= h^{1,1}-1-d$ flat directions.

Notice that the combinations of two forms $\tilde{c}_j$, dual to the axions $c_j$, which couple to the
anomalous $U(1)$s are also given by $q_{aj}\tilde{c}^j$. Therefore the $d$ combinations of axions
which are eaten up are the same as the $d$ combinations of K\"ahler moduli which are fixed by the FI-terms.
\end{enumerate}
As explained in \cite{arXiv:1110.3333}, the type IIB LARGE Volume Scenario provides a
promising way to address these issues, allowing the construction of explicit Calabi-Yau examples
with magnetised D-branes and stabilised moduli which are chiral and globally consistent.

The scalar potential receives several contributions which have a different scaling with the
exponentially large overall volume $\vo$:
\be
V = V_D+V_F^{\rm tree} + V_F^{\rm np} + V_F^{\rm p}\,.\nn
\ee
Let us briefly describe each of these contributions:
\begin{itemize}
\item $V_D$ is D-term potential (\ref{VD}) which scales as $V_D \sim \mc{O}(\vo^{-2})$.
Given that it is the leading order effect in an expansion in inverse powers of $\vo$,
it is minimised supersymmetrically at first approximation by imposing $V_D=0$.
This condition fixes $d$ combinations of K\"ahler moduli with $d< h^{1,1} - 1$.
Therefore $d$ axions are eaten up. Notice that if $n_{\rm ax}= h^{1,1}- 1 -d = 1$,
the D-term equations force all the moduli to be of the same size: $\tau_i = \tau_*$ $\forall i=1,...,h^{1,1}-1$.
In turn, $\tau_*$ is required to be small in order to obtain a visible sector gauge coupling $g_{\rm vs}^{-2}\sim \tau_*$
of the observed order of magnitude. This observation then prevents the volume to be exponentially large. 
This is because, requiring that there is a standard model which does not wrap at least one ``small'' diagonal del-Pezzo divisor, if $d = h^{1,1} - 2$ then the large cycle will be fixed relative to the small cycles. 
This is not the case if $n_{\rm ax}\geq 2$, and so we conclude that the simplest realisation of a standard LVS
requires at least $n_{\rm ax}=2$, and so $h^{1,1}= 3 + d$, giving $h^{1,1}=4$ for $d=1$.\footnote{Note that the classic realisations of the LVS on a two-modulus strong swiss cheese have $n_{ax}=1$, which violate our condition because there is no standard model cycle.} Notice that this bound applies only for models with the visible sector in
the geometric regime. For example, in the case of a canonical strong
Swiss-cheese geometry, D-terms force the visible sector to the quiver
locus, leaving just one light axion (the volume axion).
The visible sector could be kept in the geometric regime by giving a VEV
to a open string singlet, resulting again in at least two light axions:
the volume axion and the phase of the open string mode.

\item $V_F^{\rm tree}$ is the tree-level F-term scalar potential which also scales as $V_F^{\rm tree} \sim \mc{O}(\vo^{-2})$
but it is zero due to the `no-scale structure'.

\item $V_F^{\rm np}$ scales as $V_F^{\rm np} \sim \mc{O}(\vo^{-3})$ and it is the non-perturbative
scalar potential arising from corrections to the superpotential $W$
coming from an E3-instanton or gaugino condensation on a D7-stack wrapping a single del Pezzo divisor $T_{\rm dP}$:
\be
W= W_0 + A\,e^{-\,a\,T_{\rm dP}}\,.
\ee
This effect fixes $T_{\rm dP}$ at a small size giving it a mass of the order the gravitino mass.
Hence the corresponding axion becomes very heavy \cite{LVSspectrum}:
\be
m_{a_{\rm dP}}\sim m_{\tau_{dP}} \sim \frac{W_0 \sqrt{\ln \vo}}{\vo}\,M_P\sim m_{3/2}\ln\left(\frac{M_P}{m_{3/2}}\right)\,.
\ee

\item $V_F^{\rm p}$ is the perturbative potential which is derived from the $\alpha'$ and $g_s$ corrections
to the K\"ahler potential $K$. This potential lifts all the $n_{\rm ax}$ flat directions left over by the D-terms
and the non-perturbative effects, leaving all the corresponding axions massless. This is the origin of the `LVS axiverse'.

The $\alpha'$ correction to $K$ is of the form \cite{alphaprime}:
\be
K = -2 \ln\left(\vo + \frac{\zeta}{2 g_s^{3/2}}\right)
\simeq -2\ln\vo -\frac{\zeta}{g_s^{3/2}\vo}\,,
\ee
with $\zeta \propto (h^{1,2}-h^{1,1})$. It gives rise to a scalar potential which scales as $V_F^{\alpha'} \sim \mc{O}(\vo^{-3})$
and its interplay with $W_{\rm np}$, for natural values $W_0\sim\mc{O}(1)$,
yields a supersymmetry-breaking AdS minimum at exponentially large volume \cite{LVS}:
\be
\vo \sim W_0 \,e^{\,a\,\tau_{\rm dP}}\,.
\ee
Consequently, the volume mode obtains a mass of the order \cite{LVSspectrum}:
\be
m_{\vo} \sim \frac{W_0\,M_P}{\vo^{3/2}} \sim m_{3/2} \sqrt{\frac{m_{3/2}}{M_P}} \,,
\ee
and so it is much lighter than the gravitino. This can cause a CMP if $M_{\rm soft}\sim m_{3/2}$ \cite{LVSspectrum}
whereas there is no problem in sequestered models where $M_{\rm soft}\sim m_{3/2}/ \vo \ll m_{3/2}$ \cite{sequester}.
Notice that this vacuum can be uplifted to a dS solution within a manifestly supersymmetric EFT by using
the positive contribution coming from E(-1)/D3 non-perturbative effects at a singularity \cite{NPdS}.

On the other hand, $g_s$ corrections to $K$ are subleading with respect to the $\alpha'$-corrections
due to the `extended no-scale structure' which forces their contribution to $V$
to scale as $V_F^{g_s} \sim \mc{O}(\vo^{-(3 +p)})$ with $p>0$ \cite{Loops}.
Hence the remaining $n_{\rm ax}-1$ flat direction can be lifted
at subleading order either by open string loops (if the corresponding cycle intersects a cycle wrapped by a brane
or the cycle itself is wrapped by a brane with vanishing flux so that the axion is not eaten up) \cite{K3Stab}
or by closed string loops \cite{closedgs}. \footnote{If there are no branes wrapping the cycles
which do not intersect any cycle wrapped by branes,
then there is no open string localised on these cycles, and so the open string loop corrections
should not depend on these moduli.
However the closed string loop corrections can still depend on these moduli.
See for example the short discussion in the paragraph around eq. (2.48) in \cite{ADDstrings}
where there is an example of a cycle which does not intersect the SM cycle and is not wrapped by any D7-brane
but still obtains closed string loop corrections.}
\end{itemize}
Tiny subdominant higher-order non-perturbative effects similar to the ones in (\ref{higherNP})
then slightly lift the $n_{\rm ax}$ axion directions via the generation of the following potential:
\be
V_{W_{\rm np}} (c_i) \simeq \,- \sum_{i=1}^{n_{\rm ax}} \frac{n_i a_i \tau_i W_0}{\vo^2} \,e^{- n_i a_i \tau_i} \cos \left(n_i a_i c_i\right)\,.
\ee
Notice, however, that the axions can become massive also by
non-perturbative corrections to the K\"ahler potential of the form $K_{\rm np} \simeq \vo^{-1} \sum_i e^{- n_i a_i T_i}$
which are not required to be invariant under the orientifold involution \cite{Joe}.
These effects generate a potential for the axions which looks like:
\be
V_{K_{\rm np}} (c_i) \simeq \,  \frac{W_0^2}{\vo^3} \sum_{i=1}^{n_{\rm ax}} e^{-n_i a_i  \tau_i} \cos \left(n_i a_i c_i\right)\,.
\ee
In \cite{Joe} it was determined that the size of the potential generated by the above effects
could only dominate over the QCD axion potential for a single instanton superpotential contribution but
the presence of such an instanton is incompatible with chirality.
On the other hand, the mass generated by a single K\"ahler potential contribution would be:
\be
m_a \lesssim \frac{W_0}{\vo} \,e^{- \pi \alpha_{\rm QCD}^{-1}} M_P \sim m_{3/2}\,e^{- \pi \alpha_{\rm QCD}^{-1}}\,,
\ee
where we used the fact that the relevant four-cycle volume for the QCD axion is $\tau_a = \alpha_{\rm QCD}^{-1}$.
This gives $m_a \lesssim \mc{O}(10^{-11})$ eV for $m_{3/2} \sim 100$ TeV and
the suppression would be even greater for higher order effects.
Hence we expect ALP masses to be typically smaller than those of the QCD axion.

For non-local axions,
given that they correspond in general to non-rigid cycles,
single non-perturbative effects are likely to be absent due to their zero-mode structure.
However, even if they are present, they are suppressed with respect to perturbative
corrections by a huge exponential factor. Two typical examples of non-local cycles
whose non-perturbative effects would be minuscule are the volume mode of
Swiss-cheese manifolds $\tau_b \sim \vo^{2/3}$ where $W_{\rm np}= A_b\,e^{-\,a\,T_b} \propto e^{-\,a\,\vo^{2/3}}\ll 1$ 
- in fact even for $\vo \simeq 10^4$ this factor is less than $10^{-600}$ -
and the fibre modulus $T_f$ of fibred Calabi-Yau constructions which might receive only
poly-instanton contributions of the form $W_{\rm np}= A_s\,e^{-\,a_s \left(T_s + A_f\, e^{-\,a_f\,T_f }\right)}
\,\,\Rightarrow\,\, W_{\rm np}\left(T_f\right)\sim - \,a_s\,A_s\,A_f\,e^{-\,a_s T_s}\,e^{-\,a_f\,T_f } \ll 1$ \cite{ADDstrings,poly}.
On the other hand, non-local axions may obtain small masses via a QCD-like effect in the hidden sector,
as we shall discuss in section \ref{SEC:MIXINGCPDM}; these are of the order $m_{\rm non-local} \sim \Lambda^2/M_P$,
where $\Lambda$ is the relevant strong coupling scale. 
These last two mass contributions are very model dependent.

In summary, generic non-local axions are effectively massless (although in certain cases some may obtain non-negligible masses, still typically smaller than local axion masses) while local axions obtain masses of the form:
\begin{align}
m_{\rm local} \sim e^{- n \pi\tau_{\rm local}} \times \left\{ \begin{array}{cl}  M_P  & \text{Superpotential terms or QCD-like masses} \\
m_{3/2} & \text{K\"ahler potential terms} \end{array}\right.
\end{align}
Therefore the axions acquire a mass spectrum which is logarithmically hierarchical giving rise to the `LVS axiverse'.
As pointed out above, the LVS requires $n_{\rm ax}\geq 2$, implying that the simplest
realisation of the LVS axiverse involves at least two light axions with a
QCD axion candidate plus one ALP. More generic models for $h^{1,1}$ very large will
include an arbitrarily large number of light axions.

\subsection{The QCD axion in the LVS axiverse}
\label{SEC:MIXINGCPDM}

In order to have a phenomenologically viable axiverse,
it is crucial to identify potential QCD axion candidates among all the light axions. A very important observation
related to this task is the following. If we restore the QCD vacuum angle $\theta$,
we have:
\be
\mc{L} \supset \frac{g_3^2}{32\pi^2} \left( \theta - \frac{a \,\aC[a,3]}{f_a}\right)
\tr\left(F_3 \wedge F_3\right),
\ee
where the coefficient $\aC[a,3]$ gives the effective QCD anomaly. Since QCD is periodic under shifts of $\theta$ by $2\pi$,
we have the standard axion situation only if $\aC[a,3]\simeq \mc{O}\left(1\right)$.
However, given that $a$ only varies through $[0,2\pi f_a)$, if $\aC[a,3]\ll 1$,
like in some cases described in sections \ref{ScCY} and \ref{fibCY} where $\aC[a,3]\simeq \mc{O}\left(\vo^{-\alpha}\right)\ll 1$
for $\alpha>0$, a vacuum angle $\theta > 2\pi \aC[a,3] \simeq 2\pi\,\vo^{-\alpha}$
cannot be absorbed into the expectation value of $a$.

This implies that \emph{axions with a volume suppressed anomaly coefficient $\aC[a,3]$ cannot solve the strong CP problem},
and so they cannot be the QCD axion. Let us investigate which axions can play the r\^ole of the
QCD axion in the case when the SM is on a local or a non-local cycle.

\subsubsection{SM on a local cycle}
\label{sec:smlocal}

In this case the visible sector is localised on a small rigid divisor
which in sections \ref{ScCY} and \ref{fibCY} we have denoted as $\tau_s$.
The local axions $a_s$ are natural QCD axion candidates since they have $\aC[s,s]\simeq \mc{O}\left(1\right)$
and an intermediate scale decay constant of the order the string scale $f_{a_s}\simeq M_s/\sqrt{4\pi} \sim 10^{10}$ GeV
for $\vo \sim 10^{14}$ which gives rise to TeV-scale SUSY for $W_0 =1$, or $f_{a_s}$ somewhat larger for smaller $W_0$. This has already been pointed out in \cite{Joe}
but without providing an explicit chiral example where the local axion is not eaten up by an anomalous $U(1)$.
In sections \ref{RightQCDaxion} and \ref{ManyLocalAxions} we shall present two explicit examples
where the SM is on a local cycle and the QCD axion is given by a local axion which is not eaten up
by any anomalous $U(1)$ and does not receive large mass contributions from single instanton effects.

This is a crucial issue since non-local axions, like the volume axion $a_b$ of section \ref{ScCY}, cannot
be the QCD axion when the standard model is built on a local cycle. In fact, consider the case where the local axion $c_s$
coupling to QCD is eaten by an anomalous $U(1)$, and the only other axion
that can couple is the non-local one $c_b$. The theory can be written as:
\bea
\mc{L} &\simeq& \frac{1}{8\vo} \left( \partial_\mu c_s + \frac{M_P}{\pi}\, q A_\mu\right)^2
+ \left( \partial_\mu c_s + \frac{M_P}{\pi} \,q A_\mu\right) \frac{\partial^\mu c_b}{4\vo^{5/3}}
+ \frac{\left(\partial_\mu c_b \right)^2}{8\vo^{4/3}} \nn \\
&&+ \frac{g^2}{4\pi M_P} \,c_s \tr \left(F \wedge F\right),
\eea
which can be diagonalised via:
\be
c_b \simeq 2\,\vo^{2/3} a_b, \qquad c_s \simeq 2\,\vo^{1/2} a_s - 2 \,a_b\,,
\label{EQ:CPDMREDFINITIONS}
\ee
giving:
\be
\mc{L} \simeq \frac{1}{2} \left( \partial_\mu a_s + \frac{M_P}{2\sqrt{\vo} \pi} \,q A_\mu\right)^2 + \frac{1}{2}  \left(\partial_\mu a_b \right)^2
+ \frac{2g^2}{4\pi M_P} \left(\sqrt{\vo} a_s - a_b\right) \tr (F \wedge F). \nn
\ee
We can then make a chiral rotation of the fermions charged under the massive $U(1)$
to eliminate the axion $a_s$ at low energies, leaving only $\frac{1}{2}  \left(\partial_\mu a_b \right)^2
-\frac{2g^2}{4\pi M_P}  a_b \tr (F \wedge F) $. 

Notice that the axion $a_s$, which is eaten up,
corresponds to the axion $c_s$, which is gauged, only at leading order
since the kinetic terms induce also a subleading mixing with $c_b$. In fact,
the canonical normalisation (\ref{EQ:CPDMREDFINITIONS}) implies:
\be
2\,\vo^{1/2} a_s \simeq c_s + \vo^{-2/3} c_b\,.
\ee
It is this subleading mixing that induces the coupling of $a_b$ to QCD.
However, as can be seen in (\ref{Csiso}), the non-local axion $a_b$ has $\aC[b,s]\simeq \mc{O}\left(\vo^{-3/2}\right)\ll 1$,
and so it cannot play the r\^ole of the QCD axion. Moreover, there is necessarily an anomalous \emph{global} $U(1)$ symmetry left over. When this remaining global symmetry is spontaneously broken, an additional axion will be introduced, and only one linear combination of this new axion and $a_b$ is made massive by QCD. 

If, on the other hand, the strongly coupled gauge group was not QCD, and the global $U(1)$ was not spontaneously broken (but merely anomalous - and hence broken to a discrete subgroup by the strong coupling effect) then $a_b$ could obtain a mass, of order  
\be
m_{a_b} \sim \Lambda^2/M_P
\ee
where $\Lambda$ is the strong coupling scale. The dark matter density produced through misalignment would be set by sending $f \rightarrow M_P$, $\Theta_i \rightarrow \Theta_i/\vo^{2/3}$ (where $\Theta_i$ is the initial misalignment angle) in the usual axion calculation, and would therefore be comfortably tiny. This is at first rather puzzling, since the initial Lagrangian is independent of the expectation value of $c_b$, and $a_b \propto c_b$. However, we note that a shift of $a_b$ induces a shift in $c_s$ - and this is where the mass generation arises. That both axions become massive appears because we have two different mass terms  - and a \emph{mass mixing} term (the combination eaten by the gauge field is $a_s$, which depends on both $c_s$ and $c_b$). 


\exclude{
As an aside, the above coupling would still generate a mass for $a_b$;
if there is some hidden gauge group with strong-coupling scale $\Lambda$ the mass would be:
\be
m_{a_b} \sim \Lambda^2/M_P.
\ee
The dark matter density produced through misalignment would be set by sending $f \rightarrow M_P$, $\Theta_i \rightarrow \Theta_i/\vo^{2/3}$ (where $\Theta_i$ is the initial misalignment angle) in the usual axion calculation. For QCD the result would be:
\be
\frac{\Omega_{\rho_2} h^2}{0.112} \sim  \gamma \frac{10^8}{\vo^{4/3}} \left( \frac{\Theta_i}{\pi} \right)^2, \nn
\ee
where $\gamma$ is the entropy dilution factor, equal to one in the standard cosmology. Hence the amount of dark matter produced is comfortably tiny. For a hidden gauge group, the result depends upon the hidden matter content and whether it is thermalised.
}

On the other hand, in the presence of a non-perturbative superpotential of the form:
\be
W_{np} = A_s\, e^{- a_s T_s} + A_b\, e^{- a_b T_b}\,,
\ee
the local axion $c_s$ is made more massive than the QCD axion,
while the non-local axion $c_b$ develops a non-perturbative potential only at subleading order
and its coupling to QCD is extremely small. In fact, the Lagrangian:
\be
\mc{L} \simeq \frac{\left( \partial_\mu c_s\right)^2}{8\vo}
+ \frac{\partial_\mu c_s\partial^\mu c_b}{4\vo^{5/3}}  + \frac{\left(\partial_\mu c_b\right)^2}{8\vo^{4/3}}
+ \frac{g^2}{4\pi M_P} \,c_s \tr \left(F \wedge F\right) - V \left(c_s,c_b\right)
\ee
would be diagonalised by a transformation which looks like:
\bea
\frac{c_b}{\tau_b}    &\simeq&    \mc{O}\left(1\right)  a_b   +  \mc{O}\left(\vo^{-1/2}\right)  a_s\,, \\
\frac{c_s}{\tau_s}    &\simeq&    \mc{O}\left(\vo^{1/2}\right)  a_s  +  \mc{O}\left(e^{- a_b \tau_b}\right)   a_b\,,
\eea
giving:
\be
\mc{L} \sim \frac 12 \left[ \left( \partial_\mu a_s\right)^2 + \left(\partial_\mu a_b \right)^2\right]
+ \frac{g^2}{2\pi M_P} \left(\sqrt{\vo} a_s
+ \mc{O}\left(e^{- a_b \tau_b}\right)   a_b \right) \tr (F \wedge F) - V\left(a_s,a_b\right), \nn
\ee
and so the coupling of $a_b$ to $F \wedge F$ is suppressed by $e^{- a_b \tau_b}\ll 1$ (less than $10^{-600}$ for $\vo \simeq 10^4$)
for $\tau_b\simeq \vo^{2/3}\gg 1$.
This implies that the anomaly coefficient $\aC[b,s]$ in this case would scale as $\aC[b,s] \sim  e^{- a_b \tau_b}\,\vo^{-2/3}\ll 1$,
and so it is in practice zero for all intents and purposes. Such a light non-local axion would effectively produce no dark matter.

Thus we have shown that non-local axions cannot behave as the QCD axion if the SM is on a
local cycle. There is however an exception to this conclusion.
If the extra dimensions are anisotropic, as shown in (\ref{Csaniso}), the fibre axion $a_f$ has
$\aC[f,s]\simeq \mc{O}\left(1\right)$, and so it can be the QCD axion in the presence of a dilution
mechanism as its decay constant is of order $f_{a_f} \simeq M_P / \tau_s \sim M_{\rm GUT} \sim 10^{16}$ GeV.

\subsubsection{SM on a non-local cycle}
\label{sec:smnonlocal}

In this case the SM is localised on a divisor which controls the overall size of the Calabi-Yau,
and so non-local axions can play the r\^ole of the QCD axion. In fact,
considering the Swiss-cheese example of section \ref{ScCY} or isotropic fibred manifolds
of section \ref{fibCY}, the visible sector is now supported on $\tau_b$.
As can be seen from (\ref{Csiso}) and (\ref{Csaniso}), now $\aC[b,b]\simeq \mc{O}\left(1\right)$,
and so the non-local axion $a_b$ can now be the QCD axion. However, given that the visible sector
gauge coupling scales as $g_{\rm vs}^{-2}\simeq \tau_b$, the volume $\vo \simeq \tau_b^{3/2}$ has to be
set of the order $\vo \sim 10^4$, while $W_0$ has to be fine-tuned small to obtain TeV-scale soft terms.
In turn, the axion decay constant $f_{a_b} \sim M_P / \vo^{2/3} \sim M_{\rm GUT}\sim 5 \cdot 10^{15}$ GeV, and so
a dilution mechanism is needed. We shall present an explicit example of this situation in section \ref{GUTmodel}.

Alternatively, if the Calabi-Yau has an anisotropic shape, the SM can be localised on
the fibre divisor $\tau_f$ and the volume can be set exponentially large $\vo \sim 10^{14}$
without any need to tune $W_0$ since the visible sector gauge coupling does not depend on
the overall volume $g_{\rm vs}^{-2}\simeq \tau_f \sim \tau_s$. Moreover, as shown in (\ref{Csaniso}),
the anomaly coefficient of the fibre axion $a_f$ is of order unity, $\aC[f,f]\simeq \mc{O}\left(1\right)$,
allowing $a_f$ to play the r\^ole of the QCD axion provided that it can be diluted by some mechanism
as $f_{a_f}\sim M_P / \tau_s \sim M_{\rm GUT}\sim 10^{16}$ GeV.
We shall present an explicit example of this situation in section \ref{SU(3)SU(2)model}.

\subsection{Cosmology of the LVS axiverse}
\label{SEC:COSMO}

The cosmology of the LVS axiverse takes a very different form
depending on whether the value of the Hubble scale during inflation $H_{\rm inf}$
is larger or smaller then the gravitino mass $m_{3/2}$.
The standard case in string cosmology is $H_{\rm inf} \leq m_{3/2}$,
creating a well-known tension between inflation and supersymmetry breaking since
the requirement of generating enough density perturbations sets, in general, $H_{\rm inf}\sim M_{\rm GUT}$,
forcing $m_{3/2}$ to be well above the TeV scale \cite{KL}.
Given that the moduli tend to develop a mass of the order $m_{\rm mod}\sim m_{3/2}$,
the cosmological moduli problem and its solution are closely related to the above mentioned tension
between inflation and TeV-scale supersymmetry.
The best available solutions to this problem are:
\begin{enumerate}
\item The value of $m_{3/2}$ after inflation is very different from the value of $m_{3/2}$ during inflation,
so that the relation $H_{\rm inf} \leq m_{3/2}$ does not hold today where we can instead have $m^{\rm today}_{3/2}< H_{\rm inf}$
allowing TeV-scale SUSY for $\vo \sim 10^{14}$. This is achieved if the volume mode is the inflaton \cite{varyingVol}.
However this model is very fine-tuned and reheating is a big problem since the inflaton at the end of inflation becomes extremely light.

\item The value of $m_{3/2}$ after inflation is the same as the value of $m_{3/2}$ during inflation
but one has a sequestered scenario where the soft terms are suppressed
with respect to $m_{3/2}$ by some power of the volume \cite{sequester}.
Hence $M_{\rm soft}$ can be of order TeV even if $m_{3/2}$ is very large.
The most promising sequestered models have $M_{\rm soft}\sim m_{3/2}^2 / M_P \sim M_P/\vo^2$,
and so TeV-scale SUSY requires $\vo \sim 10^7$ instead of the `traditional' $\vo \sim 10^{14}$.
\end{enumerate}
Let us now analyse the two different cases separately.

\subsubsection{Case $H_{\rm inf} > m_{3/2}$}

In this case, as we have pointed out above, we can set $\vo \sim 10^{14}$,
so to obtain $M_{\rm soft}\sim m_{3/2} \sim \mc{O}\left({\rm TeV}\right)$,
\footnote{Due to the fact that the SM cycle is stabilised perturbatively, we expect $M_{\rm soft}\sim m_{3/2}$
instead of the suppressed case $M_{\rm soft}\sim m_{3/2}/\ln\left(M_P/m_{3/2}\right)$
which is typical of scenarios with the SM cycle fixed by non-perturbative effects.}
and the string scale turns out to be intermediate $M_s = M_P/\sqrt{4\pi\vo} \sim 5\cdot 10^{10}$ GeV.
This is the case considered in subsection \ref{sec:smlocal}, and found in sections \ref{RightQCDaxion}
and \ref{ManyLocalAxions} where the SM is on a local cycle
and the QCD axion is given by a local axion.
We stress again that in the literature there is no explicit inflationary model with a successful
reheating process that realises this scenario \cite{Reheating}. We shall however assume that this
is possible and proceed to analyse the cosmological implications.
\begin{enumerate}
\item The complex structure moduli and the dilaton do not play any significant cosmological r\^ole
since they are trapped at their minima, i.e. their initial displacement is tiny \cite{astroLVS}.

\item The non-local K\"ahler moduli are all very light and suffer from the CMP. In particular,
the mass of the volume mode is of the order $m_{\tau_b}= m_{3/2}/ \sqrt{\vo}\simeq 0.1$ MeV,
while that of the fibre modulus is $m_{\tau_f}= m_{3/2}/ \vo^{2/3}\simeq 0.5$ keV for
$m_{3/2} = \kappa \,M_P / \vo \simeq 1$ TeV, where $\vo = 2\cdot 10^{14}$ and $\kappa \equiv \sqrt{g_s/(4\pi)}W_0$
with $g_s=0.1$ and $W_0=1$.

\item The non-local axions associated with the base and the fibre divisors
behave as very light and almost decoupled ALPs
with a decay constant of the order the 10D KK scale,
$f_{a_b}\sim f_{a_f}\sim M^{\rm 10D}_{\rm KK} \sim M_P/(4\pi \vo^{2/3}) \sim 10^8$ GeV.
Notice that these non-local ALPs do not create problems with dark matter overproduction
since they are very light and their decay constants are very low (see appendix \ref{APP:COSMIC}).

\item The del Pezzo modulus $\tau_{\rm dP}$ and the corresponding axion $a_{\rm dP}$
fixed by non-perturbative effects, obtain a large mass of the order
$m_{\tau_{\rm dP}}\sim m_{a_{\rm dP}}= m_{3/2} \ln\vo \simeq 35$ TeV.
However  this modulus suffers from the CMP since it has a very small coupling with the SM,
of the order $1/(M_P \sqrt{\vo})$ \cite{Reheating}, reflecting the fact that this cycle blows up a point
which is geometrically separated from the point blown up by the SM cycle.
Thus $\tau_{\rm dP}$ and $a_{\rm dP}$ would decay after BBN at a temperature of the order:
\be
T_{\tau_{\rm dP}}\sim T_{a_{\rm dP}} \simeq \sqrt{\Gamma_{\tau_{\rm dP}}M_P}
\simeq\left(\frac{m_{\tau_{\rm dP}}}{M_P\vo}\right)^{1/2} m_{\tau_{\rm dP}}\simeq 0.1\,{\rm eV}.
\ee
Notice that $\tau_{\rm dP}$ and $a_{\rm dP}$ would have a much stronger coupling, of the order $1/M_s$,
to hidden sector degrees of freedom living on this del Pezzo divisor, and so the CMP would be solved
if they could decay to hidden particles. However, if the hidden sector consists of an E3-instanton or
a pure $\mc{N}=1$ SYM theory that develops a mass gap, no light hidden degree of freedom is present.

\item Some of the local K\"ahler moduli supporting the SM
are fixed by the D-terms which induce masses of the order $M_s$, and so these moduli disappear from the EFT.
The directions left massless by the D-terms (the ones corresponding to the uneaten axions)
are instead fixed by string loops and obtains a mass of the order $m_{\tau_{\rm SM}}\simeq \alpha_{\rm SM} m_{3/2}\simeq 40$ GeV
for $\alpha_{\rm SM}\simeq 1/25$. This scaling can be understood from the fact that
the $g_s$ potential is of the order $V_{g_s}(\tau_{\rm SM}) \simeq 1/(\vo^3\sqrt{\tau_{\rm SM}})$ \cite{arXiv:1110.3333,Loops},
while the mass can be estimated as $m_{\tau_{\rm SM}}^2\sim K^{-1}_{\tau_{\rm SM},\tau_{\rm SM}} V_{\tau_{\rm SM},\tau_{\rm SM}}
\sim 1/(\vo\tau_{\rm SM})^2 \simeq (m_{3/2}\alpha_{\rm SM})^2$.
These moduli couple locally as $1/M_s$ but, as shown in \cite{K3norm}, they decay
before dominating the energy density. Hence their decay does not reheat
the Universe and cannot dilute any unwanted relic. Therefore the best way
to solve the CMP seems to be the presence of a late time period of thermal
inflation \cite{KiwoonChoi}.

\item Not all the local closed string axions are eaten up,
and so one of them can be the QCD axion with an intermediate scale decay constant $f_{a_s}\sim M_s/\sqrt{4\pi}\sim 10^{10}$ GeV
and an order unity anomaly coefficient $\aC[s,s]\sim \mc{O}(1)$.
Notice, however, that this axion does not constitute a relevant component of dark matter, unless one moderately
fine-tunes the parameters such that the string scale and correspondingly the decay constant is increased to $f_{a_s}\sim M_s/\sqrt{4\pi}\sim 10^{12}$ GeV (see appendix \ref{APP:COSMIC}).
Thus, the constraints from isocurvature fluctuations (see appendix \ref{Iso}) do not apply to this model.
\end{enumerate}

Different considerations apply to the case considered in subsection \ref{sec:smnonlocal} and found in section \ref{GUTmodel} where
the SM is supported on the non-local cycle $\tau_b$ controlling the overall volume of the Calabi-Yau.
As we have already stressed, the volume cannot be set very large, otherwise the visible sector
gauge coupling would be too small, whereas $W_0$ has to be fine tuned small in order to
obtain TeV-scale soft terms. For $\vo \simeq 10^4$, $g_s\simeq 0.1$ and $W_0\simeq 1.5\cdot 10^{-10}$,
we would obtain $M_{\rm soft}\sim m_{3/2} \simeq 3$ TeV. The string scale turns out to be
quite high: $M_s \sim 5\cdot 10^{15}$ GeV. Let us present a brief analysis of the cosmology of this scenario:
\begin{enumerate}
\item The non-local K\"ahler modulus $\tau_b$ is light and suffers from the CMP given that its mass
is of the order $m_{\tau_b}= m_{3/2}/ \sqrt{\vo}\simeq 30$ GeV.

\item The non-local axion $a_b$ associated with the large cycle can play the r\^ole of the QCD axion
since $\aC[b,b]\simeq \mc{O}\left(1\right)$. Its decay constant is however rather high
$f_{a_b} \sim M_P / (4\pi\vo^{2/3}) \sim 5\cdot 10^{14}$ GeV, and so it would
naively lead to dark matter overproduction (without tuning the initial misalignment angle, or, as is in fact the case, entropy dilution).

\item The del Pezzo modulus $\tau_{\rm dP}$ and the corresponding axion $a_{\rm dP}$
fixed by non-perturbative effects obtain a larger mass of the order
$m_{\tau_{\rm dP}}\sim m_{a_{\rm dP}}= m_{3/2} \ln\vo \simeq 30$ TeV.
Hence, given that this modulus has an ordinary Planck-strength coupling to the SM,
it does not suffer from the CMP since it would decay at a temperature of the order:
\be
T_{\tau_{\rm dP}} \sim T_{a_{\rm dP}} \simeq \sqrt{\Gamma_{\tau_{\rm dP}}M_P}
\simeq\left(\frac{m_{\tau_{\rm dP}}}{M_P}\right)^{1/2} m_{\tau_{\rm dP}}\simeq 5\,{\rm MeV},
\ee
which is above the BBN temperature but below the freeze-out temperature of thermal dark matter.
Thus, dark matter will be produced again non-thermally by the decay of $\tau_{\rm dP}$ which
will also dilute the modulus $\tau_b$ suffering from the CMP and the QCD axion $a_b$.
Contrary to the previous case, $a_b$ would constitute a relevant component of dark matter,
and so the constraints from isocurvature fluctuations would apply to this model.
Conventional haloscope searches exploiting
microwave cavities, however, would not be sensitive to $a_b$ dark matter
due to its tiny mass and extremely small coupling.
Molecular interferometry may be a future option to search for it
if one allows for a bit of fine-tuning in $W_0$ such that $f_{a_b} \simeq 10^{16}$ GeV.
\end{enumerate}
We finally briefly analyse the model of section \ref{SU(3)SU(2)model}, where the
Calabi-Yau has an anisotropic shape and the SM is localised on
the fibre divisor $\tau_f$. In this case the visible sector gauge coupling does not depend on
the overall volume $g_{\rm vs}^{-2}\simeq \tau_f \sim \tau_s$, and so we can set $\vo \simeq 10^{14}$
and $W_0\simeq \mc{O}(1)$. The two non-local K\"ahler moduli, $\tau_f$ and $\tau_b$ are very light,
with masses again of the order $m_{\tau_b}\simeq 1$ MeV and $m_{\tau_f}\simeq 1$ keV.
The axion of the fibre divisor, $a_f$, behaves as the QCD axion since $\aC[f,f]\simeq \mc{O}\left(1\right)$,
but its decay constant is very high, $f_{a_f}\simeq M_P/(4\pi\tau_f)\simeq 10^{16}$ GeV for $\tau_f\simeq \mc{O}(10)$,
even if the string scale is intermediate.
The axion of the base divisor, $a_b$, is a very light and decoupled ALP ($C_{bf} \sim \vo^{-2}$) with $f_{a_b} \simeq M_P/(4\pi\tau_b)\simeq 5$ TeV
for $\tau_b = \vo / \sqrt{\tau_f}\simeq 5\cdot 10^{13}$.
The del Pezzo modulus and its axion have a mass of the order $m_{\tau_{\rm dP}}\sim m_{a_{\rm dP}}\simeq 35$ TeV
and couple to the SM as $1/M_P$. Therefore they would decay at $T_{\tau_{\rm dP}} \sim T_{a_{\rm dP}} \simeq 5$ MeV,
producing non-thermal dark matter and diluting the abundance of both the light moduli and the QCD axion.

It is worth pointing out that the cosmology of an axiverse with a high string scale was also discussed in \cite{Acharya:2008bk,Acharya:2010zx}.
As we have seen, our scenarios have some similarities and crucial differences to that case.

\subsubsection{Case $H_{\rm inf} \leq m_{3/2}$}

In this case, inflation can be reconciled with TeV-scale SUSY if the SM
is sequestered from the SUSY-breaking sector. This can be achieved if the SM
is built via fractional D3-branes at a singularity obtained by shrinking
a blow-up divisor. The most promising scenario involves the following
hierarchy $M_{\rm soft}\simeq m_{3/2}/\vo$ which leads to $M_{\rm soft}\simeq 2$ TeV
for $m_{3/2}\simeq 10^{10}$ GeV with $g_s=0.1$, $W_0=1$ and $\vo\simeq 10^7$.
This is also the value of the volume which gives the right amount of density perturbations
in the inflationary model of \cite{KahlerInfl} where the inflaton is a blow-up mode
fixed by non-perturbative effects. Therefore, contrary to the case $H_{\rm inf}>m_{3/2}$,
we now have a more robust construction with an explicit realisation of
inflation and reheating \cite{Reheating}. Notice that the string scale
turns out to be relatively high: $M_s = M_P/\sqrt{4\pi\vo} \simeq 10^{14}$ GeV.

Let us now present an outline analysis of the cosmological implications of this scenario.
\begin{enumerate}
\item The del Pezzo modulus $\tau_{\rm dP}$, and the corresponding axion $a_{\rm dP}$
fixed by non-perturbative effects, play the r\^ole of the inflaton and are very light during inflation.
After the end of inflation, when the classical condensate oscillates coherently around the minimum,
they obtain a large mass of the order $m_{\tau_{\rm dP}}\sim m_{a_{\rm dP}}= m_{3/2} \ln\vo \simeq 5\cdot 10^{11}$ GeV.
Given that these moduli couple to the SM as $1/(M_P \sqrt{\vo})$, their perturbative decay
gives rise to reheating at a temperature:
\be
T_{\tau_{\rm dP}}\sim T_{a_{\rm dP}} \simeq \sqrt{\Gamma_{\tau_{\rm dP}}M_P}
\simeq\left(\frac{m_{\tau_{\rm dP}}}{M_P\vo}\right)^{1/2} m_{\tau_{\rm dP}}\simeq 40\,{\rm TeV}.
\ee
Notice that $\tau_{\rm dP}$ and $a_{\rm dP}$ have a much stronger coupling, of the order $1/M_s$,
to hidden sector degrees of freedom living on this del Pezzo divisor, and so they would dump all
their energy to hidden, instead of visible, particles. However, this problem is absent
if the hidden sector consists of an E3-instanton or a pure $\mc{N}=1$ SYM theory that develops a mass gap
since no light hidden degree of freedom would be present \cite{Reheating}.

\item The non-local K\"ahler moduli are all very heavy in this case and do not suffer from the CMP.
In fact, the mass of the volume and fibre moduli are $m_{\tau_b}= m_{3/2}/ \sqrt{\vo}\simeq 5\cdot 10^6$ GeV
and $m_{\tau_f}= m_{3/2}/ \vo^{2/3}\simeq 500$ TeV. Given that they both couple to the SM with
Planck-strength they would decay respectively at:
\be
T_{\tau_b} \simeq \sqrt{\Gamma_{\tau_b}M_P}
\simeq\left(\frac{m_{\tau_b}}{M_P}\right)^{1/2} m_{\tau_b}\simeq 10\,{\rm GeV},
\ee
and:
\be
T_{\tau_f}\simeq \sqrt{\Gamma_{\tau_f}M_P}
\simeq\left(\frac{m_{\tau_f}}{M_P}\right)^{1/2} m_{\tau_b}\simeq 100\,{\rm MeV}.
\ee
Since $T_{\tau_f}$ is above $T_{\rm BBN}\simeq \mc{O}(1)$ MeV
but below $T_{\rm freeze-out}\sim \mc{O}(10)$ GeV, non-thermal dark matter will be
produced by the decay of $\tau_f$ which will also dilute the abundance of any axion-like particle.\footnote{Ref. \cite{Cicoli:2012aq,Higaki:2012uy} showed that the decay of $\tau_b$
gives rise to dark radiation made of light volume axions.}

\item If the Calabi-Yau is isotropic,
the non-local axions associated with the base and the fibre divisors cannot be the QCD axion and
behave as very light and almost decoupled ALPs
with a decay constant of the order the 10D KK scale,
$f_{a_b}\sim f_{a_f}\sim M^{\rm 10D}_{\rm KK} \sim M_P/(4\pi \vo^{2/3}) \sim 5\cdot 10^{12}$ GeV.
Notice that these non-local ALPs do not create problems with dark matter overproduction
since they are very light and their decay constants are very low.
On the other hand, if the Calabi-Yau is anisotropic, the fibre axion $a_f$
can be the QCD axion since $\aC[f,f]\simeq \mc{O}\left(1\right)$,
but its decay constant is very high: $f_{a_f}\simeq 10^{16}$ GeV. However this is not
a problem since $a_f$ will be diluted by the decay of the non-local K\"ahler moduli.
The axion of the base divisor, $a_b$, is instead again a very light and decoupled ALP.

\item In the case of fractional D3-branes at del Pezzo singularities,
all the local closed string axions are eaten up, and so cannot be the QCD axion.
Correspondingly, all the local K\"ahler moduli are fixed by D-terms, and
so obtain masses of the order $M_s$ and disappear from the EFT.
In fact, each dP$_n$ has $n$ two-cycles which can be trivial or non-trivial within the Calabi-Yau
plus one two-cycle corresponding to the canonical class which is always also
a two-cycle of the Calabi-Yau dual to the del Pezzo divisor.
Moreover each dP$_n$ has exactly two anomalous $U(1)$s which become massive by eating
the axions given by the reduction of $C_2$ on the canonical two-cycle
and $C_4$ on the del Pezzo divisor \cite{Buican:2006sn}.
Given that these are the only two local cycles, all the local closed string axions are eaten up.
In fact, if some of the $n$ two-cycles are non-trivial within the Calabi-Yau,
they would correspond to non-local cycles, and so to non-local axions.
Two possible way-outs could be the following:
\begin{itemize}
\item Consider more complicated singularities whose resolution involves more than two local cycles.
In this case, some local closed string axions might be left over, and they could play the r\^ole
of the QCD axion. The corresponding decay constant would be rather high,
$f_{a_s}\simeq M_s/\sqrt{4\pi} \simeq 5\cdot 10^{13}$ GeV, but the abundance of the QCD axion would be
diluted by the decay of the non-local K\"ahler moduli.

\item Focus on the phase of an open string axion $\phi$ to realise the QCD axion in a more model-dependent way.
In this case, one should check that the D-terms,
which give a VEV to the radial part of $\phi$ that sets the axion decay constant, do not resolve the singularity
obtained by setting the Fayet-Iliopoulos term $\xi$ to zero.
In fact, the D-term potential used to fix the blow-up mode $\tau_{\rm blow}$ at the singularity looks like:
\be
V_D \simeq g^2 \left( |\phi|^2 - \xi \right)^2,\qquad\text{with}\qquad \xi = \frac{\tau_{\rm blow}}{\vo}\,. \nn
\ee
Hence, if $\langle|\phi|\rangle = 0$, $V_D=0$ for $\xi=0$ implying $\langle\tau_{\rm blow}\rangle=0$.
However, if $\phi$ is a SM singlet which can acquire a non-zero VEV,
$V_D=0$ would set $\langle|\phi|\rangle = \sqrt{\xi} \simeq \langle\sqrt{\tau_{\rm blow}}\rangle\, M_s$,
creating a tension between $\langle\tau_{\rm blow}\rangle=0$ and $\langle|\phi|\rangle \neq 0$.
However $\tau_{\rm blow}$ could still have a non-zero VEV, as long as it is below the string scale,
since we would still be at the quiver locus if $\langle\tau_{\rm blow}\rangle = \vo^{-2\alpha}$ with $\alpha>0$.
If this kind of VEV were induced by the F-terms,
the VEV of $|\phi|$ would become $\langle|\phi|\rangle \simeq M_s / \vo^\alpha$.
Given that this VEV also sets the axion decay constant $f_a$,
the volume suppression factor might bring $f_a$ at the intermediate scale.
At that point, the QCD axion would not form a relevant component of dark matter
due to the dilution induced by the decay of the non-local K\"ahler moduli,
and the constraints from isocurvature fluctuations would not apply to this case.
\end{itemize}
\end{enumerate}

\section{Explicit LVS models}
\label{LVSexamples}

Not all the axions of the examples studied in section \ref{AxionDC} will survive in the EFT since some of them
will become heavy due to non-perturbative effects whereas others will be eaten up by anomalous $U(1)$s.
According to the general discussion of the axiverse and moduli stabilisation performed in section \ref{AxionsMS},
only $n_{\rm ax}= h^{1,1} -d -1$ axions remain light. Moreover, the LVS requires $n_{\rm ax}\geq 2$
guaranteeing the presence of a QCD axion candidate and at least one ALP.
However, the couplings of the $n_{\rm ax}$ light axions depends crucially on
the Calabi-Yau topology and the particular choice of brane set-up and fluxes.
In particular, it is important to determine under what circumstances one of these $n_{\rm ax}$ light
axions can have an intermediate scale decay constant.

This axion should be associated to a local blow-up mode but if we consider the
simplest case of a GUT model with $SU(5)$ that lies on a diagonal del Pezzo four-cycle $\tau_{\rm dP}$
that does not intersect any other divisor, this axion is definitely eaten up.
In fact, even if the D-term induced shrinking of $\tau_{\rm dP}$ is avoided by giving a non-zero VEV to a SM singlet,
since $SU(5)$ descends from a $U(5)$ gauge group, we must make the $U(1) \subset U(5)$ massive via a diagonal flux upon the stack,
which gauges the axion corresponding to $\tau_{\rm dP}$.

The way to circumvent the above issue is to introduce additional local cycles that intersect the GUT/MSSM stack.
This, then, introduces the additional problem of how to stabilise the corresponding moduli.
In \cite{Joe}, this was performed by an instanton superpotential that included two blow-up moduli,
and then their relative sizes were fixed by loop corrections, leaving a massless axion.
However, there may be difficulties with that approach due to chirality, since it required an instanton coupling to the SM cycle.
Moreover, the massless axion left over is likely to be eaten up as in the example above with a single del Pezzo divisor.

We shall instead advocate a different strategy, where D-term constraints are used to fix the relative sizes of local cycles,
and then the overall size of these is fixed by loop corrections following the general strategy outlined in \cite{arXiv:1110.3333}.
In this way, we are sure that the massless axion left over is definitely not eaten up by any anomalous $U(1)$,
and so it can play the r\^ole of QCD axion with an intermediate scale decay constant.

On top of these local axions, as we discussed in previous sections,
the LVS ensures the presence of at least one additional non-local axion associated to the
overall volume. This axion is also a QCD axion candidate, albeit with a GUT-scale decay constant.
Due to the large size of this decay constant (and also of the decay constants of other possible non-local ALPs),
severe constraints from overproduction of dark matter have to be considered with care
(for a complete set of constraints upon such ALPs see \cite{Arias:2012mb}).
However, as explained in section \ref{SEC:COSMO}, these constraints can be avoided
in the presence of a late out-of-equilibrium decay of heavy moduli which dilute the non-local axions.

In the following sections, we shall discuss several explicit examples with different topologies and
choices of brane set-up and fluxes which lead to different phenomenological features:
\begin{itemize}
\item We shall first focus on a fibred example with $h^{1,1} = 4$ taken from \cite{arXiv:1110.3333}.
\begin{itemize}
\item In section \ref{GUTmodel} we shall analyse an isotropic GUT-like model with $d=2$ D-term constraints leading to
$n_{\rm ax}=1$ light axion with $f_a \simeq M^{\rm 10D}_{\rm KK}/(4\pi)\simeq 5\cdot 10^{14}$ GeV which is associated to the overall volume.
As we have seen in section \ref{SEC:COSMO}, this axion can be the QCD axion in the presence of a dilution mechanism (or tuning of the initial misalignment angle).

\item On the other hand, in section \ref{SU(3)SU(2)model}, we shall study an anisotropic $SU(3)\times SU(2)$ model with
$d=1$ D-term equation leading to $n_{\rm ax}=2$ light axions: the fibre axion $a_f$ with $f_{a_f} \simeq 10^{16}$ GeV,
and the base axion $a_b$ with $f_{a_b} \simeq M^{\rm 6D}_{\rm KK}/(4\pi) \simeq 5$ TeV. As explained in section \ref{SEC:COSMO},
$a_f$ can be the QCD axion in the presence of a dilution mechanism, whereas $a_b$ is a very light and almost decoupled ALP.
\end{itemize}
\item In section \ref{RightQCDaxion} we shall then consider another fibred Calabi-Yau example with $h^{1,1}=5$ taken from \cite{ModReheating},
which allows a chiral model with $n_{\rm ax}=3$ light axions.
One of them is a local axion $a_s$ which plays the r\^ole of the QCD axion
with an intermediate scale decay constant: $f_{a_s} \simeq M_s/\sqrt{4\pi} \sim 10^{10}$ GeV,
whereas the other two light axions, $a_f$ and $a_b$, are non-local and
have $f_{a_f} \sim f_{a_b} \simeq M^{\rm 10D}_{\rm KK}/(4\pi) \sim 10^8$ GeV.
They behave as very light and almost decoupled ALPs.

\item Finally, in section \ref{ManyLocalAxions}, we shall outline the general conditions
under which the EFT can have more than one light axion with an intermediate scale decay constant.
\end{itemize}

\subsection{A GUT-like model with a QCD axion with large decay constant}
\label{GUTmodel}

Here we shall study the model of section 4 of \cite{arXiv:1110.3333} and derive the
low-energy properties of the closed string axions.
This type IIB model involves intersecting magnetised D7-branes,
tadpole and Freed-Witten anomaly cancellation and all the geometric moduli
stabilised in a way compatible with the presence of chirality.
The internal space is an orientifold of a K3 fibred Calabi-Yau three-fold
with $h^{1,1}_+=4$ and $h^{1,1}_-=0$ realised as a hypersurface embedded in a toric variety.
The relevant divisors for model building are the K3 fibre $D_1$, two intersecting rigid four-cycles
$D_4$ and $D_5$ and a del Pezzo divisor $D_7$.
Expanding the K\"ahler form $J$ in this basis as:
\be
J= t_1 \hat{D}_1 + t_4 \hat{D}_4 + t_5 \hat{D}_5 + t_7 \hat{D}_7\,,
\ee
the only non-vanishing intersection numbers are:
\be
k_{145}=k_{447}=k_{777}=2,\qquad k_{155}=k_{444}=k_{455}=k_{477}=-2,\qquad k_{555}=4\,,
\ee
and so the overall volume reads:
\be
\vo=\frac 16\sum_{i,j,k} k_{ijk} t_i t_j t_k = \left(t_1-t_5\right) \left(2 t_4 - t_5\right) t_5 + t_4 t_5^2
-\frac 13 \,t_5^3 - \frac 13 \left(t_4 - t_7\right)^3.
\label{vol}
\ee
The volume of the basis divisors,
which are the real part of the K\"ahler moduli $T_j=\tau_j + {\rm i}\, c_j$, are given by:
\bea
\tau_1&=& \left(2 t_4 - t_5\right) t_5,\qquad
\tau_4=2 t_1 t_5 - t_5^2 - \left(t_4 - t_7\right)^2,\nn \\
\tau_5&=&2 \left(t_1 - t_5\right) \left(t_4 - t_5\right),
\qquad \tau_7 = \left(t_4 - t_7\right)^2.
\eea
The $\mathbb{P}^1$ base of the fibration is given by the intersection of the two rigid cycles $D_4$ and $D_5$:
\be
t_{\rm base}={\rm Vol}(\mathbb{P}^1)={\rm Vol}\left(D_4\cap D_5\right)=\sum_i k_{45i}t_i = 2\left(t_1-t_5\right),
\ee
and so the expression for the volume (\ref{vol}) can also be rewritten as:
\be
\vo=\frac 12 \,t_{\rm base}\tau_1 + t_4 t_5^2 -\frac 13 \,t_5^3 - \frac 13 \tau_7^{3/2}.
\label{vol2}
\ee
The GUT-like model is realised by wrapping $N_a=5$ D7-branes around $D_4$ and $N_b=2$ D7-branes
around $D_5$ (this number has to be even in order to cancel the K-theory charges).
Given that each divisor is transversally invariant under the orientifold,
we obtain symplectic gauge groups $Sp(2 N_a)$ and $Sp(2 N_b)$ which are broken down to unitary groups
by switching on appropriate magnetic fluxes $\mc{F}_a$ and $\mc{F}_b$ on the world-volume of the D7-branes,
so that $Sp(10) \to SU(5) \times U(1)$ and $Sp(4) \to U(1) \times U(1)$.
As explained in the previous sections, the anomalous $U(1)$s become massive via the St\"uckelberg mechanism
by eating up a closed string axion. The $U(1)$ charges of the K\"ahler moduli are given by:
\be
q_{a1} =2, \quad  q_{a4}=1, \quad  q_{a5}=-4, \quad q_{a7}=q_{b4}=-1, \quad q_{b1}=9, \quad q_{b5}=-8, \quad q_{b7}=0\,, \nn
\ee
and so the combinations of two-forms $\tilde{c}_i$ dual to the axions $c_i$ which are eaten are:
\bea
\tilde{c}_a &=& \sum_i q_{ai} \tilde{c}_i = 2 \tilde{c}_1 + \tilde{c}_4 -4 \tilde{c}_5 - \tilde{c}_7 \nn \\
\tilde{c}_b &=& 9 \tilde{c}_1 - \tilde{c}_4 -8 \tilde{c}_5\,.
\label{Comb}
\eea
Therefore the visible sector gauge group at low-energy is $SU(5)\times U(1)$ where the massless $U(1)$ factor turns out
to be very weakly coupled, and so it behaves as a \emph{dark force} \cite{HiddenPhotons}.
The gauge fluxes $\mc{F}_a$ and $\mc{F}_b$ induce also moduli-dependent Fayet-Iliopoulos (FI) terms which fix two K\"ahler moduli at:
\be
\tau_4=\frac{3}{19}\,\tau_1-\tau_7\,, \qquad \tau_5=\frac{18}{19}\,\tau_1\,.
\label{DtermStab}
\ee
Notice that the combinations of axions that are eaten up are the same as the combinations of K\"ahler moduli which are fixed by the FI-terms:
\be
c_a = \frac{3}{19}\,c_1 -c_4 -c_7, \qquad
c_b = \frac{18}{19}\,c_1-c_5\,.
\label{Combax}
\ee
Given that the two moduli fixed by the D-terms obtain an $\mc{O}(M_s)$ mass, the EFT
can be studied just in terms of the two remaining K\"ahler moduli $\tau_1$ and $\tau_7$.
Thus we can simply integrate out the fixed moduli and work with an `effective' volume form to determine the masses of the light axions.
This is particularly pertinent given that D-term corrections to the potential typically dominate over all others.

The D-term stabilisation (\ref{DtermStab}) gives a volume of the form:
\be
\vo=\alpha\left(\tau_1^{3/2}-\gamma\tau_7^{3/2}\right)\,,
\label{vol2Dterms}
\ee
where $\alpha= 86/(57\sqrt{19})$ and $\gamma= 1/(3\alpha)$. The form of the volume (\ref{vol2Dterms})
is very similar to (\ref{volSwiss}), and so the canonical normalisation takes the same form as
the simplest Swiss-cheese example of section \ref{ScCY}:
\bea
\frac{c_1}{\tau_1} &\simeq & \mc{O}\left(1\right)\,a_1 +\mc{O}\left(\tau_7^{3/4}\,\vo^{-1/2}\right) \,a_7\,, \\
\frac{c_7}{\tau_7} &\simeq& \mc{O}\left(1\right)\,a_1 +\mc{O}\left(\tau_7^{-3/4}\,\vo^{1/2}\right) a_7\,.
\eea
In order to stabilise $\tau_1$ and $\tau_7$, the authors of \cite{arXiv:1110.3333} introduced also $N_{gc}=4$
D7-branes around the del Pezzo divisor $D_7$ which give rise to a pure $\mc{N}=1$ $Sp(8)$ hidden sector that undergoes
gaugino condensation. The interplay between this non-perturbative effect and the leading order $\alpha'$ correction to the
K\"ahler potential fixes both $\tau_1$ and $\tau_7$ following the standard LVS procedure. The minimum is at:
\be
\tau_7 \sim g_s^{-1} \qquad \text{and} \qquad
\vo\sim W_0 \,e^{\frac{2\pi  \tau_7}{5}} \sim W_0 \,e^{\frac{2\pi}{5}\frac{1}{g_s}}\,.
\ee
For values of $W_0$ of order unity, the volume, and so $\tau_1$, would be exponentially large
resulting in an exponentially small value of the visible sector gauge coupling since from (\ref{DtermStab})
we realise that $\alpha_{\rm GUT}^{-1} = \tau_4$ is proportional to $\tau_1$.
Thus this phenomenological requirement tells us that we need to fine tune $W_0\ll 1$.
This tuning is actually also needed to obtain TeV-scale SUSY since,
for $\vo \sim \mc{O}(10^4)$, the gravitino mass (and the soft terms
generated via gravity mediation) turns out to be of the order the TeV scale
only if $W_0 \sim \mc{O}(10^{-10})$.
Moreover the non-perturbative effects on $D_7$ generate a potential for the axion $c_7$, and so
the canonically normalised axion $a_7$ obtains a mass of the order $m_{3/2}$. This axion is
definitely too heavy for playing the r\^ole of the QCD axion, and so we shall focus on the remaining
axion $a_1$ which remains massless since the corresponding saxion, i.e. $\tau_1$, is fixed
purely perturbatively via $\alpha'$ corrections.

The couplings of $a_1$ to the gauge bosons living on $D_4$ and $D_5$ scale as:
\bea
\mc{L} &\supset&  \frac{\left(\frac{3}{19}\,c_1-c_7\right) }{M_P}\, g_{SU(5)}^2\tr(F_{SU(5)} \wedge F_{SU(5)})
+  \frac{18}{19}\frac{c_1}{M_P} \,g_{U(1)}^2\tr(F_{U(1)}\wedge F_{U(1)})  \nn\\
&\simeq & \mc{O}\left(\frac{1}{M_P}\right)\,\rhoo_1 \tr(F_{SU(5)} \wedge F_{SU(5)}) + \mc{O}\left(\frac{1}{M_P}\right)\,\rhoo_1 \tr(F_{U(1)} \wedge F_{U(1)}).
\eea
The volume scaling of these couplings can again be understood in terms of the locality argument of (\ref{EQ:LOCALITY}).
In fact, the two rigid divisors $D_4$ and $D_5$ are non-local effects since they control the size of the overall volume by
defining the $\mathbb{P}^1$ base of the fibration: $t_{\rm base}={\rm Vol}\left(D_4\cap D_5\right)$. Hence we correctly
found that the massless axion couples to these gauge bosons with Planck strength.

The axion decay constant instead scale as the 10D KK scale:
\be
f_{a_1}\simeq \frac{M_P}{4\pi\tau_1} \simeq \frac{M^{\rm 10D}_{\rm KK}}{4\pi} \simeq 5\cdot 10^{14}\,{\rm GeV}\,,
\ee
while the anomaly coefficient is of order unity, $\aC[1,1]\simeq \mc{O}(1)$, and so $a_1$
can play the r\^ole of the QCD axion. As explained in section \ref{SEC:COSMO}, the abundance
of the axion $a_1$ will be diluted by the late out-of-equilibrium decay of $\tau_7$ and $a_7$.

\subsection{An $SU(3) \times SU(2)$ model with a QCD axion with large decay constant plus a light ALP}
\label{SU(3)SU(2)model}

Here we study the second example of \cite{arXiv:1110.3333} which has two massless axions.
This is another globally consistent chiral model
with magnetised D7-branes and moduli stabilisation within type IIB Calabi-Yau compactifications.
The internal manifold is the same described in section \ref{GUTmodel}. However this time the gauge group
is not GUT-like since the brane fluxes are chosen in such a way to give rise to just one D-term condition,
so that the volume of the Calabi-Yau is no longer proportional to an inverse power of the visible sector
gauge coupling. In turn, $W_0$ can be set of order unity and the volume can be exponentially large lowering the
string scale to the intermediate value $M_s \simeq 5\cdot 10^{10}$ GeV and obtaining TeV-scale supersymmetry without any fine-tuning.
Thus there is no space anymore for GUT theories but only for an MSSM-like construction.
This is achieved by wrapping $N_a=3$ D7-branes around $D_4$ and $N_{K3}=1$ D7-branes
around the K3 fibre $D_1$. The corresponding gauge groups are $Sp(6)$, which is broken down to $SU(3) \times U(1)$
by switching on appropriate magnetic flux $\mc{F}_a$ on $D_4$, and $Sp(2)\cong SU(2)$.
Notice that the total gauge flux on $D_1$ is zero.

The anomalous $U(1)$ becomes massive via the St\"uckelberg mechanism
by eating up a closed string axion. The $U(1)$ charges of the K\"ahler moduli are given by:
\be
q_{a1} =2, \quad  q_{a4}=1, \quad  q_{a5}=-4, \quad q_{a7}=-1\,, \nn
\ee
and so the combination of two-forms (dual to the axions) which is eaten up is:
\be
\tilde{c}_a = \sum_i q_{ai} \tilde{c}_i = 2 \tilde{c}_1 + \tilde{c}_4 -4 \tilde{c}_5 - \tilde{c}_7\,. \nn
\ee
Therefore the visible sector gauge group at low-energy is $SU(3)\times SU(2)$.
The gauge flux $\mc{F}_a$ induces also a moduli-dependent FI term which fixes one K\"ahler modulus at:
\be
\tau_4=3\left(\tau_1-\tau_5\right)-\tau_7\,,
\label{DtermStabn}
\ee
and so the combination of axions eaten is:
\be
c_a = 3 c_1 - c_4 -3 c_5 - c_7\,.
\ee
Given that the modulus fixed by the D-term has an $\mc{O}(M_s)$ mass, the EFT
can be studied just in terms of the three remaining K\"ahler moduli $\tau_1$, $\tau_5$ and $\tau_7$.
The D-term stabilisation (\ref{DtermStabn}) gives a volume of the form:
\be
\vo=\frac 13 \left( \sqrt{\tau_s}\, \tau_b - \,\tau_7^{3/2}\right)\,,
\label{vol1Dterm}
\ee
where the variables $\tau_1$ and $\tau_5$ have been traded for $\tau_s$ and $\tau_b$
defined as:
\be
\tau_s\equiv\tau_1-\tau_5, \qquad \tau_b\equiv\frac{10\tau_1-\tau_5}{2}\,,
\ee
where the label $s$ stays for `small' and the label $b$ for `big'.
This redefinition is suggested by the D-term condition (\ref{DtermStabn}) which
implies that the combination $\left(\tau_1-\tau_5\right)$
has to be fixed small in order to obtain a visible sector
gauge coupling $\alpha_{\rm vis}^{-1}=\tau_4$ which is not too small
The form of the volume (\ref{vol1Dterm})
is very similar to (\ref{volFib}) once we identify the base $t_b$ with $\tau_b/\sqrt{\tau_s}$ and the fibre
$\tau_f$ with $\tau_s$. Given that we are interested in the regime where $\tau_s$ is much smaller
than $\tau_b$, we shall consider the anisotropic limit where the size of the base $\sqrt{t_b}$ is much bigger
than the size of the fibre $\tau_f^{1/4}$. Hence the canonical normalisation takes the same form as
the fibred example of section \ref{fibCY}:
\bea
\frac{c_s}{\tau_s} &\simeq & \mc{O}\left(1\right) a_s + \mc{O}\left(\tau_7^{3/2} \vo^{-1}\right) a_b
+\mc{O}\left(\tau_7^{3/4}\vo^{-1/2}\right) \,a_7\,,
\label{cfn} \\
\frac{c_b}{\tau_b} &\simeq & \mc{O}\left(\tau_7^{3/2}\vo^{-1}\right) a_s + \mc{O}\left(1\right) a_b
+\mc{O}\left(\tau_7^{3/4}\vo^{-1/2}\right) \,a_7\,,
\label{cbn} \\
\frac{c_7}{\tau_7} &\simeq& \mc{O}\left(1\right) a_s + \mc{O}\left(1\right)\,a_b +\mc{O}\left(\tau_7^{-3/4} \vo^{1/2}\right) a_7\,.
\label{csn}
\eea
In order to stabilise $\vo \simeq \sqrt{\tau_s}\tau_b$ and $\tau_7$, the authors of \cite{arXiv:1110.3333}
introduced also $N_{gc}=3$ D7-branes around the del Pezzo divisor $D_7$
which give rise to a pure $\mc{N}=1$ $Sp(6)$ hidden sector that undergoes
gaugino condensation. The interplay between this non-perturbative effect and the leading order $\alpha'$ correction to the
K\"ahler potential fixes both the combination $\vo \simeq \sqrt{\tau_s}\tau_b$ and $\tau_7$ following the standard LVS procedure.
The minimum is at:
\be
\tau_7 \sim g_s^{-1} \quad \text{and} \quad
\vo\sim W_0 \,e^{\frac{\pi  \tau_7}{2}} \sim W_0 \,e^{\frac{\pi}{2}\frac{1}{g_s}}\,.
\ee
For natural values of $W_0$ of order unity and a volume of order $\vo \sim \mc{O}(10^{14})$,
the gravitino mass (and the soft terms
generated via gravity mediation) is around the TeV scale and the string scale is intermediate:
$M_s \simeq 5\cdot 10^{10}$ GeV.

Moreover the non-perturbative effects on $D_7$ generate a potential for the axion $c_7$, and so
the canonically normalised axion $a_7$ obtains a mass of the order $m_{3/2}$. This axion is
definitely too heavy for playing the r\^ole of the QCD axion, and so we shall focus on the remaining
massless axions $a_s$ and $a_f$. These axions remain massless since the corresponding saxions
are fixed purely perturbatively via $\alpha'$ and string loop corrections to the K\"ahler potential.
The couplings of $a_s$ and $a_f$ to the gauge bosons living on $D_4$ and $D_1$ scale as:
\bea
\mc{L} &\supset&  \frac{\left(3\,c_s-c_7\right) }{M_P}\, g_{SU(3)}^2\tr(F_{SU(3)} \wedge F_{SU(3)})
+  \frac 19\frac{\left(2 c_b - c_s\right)}{M_P} \,g_{SU(2)}^2\tr(F_{SU(2)}\wedge F_{SU(2)})  \nn\\
&\simeq & \mc{O}\left(\frac{1}{M_P}\right)\,a_s \,\tr(F_{SU(3)} \wedge F_{SU(3)})
+ \mc{O}\left(\frac{1}{M_P}\right)\,a_b \,\tr(F_{SU(3)} \wedge F_{SU(3)}) \nn \\
&+& \mc{O}\left(\frac{\tau_7^{3/2}}{\vo M_P}\right)\,a_s\, \tr(F_{SU(2)} \wedge F_{SU(2)})
+ \mc{O}\left(\frac{1}{M_P}\right)\,a_b\, \tr(F_{SU(2)} \wedge F_{SU(2)}).
\eea
The volume scaling of these couplings can again be understood in terms the locality argument of (\ref{EQ:LOCALITY}).
In fact, both the K3 fibre $\tau_1$ and the rigid divisor $D_4$ are non-local effects since they control the size of the overall volume
(recall that $D_4$ controls the size of the $\mathbb{P}^1$ base of the fibration). Hence we correctly
found that the massless axions couple to these gauge bosons with Planck strength or weaker.

In the anisotropic limit $\tau_b \gg \tau_s \sim \tau_7 \simeq \mc{O}(10)$,
the axion decay constants become:
\be
f_{a_s}\simeq \frac{M_P}{4\pi\tau_s}\simeq 10^{16}\,{\rm GeV}\,,\qquad f_{a_b}\simeq
\frac{M_P}{4\pi\tau_b} \simeq \frac{M^{\rm 6D}_{\rm KK}}{4\pi}\simeq 5\,{\rm TeV}\,,
\ee
while the anomaly coefficients look like (see (\ref{Csaniso})):
\be
\aC[s, SU(3)] \sim \aC[s, SU(2)] \sim \aC[b, SU(2)] \simeq \mc{O}\left(1\right),
\quad \aC[b, SU(3)] \simeq \mc{O}\left(\vo^{-1}\right). \nn
\ee
As described in section \ref{SEC:COSMO}, $a_s$ can be the QCD axion with an
$\mc{O}(M_{\rm GUT})$ decay constant. This value is not in the allowed window
but it might still be viable in the presence of the
late out-of-equilibrium decay of the heavy moduli $\tau_7$ and $a_7$
which would dilute the axion $a_s$ \cite{Acharya:2010zx}.
The other axion $a_b$ is instead a light ALP with a very low decay constant which does
not create any problem with dark matter overproduction. We finally mention that the
$SU(2)$ gauge group is a hyperweak interaction since it is realised via D7-branes wrapping
an exponentially large four-cycle.

\subsection{A chiral model with a QCD axion with intermediate decay constant plus two light ALPs}
\label{RightQCDaxion}

In this section we shall present an explicit type IIB model with the presence of a closed string modulus which
is a perfect QCD axion candidate since it is not eaten up by any anomalous $U(1)$,
does not develop any potential by non-perturbative effects in $g_s$
and has a decay constant at the intermediate scale.

The model is very similar to the one used in \cite{ModReheating} to obtain a modulated reheating scenario
with the production of detectable non-Gaussianities of local form. It is based on
an orientifold of a Calabi-Yau three-fold with a K3 or a $T^4$ fibration structure and $h^{1,1}_+=5$ and $h^{1,1}_-=0$.
Again, this internal space can be realised as a hypersurface embedded in a toric variety \cite{FibCY}.
The relevant divisors for model building are the K3 or $T^4$ fibre $D_1$, the divisor $D_2$
controlling the $\mathbb{P}^1$ base of the fibration, a del Pezzo four-cycle $D_3$ and two
rigid divisors $D_4$ and $D_5$ which intersect each other but have no intersections
with the other four-cycles. The K\"ahler form $J$ can be expanded in this basis as:
\be
J= t_1 \hat{D}_1 +t_2 \hat{D}_2-t_3 \hat{D}_3-t_4 \hat{D}_4-t_5 \hat{D}_5\,,
\ee
where we took a minus sign for the rigid cycles in order to have positive dual two-cycles.
All the non-vanishing intersection numbers are positive and are:
\be
k_{122}\,, \quad k_{333}\,, \quad k_{444}\,, \quad k_{445}\,, \quad k_{455}\,, \quad k_{555}\,,
\ee
with the relation $k_{455}^2=k_{445} k_{555}$.
Thus the overall volume reads:
\be
 \vo\,=\,\alpha \left[ \sqrt{\tau_1} \tau_2-\gamma_3 \tau_3^{3/2}- \gamma_5 \tau_5^{3/2}
 -\gamma_4 \left(\tau_4- x\,\tau_5\right)^{3/2} \right],
 \label{volQCDax}
\ee
where $\alpha$, $\gamma_3$, $\gamma_4$, $\gamma_5$ and $x$ are $\mc{O}(1)$ constants which depend on the
intersection numbers (in particular $x=k_{445}/k_{455}$).
In \cite{ModReheating}, the authors consider a particular set-up
suitable to build a phenomenologically viable model of modulated reheating.
The visible sector is built via two stacks of intersecting D7-branes wrapping two different
combinations of the rigid divisors $D_4$ and $D_5$.
Moreover an E3-instanton wrapped around $D_5$ gives rise to poly-instanton corrections to the superpotential
since $D_5$ is assumed to be a rigid four-cycle with Wilson lines, {\em i.e.}~$h^{2,0}(D_5)=0$ while $h^{1,0}(D_5)= 1$ \cite{poly}.
The gauge fluxes are chosen in such a way to yield chirality at the intersection
between the two visible sector stacks but with the generation of just one FI-term.
Furthermore, no chiral matter at the intersection
of the visible sector with the E3-instanton is induced. This set-up allows to keep the field $\tau_5$
light since it develops a potential only via tiny poly-instanton effects. This is a crucial condition to
obtain a viable modulated reheating scenario. More precisely, three effects develop a potential for
$\tau_4$, $\tau_5$ and the absolute value of a open string field $|\phi|$ which is a visible sector singlet:
\begin{enumerate}
\item D-terms fix $|\phi|$\,,
\item String loop corrections to the K\"ahler potential fix the combination $\left(\tau_4- x\,\tau_5\right)$\,,
\item Poly-instanton corrections to the superpotential fix $\tau_5$\,.
\end{enumerate}
Moreover, as in standard LVS, $\tau_3$ is fixed non-perturbatively, the volume by the interplay of non-perturbative
effects on $D_3$ and $\alpha'$ corrections to the K\"ahler potential, while the fibre modulus $\tau_1$ is stabilised by
string loops.

Due to the locality argument of (\ref{EQ:LOCALITY}) only the axions of $\tau_4$, $\tau_5$ and $\phi$ (the corresponding phase)
couple to the visible sector as $1/M_s$, and so can be good QCD axion candidates if the string scale is intermediate.
However, the axion of $\tau_5$ cannot be the QCD axion since it develops a non-perturbative
superpotential. On the other hand, a combination of the axion of $\tau_4$ and the phase of $\phi$ will be eaten up by the anomalous $U(1)$
while the other can play the r\^ole of the QCD axion.
If one is not interested in keeping $\tau_5$ very light, then one could drop the assumption that $D_5$ has Wilson lines
resulting in the absence of poly-instanton corrections. Then also this divisor would be frozen by string loop effects, and
the corresponding axion would also remain massless and have the right coupling to QCD.

In the analysis of \cite{ModReheating}, the open string field $\phi$ plays a crucial r\^ole
because in its absence it is impossible to obtain a light modulating field.
In fact, a combination of $\tau_4$ and $\tau_5$ would be
fixed by D-terms while another would be stabilised by string loops which beat the tiny poly-instanton effects
and generate a mass which is not below the Hubble constant (a crucial condition to find a viable modulated reheating).
However, given that we are not interested in modulated reheating, we shall explore exactly this simpler possibility
since it involves no open string fields, and so we consider it to be more model-independent.

In our set-up the visible sector is realised by two stacks $N_a$ and $N_b$ of intersecting D7-branes
wrapped around the divisors $D_4$ and $D_5$: $D_a = D_4$ and $D_b = D_5$.
We also turn on the following world-volume fluxes:
\be
F_a = \left(f_4 +\frac 12\right) \hat D_4+ f_5 \hat D_5\,, \qquad
F_b = g_4 \hat D_4 + \left(g_5+\frac 12\right) \hat D_5 \,,
\ee
where the half-integer contributions come from the cancellation of the Freed-Witten anomalies
\footnote{The value of the $B$-field is fixed in order to cancel the total flux $\mc{F}_3=F_3-B$ on $D_3$ so that
the instanton wrapped on this del Pezzo contributes to the superpotential. Given that $D_3$ does not intersect with
$D_4$ and $D_5$, the gauge flux $F$ and the total flux $\mc{F}=F-B$ on $D_4$ and $D_5$ give rise to the same phenomenological
features.}.
The $U(1)$-charges of the K\"ahler moduli induced by these fluxes can be written one in terms of the other as:
\be
x q_{a\,5} = q_{a\,4}- \left(k_{444}-\frac{k_{445}^2}{k_{455}}\right) \left(f_4 +\frac{1}{2}\right), \qquad
x\, q_{b\,5} = q_{b\,4}.
\label{chargerelations}
\ee
The moduli-dependent FI-terms take the form:
\bea
 \xi_a&=&\frac{1}{4\pi\vo}\int J\wedge F_a\wedge \hat{D}_a
 =\frac{1}{4\pi}\left(q_{a\,4} \, \frac{\partial K}{\partial \tau_4} + q_{a\,5} \, \frac{\partial K}{\partial \tau_5}\right), \label{FIvis} \\
 \xi_b&=&\frac{1}{4\pi\vo}\int J\wedge F_b\wedge \hat{D}_b
 =\frac{1}{4\pi}\left(q_{b\,4} \,\frac{\partial K}{\partial \tau_4} + q_{b\,5} \, \frac{\partial K}{\partial \tau_5}\right). \label{FIint}
\eea
The chiral intersections between the stacks of D7-branes on $D_4$ and $D_5$ depend on the gauge fluxes in the following way:
\be
 I_{ab}=\int \left(F_a - F_b\right) \wedge \hat D_a \wedge \hat{D}_b
 =q_{a\,5}-q_{b\,4} . \nonumber
\ee
Setting $g_4=- k_{555}/(2 k_{455})$ and $g_5=0$, we obtain $q_{b\,4}=q_{b\,5}=0$
which implies $\xi_b=0$, and so we are left over with just one FI-term.
Consequently, just one closed string axion will be eaten up.
Moreover, we obtain $I_{ab}=q_{a\,5}$ which implies $q_{a\,5}\neq 0$
in order to have chiral matter.

From (\ref{chargerelations}) we notice that we can choose the fluxes in such a way
to have $q_{a\,5}\neq 0$ and $q_{a\,4}= 0$ at the same time, so that the
FI-term (\ref{FIvis}) simplifies to \footnote{The choice of fluxes to produce $q_{a\,4}= 0$
is just meant to simplify the system since the same considerations would apply also to the more general case with $q_{a\,4}\neq 0$.}:
\be
\xi_a=\frac{q_{a\,5}}{4\pi} \, \frac{\partial K}{\partial \tau_5}
= \frac{q_{a\,5}}{4\pi} \frac{3 \alpha \left(\gamma_5 \sqrt{\tau_5}- \gamma_4 \,x\sqrt{\hat\tau_4}\right)} {\vo}\,,
\ee
where we introduced the combination $\hat\tau_4$ defined as $\hat\tau_4\equiv \tau_4-x\,\tau_5$.
In the absence of open string singlets which can develop a non-zero VEV, the vanishing
of the D-terms implies $\xi_a=0$, and so one combination of $\tau_4$ and $\tau_5$ is fixed at:
\be
\tau_5 = \lambda \,\hat\tau_4\qquad\text{where}\qquad\lambda\equiv \left(\frac{\gamma_4\,x}{\gamma_5}\right)^2\,.
\label{FixD}
\ee
This combination can be fixed within the regime of validity of the EFT
due to the intersection between $D_4$ and $D_5$. In fact, in the absence of this intersection for $x=0$, (\ref{FixD})
would imply the shrinking of the rigid divisor $D_5$ (see also \cite{arXiv:1110.3333} for related issues).
Moreover, (\ref{FixD}) implies that the combination of axions eaten is:
\be
c_a = c_5 - \lambda\, \hat{c}_4\,, \nn
\ee
where $\hat{c}_4 \equiv c_4 -x\,c_5$. Given that the modulus fixed by the D-term receives an $\mc{O}(M_s)$ mass, the EFT
can be studied in terms of the four remaining K\"ahler moduli $\tau_1$, $\tau_2$, $\tau_3$ and $\hat\tau_4$.
Hence, after the D-term stabilisation (\ref{FixD}), the volume (\ref{volQCDax}) simplifies to:
\be
\vo\,=\,\alpha \left( \sqrt{\tau_1} \tau_2-\gamma_3 \tau_3^{3/2}- \lambda_4 \hat\tau_4^{3/2} \right),
\quad\text{where}\quad\lambda_4\equiv \gamma_5 \lambda^{3/2}+\gamma_4\,.
\label{volQCD}
\ee
The form of the volume (\ref{volQCD})
is very similar to (\ref{volFib}) once we identify the base $t_b$ with $\tau_2/\sqrt{\tau_1}$ and the fibre
$\tau_f$ with $\tau_1$ (setting $\alpha=1$ without loss of generality). Hence the canonical normalisation,
in the limit $\vo\simeq \sqrt{\tau_1}\tau_2\gg \tau_3 \sim \hat\tau_4$, takes the same form as
the fibred example of section \ref{fibCY}:
\bea
\frac{c_1}{\tau_1} &\simeq & \mc{O}\left(1\right) a_1 + \mc{O}\left(\frac{\hat\tau_4^{3/2}}{\vo}\right) a_2
+\mc{O}\left(\frac{\tau_3^{3/4}}{\vo^{1/2}}\right) \,a_3 +\mc{O}\left(\frac{\hat\tau_4^{3/4}}{\vo^{1/2}}\right) \,a_4\,,
\label{cfQCD} \\
\frac{c_2}{\tau_2} &\simeq & \mc{O}\left(\frac{\hat\tau_4^{3/2}}{\vo}\right)a_1 + \mc{O}\left(1\right) a_2
+\mc{O}\left(\frac{\tau_3^{3/4}}{\vo^{1/2}}\right) \,a_3 +\mc{O}\left(\frac{\hat\tau_4^{3/4}}{\vo^{1/2}}\right) \,a_4\,,
\label{cbQCD} \\
\frac{c_3}{\tau_3} &\simeq& \mc{O}\left(1\right) a_1 + \mc{O}\left(1\right)\,a_2
+\mc{O}\left(\frac{\vo^{1/2}}{\tau_3^{3/4}}\right) a_3 +\mc{O}\left(\frac{\hat\tau_4^{3/4}}{\vo^{1/2}}\right) a_4\,.
\label{cs1QC} \\
\frac{\hat{c}_4}{\hat\tau_4} &\simeq& \mc{O}\left(1\right) a_1 + \mc{O}\left(1\right)\,a_2
+\mc{O}\left(\frac{\tau_3^{3/4}}{\vo^{1/2}}\right) a_3 +\mc{O}\left(\frac{\vo^{1/2}}{\hat\tau_4^{3/4}}\right) a_4\,.
\label{cs2QCD}
\eea
Notice that the mixing terms between the axions $a_3$ and $a_4$ associated to the blow-up modes
$\tau_3$ and $\hat\tau_4$ are very suppressed due to
the fact that these two rigid divisors do not intersect each other, and so they correspond to the resolution of
singularities which are localised in different regions of the Calabi-Yau background \cite{Reheating}.

The overall volume and $\tau_3$ are fixed by the leading order F-term scalar potential
following the standard LVS procedure which requires non-perturbative effects on $D_3$
and the leading order $\alpha'$ correction to the
K\"ahler potential. The minimum is at:
\be
\tau_3 \sim g_s^{-1} \qquad \text{and} \qquad
\vo\sim W_0 \,e^{\frac{2 \pi  \tau_7}{N_3}}\,,
\ee
where $N_3=1$ in the case of an E3-instanton whereas $N_3$ is the rank of the condensing gauge group in
the case of gaugino condensation on D7-branes.
For natural values of $W_0\simeq \mc{O}(1)$ and a volume of order $\vo \simeq \mc{O}(10^{14})$,
the gravitino mass (and the soft terms generated via gravity mediation)
is around the TeV scale and the string scale is intermediate:
$M_s \simeq 5\cdot 10^{10}$ GeV.

The non-perturbative effects on $D_3$ generate a potential for the axion $c_3$, and so
the canonically normalised axion $\rhoo_3$ obtains a mass of the order $m_{3/2}$. This axion is
definitely too heavy for playing the r\^ole of the QCD axion, and so we shall focus on the remaining
massless axions $a_1$, $a_2$ and $a_4$.
These axions remain massless since the corresponding saxions, i.e. $\vo$, $\tau_1$ and $\hat\tau_4$,
are fixed purely perturbatively via $\alpha'$ and string loop corrections to the K\"ahler potential \cite{arXiv:1110.3333,K3Stab}.
As explained in \cite{K3Stab}, in order to fix $\tau_1$ via $g_s$ effects, two stacks of D7-branes have to
be wrapped around $D_1$ and $D_2$, whereas the string loops which stabilise $\hat\tau_4$ are
sourced by the visible sector stacks on $D_4$ and $D_5$. In fact, following the systematic analysis of
string loop corrections performed in \cite{Loops}, the leading order behaviour of the effective
scalar potential generated by $g_s$ effects is:
\be
V_{(g_s)}=\left(\frac{\mu_1}{\sqrt{\hat\tau_4}} -\frac{\mu_2}{\sqrt{\hat\tau_4} - \mu_3} \right)
\frac{W_0^2}{\vo^3}+\mc{O}\left(\frac{1}{\vo^4}\right),
\label{Vloop}
\ee
where $\mu_1$ and $\mu_2$ are positive constants which depend on the complex structure moduli,
while $\mu_3$ is a positive constant that depends on the VEV of stabilised moduli like $\tau_3$
once we include the intersection of the branes wrapping $D_4$ and $D_5$ with the O7-planes
and additional Whitney-type branes for tadpole cancellation (see \cite{arXiv:1110.3333}).
For natural $\mc{O}(1)$ values of the coefficients of the string loop corrections,
the potential (\ref{Vloop}) can develop a minimum at:
\be
\hat\tau_4 = \frac{\mu_1 \,\mu_3^2}{\left(\sqrt{\mu_1} + \sqrt{\mu_2}\right)^2}\sim \mc{O}(10).
\ee
This minimum is indeed located at small $\hat\tau_4$ since for example for $\mu_1=\mu_2=1$ and $\mu_3=10$,
we obtain $\hat\tau_4=25$.

The couplings of $a_1$, $a_2$ and $a_4$ to the gauge bosons living on $D_1$, $D_2$, $D_4$ and $D_5$ scale as
(recalling that $\tau_4 = \hat\tau_4 +x\,\tau_5$ and $\tau_5=\lambda\hat\tau_4$ $\Leftrightarrow$
$\tau_4 = \left(1+x\,\lambda\right) \hat\tau_4$): \footnote{We neglect the axionic coupling
to hidden sector gauge bosons living on $D_3$ in the case of gaugino condensation.}
\bea
\mc{L} &\supset&  \frac{c_1 }{M_P}\, g_1^2\tr(F_1 \wedge F_1)
+  \frac{c_2}{M_P} \,g_2^2\tr(F_2\wedge F_2) \nn\\
&+& \left(1+x\,\lambda\right)\frac{\hat{c}_4}{M_P}\, g_4^2\tr(F_4 \wedge F_4) + \lambda\frac{ \hat{c}_4}{M_P} \,g_5^2\tr(F_5\wedge F_5)  \nn\\
&\simeq & \left[\mc{O}\left(\frac{1}{M_P}\right)\,a_1 + \mc{O}\left(\frac{\hat\tau_4^{3/2}}{\vo M_P}\right)\,a_2
+ \mc{O}\left(\frac{\hat\tau_4^{3/4}}{\vo^{1/2} M_P}\right) \,a_4 \right] \tr(F_1 \wedge F_1) \nn \\
&+& \left[\mc{O}\left(\frac{\hat\tau_4^{3/2}}{\vo M_P}\right)\,a_1 + \mc{O}\left(\frac{1}{M_P}\right)\,a_2
+ \mc{O}\left(\frac{\hat\tau_4^{3/4}}{\vo^{1/2} M_P}\right) \,a_4 \right] \tr(F_2 \wedge F_2) \nn \\
&+& \sum_{i=4}^5\left[\mc{O}\left(\frac{1}{M_P}\right)\,a_1
+ \mc{O}\left(\frac{1}{M_P}\right)\,a_2 + \mc{O}\left(\frac{\vo^{1/2}}{\hat\tau_4^{3/4} M_P}\right)\,a_4\right]\tr(F_i \wedge F_i)\,. \nn
\eea
We stress that we managed to obtain a closed string axion, $a_4$, which is
not eaten up by any anomalous $U(1)$ and couples to the visible sector on $D_4$ and $D_5$ as $1/M_s$ opposed to
the standard $1/M_P$ gravitational couplings.

The axion decay constants instead become:
\be
f_{a_1} = \frac{M_P}{4\pi\tau_1}\,,
\qquad f_{a_2} \simeq \frac{M_P}{4\pi\tau_2}\,,
\qquad f_{a_4} \simeq \frac{M_s}{\sqrt{4\pi}}\simeq 10^{10}\,{\rm GeV}\,, \nn
\ee
whereas the anomaly coefficients look like:
\bea
\aC[1,1] &\simeq & \mc{O}\left(1\right), \qquad
\aC[2,1] \simeq \mc{O}\left(\frac{(\hat\tau_4\tau_1)^{3/2}}{\vo^2}\right),
\qquad \aC[4,1]\simeq \mc{O}\left(\frac{\tau_1\sqrt{\hat\tau_4}}{\vo}\right) \nn \\
\aC[1,2] &\simeq & \mc{O}\left(\left(\frac{\hat\tau_4}{\tau_1}\right)^{3/2}\right), \qquad
\aC[2,2]\simeq \mc{O}\left(1\right),\qquad
\aC[4,2]\simeq \mc{O}\left(\sqrt{\frac{\hat\tau_4}{\tau_1}}\right), \nn \\
\aC[1,i] &\simeq& \mc{O}\left(\frac{\hat\tau_4}{\tau_1}\right),\quad
\aC[2,i]\simeq \mc{O}\left(\frac{\hat\tau_4\sqrt{\tau_1}}{\vo}\right),
\quad \aC[4,i]\simeq \mc{O}\left(1\right),\quad\forall i=4,5.
\eea
The field $a_4$ is a perfect QCD axion candidate since
its decay constant is intermediate and $\aC[4,i]\simeq \mc{O}(1)$ $\forall i=4,5$.
Therefore we have managed to build an explicit example where a closed string axion
with these features \cite{Joe} survives in the EFT since it is not eaten up by any anomalous $U(1)$.
As pointed out in section \ref{SEC:COSMO}, this axion will not form a relevant component
of dark matter since it will be diluted by the decay of $\hat\tau_4$. Thus the constraints
from isocurvature fluctuations do not apply to this case.

Moreover, there are two further ALPs, $a_1$ and $a_2$, whose
behaviour is different in the case of a isotropic or anisotropic Calabi-Yau.
\begin{itemize}
\item {\bf Isotropic case:} in the isotropic limit $\tau_1 \sim \tau_2 \sim \vo^{2/3}$,
the decay constants of $a_1$ and $a_2$ scale as the 10D KK scale:
\be
f_{a_1}\simeq f_{a_2}\simeq \frac{M_P}{4\pi\vo^{2/3}} \simeq \frac{M^{\rm 10D}_{\rm KK}}{4\pi}\simeq 10^8\,{\rm GeV}\,\,,
\ee
and the anomaly coefficients simplify to:
\be
\aC[1,1] \simeq \aC[2,2] \simeq \mc{O}\left(1\right), \quad
\aC[2,1] \simeq \aC[1,2] \simeq \mc{O}\left(\vo^{-1}\right), \quad
\aC[1,i] \simeq \aC[2,i]\simeq \mc{O}\left(\vo^{-2/3}\right). \nn
\ee
The two non-local axions $a_1$ and $a_2$ correspond to two light and almost decoupled ALPs
that do not create any problem with dark matter overproduction. Moreover the
gauge theories on $\tau_1$ and $\tau_2$ are hyperweak.
\item {\bf Anisotropic case:} in the anisotropic limit $\tau_2 \gg \tau_1 \sim \hat\tau_4$,
the decay constants of $a_1$ and $a_2$ become:
\be
f_{a_1}\simeq \frac{M_P}{4\pi\hat\tau_4}\simeq 10^{16}\,{\rm GeV}\,,
\qquad f_{a_2}\simeq \frac{M_P}{4\pi\tau_2} \simeq \frac{M^{\rm 6D}_{\rm KK}}{4\pi}\simeq 5\,{\rm TeV}\,,
\ee
while the anomaly coefficients take the simplified form:
\be
\aC[1,1] \simeq \aC[1,2] \simeq \aC[2,2]\simeq \aC[1,i] \simeq \mc{O}\left(1\right), \quad
\aC[2,i] \simeq \mc{O}\left(\vo^{-1}\right), \quad
\aC[2,1] \simeq \mc{O}\left(\vo^{-2}\right). \nn
\ee
\end{itemize}
Now the anomaly coefficient $\aC[1,i]$ is of order unity, and so also $a_f$ can play the r\^ole
of the QCD axion but with a GUT-scale decay constant. However, there are no
problems with dark matter overproduction since, as explained in section \ref{SEC:COSMO},
the abundance of the axion $a_1$ will be diluted by the decay of $\hat\tau_4$.
Notice that in this case $\tau_1\sim\hat\tau_4$, and so the visible sector could also
be placed on the fibre divisor. However, the gauge theory on the base divisor $\tau_2$
is still a hyperweak interaction. Finally the non-local axion $a_2$ is a light (almost decoupled) ALP
with a very low decay constant of the order the TeV-scale.

\subsection{A chiral model with a QCD axion and many ALPs with intermediate decay constants}
\label{ManyLocalAxions}

The previous example has shown how it is possible to obtain models with a QCD axion in the axion window.
However, it also shows how there can be additional ALPs coupling to the visible sector with a very similar decay constant.
For this, we simply need additional local cycles intersecting the visible branes, but without additional D-term conditions.
The same method of moduli stabilisation will apply: for $n$ intersecting local cycles, $d$ of them
are fixed by D-terms (and so $d$ axions are eaten up) while the remaining $n-d$ moduli are frozen by loop corrections,
leading to $n-d$ light axions with intermediate scale decay constant.
We stress that the number of D-term conditions can be reduced either by not
wrapping some of the intersecting cycles or by setting some world-volume fluxes to zero.

To be concrete, consider the blow-up of a $\mathds{Z}_7$ singularity
which has no non-compact exceptional divisors.
Thus it will not be altered by the space in which it is embedded, and so it can be studied locally.
This is not necessary: a much larger class of examples could be found, for example, via the blow-ups of del-Pezzo singularities, but then we would need to take into account the global embedding; we leave the construction of these rich and fully realistic examples to future work.
We denote the overall volume $\vo_0 \simeq \vo$ (which could be a Swiss-cheese or a fibred Calabi-Yau)
with a small cycle $\tau_s$ stabilised at $\tau_s \sim \ln \vo$ by a gaugino condensate wrapping $D_s$.
The volume form is then:
\be
\vo = \vo_0 + \frac{1}{6}\left[8 \left( t_1^3 + t_2^3 + t_3^3\right) -6 \left(t_1^2 t_3 + t_1 t_2^2 + t_2 t_3^2\right) + 6 t_1 t_2 t_3 \right]\,,
\ee
and the K\"ahler cone is given by:
\be
t_1 - 2 t_3 > 0\,, \qquad t_3 - 2 t_2 > 0\,, \qquad t_2 - 2 t_1 > 0\,. \nn
\ee
We wrap a `visible' stack on $D_a = D_1$,
and another on $D_b = D_2$, turning a flux of $\mc{F}_a = 3\hat{D}_3 + \frac{1}{2} \hat{D}_1$, leading to $I_{ab}= 3$ and no other chiral intersections. The D-term generated stabilises $\tau_3$ in terms of $\tau_1$ and $\tau_2$ at:
\be
\tau_3 = \frac{\tau_2}{\tau_1} \left(2 \tau_2 - \tau_1\right),
\label{Dstab}
\ee
or equivalently:
\be
t_1 = - \frac{5 \tau_1 + 2 \tau_2}{7 \sqrt{2} \sqrt{\tau_1}}\,, \qquad
t_2 = - \frac{3 \tau_1 + 4 \tau_2}{7 \sqrt{2} \sqrt{\tau_1}}\,, \qquad
t_3 =- \frac{ 8\tau_2 -  \tau_1}{7 \sqrt{2} \sqrt{\tau_1}}\,. \nn
\ee
Hence we can write the volume form effectively as:
\be
\vo = \vo_0
- \frac{1}{21 \sqrt{2}\tau_1^{3/2}} \left( 5 \tau_1^{3} + 6 \tau_1^2 \tau_2 - 6 \tau_1 \tau_2^2 + 16 \tau_2^3\right).
\ee
$\vo_0$ is stabilised in the usual LVS way;
string loops then stabilise the remaining moduli $\tau_1$ and $\tau_2$ via terms in the potential of the form (\ref{Vloop}) (the full loop potential is rather complicated even in this simple case so we neglect to present it here).

The axions corresponding to moduli $\tau_1$ and $\tau_2$ are thus left massless at this level, and couple with string strength to branes $a$ and $b$, since the effective mixing matrix is (restricting indices to $i,j = 1...3$, since the mixing with $\tau_s$ and $\tau_b$ is negligible):
\be
\mc{K}_{ij} =\frac{(6 \tau_2^2 + 2 \tau_1 \tau_2 - \tau_1^2)^{-1}}{14 \sqrt{2} \,\vo \sqrt{\tau_1} }
\left(
\begin{array}{ccc}
28 \tau_2^2+12 \tau_1 \tau_2-5 \tau_1^2 & \tau_1 (10 \tau_2-3 \tau_1) & \tau_1 (\tau_1+6 \tau_2) \\
 \tau_1 (10 \tau_2-3 \tau_1) & \tau_1 (48 \tau_2-13 \tau_1) & \tau_1 (12 \tau_2-5 \tau_1) \\
 \tau_1 (\tau_1+6 \tau_2) & \tau_1 (12 \tau_2-5 \tau_1) \
& \tau_1 (11 \tau_1+24 \tau_2)
\end{array}
\right). \nn
\ee
This is non-degenerate with full mixing between the axions. For $\tau_1 = \tau_2 = \tau_*$,
the D-term relation (\ref{Dstab}) gives also $\tau_3=\tau_*$ and so the mixing matrix simplifies to:
\be
\mc{K}_{ij} =\frac{1}{14 \sqrt{2} \,\vo \sqrt{\tau_*} }
\left(
\begin{array}{ccc}
5 & 1 & 1 \\
 1 & 5 & 1 \\
 1 & 1 & 5
\end{array}
\right). \nn
\ee
Therefore the gauge couplings of groups $a$ and $b$ are equal, $g_a^2 = g_b^2 \equiv g^2\sim\tau_*^{-1}$.
Taking for example $\tau_* = 10$, the axionic couplings turn out to be:
\be
\mc{L} \supset \frac{g^2 \sqrt{\vo}}{2\pi M_P} \left[ (3.1\, \rhoo_1 - 1.8 \,\rhoo_2) \,\tr(F_a \wedge F_a)
- (2.3 \,\rhoo_1 - 2.8\, \rhoo_2) \,\tr(F_b \wedge F_b) \right],
\ee
giving a `QCD axion' $\rhoo_{\rm QCD} =  (0.86\, \rhoo_1 - 0.5 \,\rhoo_2)$
together with an ALP $\rhoo_{\rm ALP} = (0.5 \,\rhoo_1 +0.86\, \rhoo_2)$ and:
\be
\mc{L} \supset \frac{g^2 \sqrt{\vo}}{2\pi M_P} \left[ 3.6 \,\rhoo_{\rm QCD}\,\tr(F_a \wedge F_a)
-(3.6\, \rhoo_{\rm ALP}+ 0.6\,\rhoo_{\rm QCD} ) \,\tr(F_b \wedge F_b) \right].
\ee
Hence if we interpret the cycle `$b$' as a $U(1)$ stack, the ALP will couple with approximately equal strength to the photon as the QCD axion does. We believe this is a generic prediction of this class of models: it is highly non-generic to have an ALP that couples parametrically more  strongly to the photon than the QCD axion does. In order to achieve this, it would be necessary for there to be a local stack which carries a $U(1)$ which contributes to the hypercharge, but which does not intersect the QCD stack, and has no mass mixing (through D-terms) with the axion of the QCD stack. This implies a construction with at least four stacks of branes; it would be an interesting scenario to realise.

\subsubsection{Explaining astrophysical anomalies with a stringy ALP}

Since the above construction is built from intersecting branes in the geometric regime, the coupling of the local axion to matter fields should be given by the first expression in (\ref{EQ:ALPMATTERCOUPLINGS}). Hence we expect the ALP coupling to electrons to be suppressed only by a factor of the gauge coupling squared relative to the photon coupling. This is perfectly consistent with the astrophysical hints for an ALP, -- an anomalous energy loss of white dwarfs (see section \ref{APP:STELLAR} in the appendix) and an anomalous transparency of the Universe for TeV photons (see section \ref{APP:FLUXES}) -- which would be explained by:
\be
\aC[i,\gamma]/\mC[i,e] \simeq 10, \qquad f_{a_i}/\aC[i,\gamma] \simeq 10^8\ {\rm GeV},\qquad m_{\rm ALP} \lesssim 10^{-9} \div 10^{-10}\ {\rm eV}.
\ee
For models in this class - i.e. built from local constructions of typical size $\tau_*$ - we find generically:
\be
\frac{\aC[i,\gamma]}{\aC[i,e]} \sim \frac{8\pi \tau_*}{3}, \qquad \frac{f_{a_i}}{\aC[i,\gamma]} = \frac{1}{8\pi N_{i\gamma} \tau_*^{1/4}} \frac{M_P}{\sqrt{\vo}} =\frac{1}{8\pi N_{i\gamma}\tau_*^{1/4}} \sqrt{\frac{\sqrt{2} M_P m_{3/2}}{g_s^{1/2} W_0}}\,.
\ee
where $N_{i\gamma}$ is a purely numerical factor coming from the canonical normalisation of the ALP.
For $m_{3/2} = 10$ TeV, $g_s \simeq 0.1$ and $W_0 \sim 10$ we can easily arrive at the desired figure for $N_{i\gamma} \sim 10$.

The ALP mass could be generated, for example, by a single K\"ahler potential instanton, which for $m_{\rm ALP} \sim 10^{-10}$ eV would require:
\be
\tau_* \sim \frac{1}{\pi} \ln \left(\frac{g_s m_{3/2}}{m_{\rm ALP}} \right)\sim 16.
\ee
Note that there is then a small amount of tension with the ratio $\aC[i,\gamma]/\mC[i,e]$, but this is the least well predicted quantity. Hence it is possible to explain these astrophysical hints for an ALP via a natural intermediate string-scale scenario with stable moduli.

Note that such an ALP would constitute only a tiny ($\mc{O}(10^{-9})$) fraction of dark matter, and would satisfy all constraints - interestingly being just one or two orders of magnitude in mass heavier than the black hole super-radiance constraint \cite{Arvanitaki:2009fg,Arvanitaki:2010sy} (see section \ref{APP:COSMIC}) and within the reach of future experiments (see figure \ref{FIG:Constraints} and appendix \ref{APP:CONSTRAINTS}).

\section{Conclusions}
\label{conclusions}

String phenomenology holds the promise of an axiverse -- the QCD axion plus a (possibly large) number of
further ultralight axion-like particles, possibly populating each decade of mass down to the Hubble scale, $10^{-33}$~eV.
However, although a plenitude of axions is a generic prediction of string theory, there may be few or no light axions remaining once constraints such as tadpole cancellation and moduli stabilisation are taken into account (for example in the KKLT scenario each axion would be made massive by an exponential term in the superpotential) or, particularly for D-brane models, when considering which axions are eaten in making $U(1)$ symmetries massive. This last issue is particularly important, for example, in GUT models where the diagonal $U(1) \subset U(5)$ is made massive in this way; we have shown that in such cases (and more generally for any model where the diagonal $U(1) \subset U(3)$ is made massive like this) if the Standard Model is built on a local cycle, the non-local axions cannot solve the strong CP problem, and particular care must be taken to ensure that there are additional light local axions that can do the job.

We have shown that the promise of an axiverse is fulfilled in the LARGE Volume Scenario of IIB string compactifications.
In fact, the least fine-tuned (i.e. $W_0\sim \mc{O}(1)$) models of globally consistent constructions with magnetised D-branes and chirality have at least two light axions. In general, the closed string axions separate into local and non-local classes,
where the non-local axions may have small decay constants but also small couplings and vanishingly small masses, and so are unlikely to overproduce dark matter (unless the Standard Model also wraps a non-local cycle but in this case
the axions would be diluted by the decay of heavy K\"ahler moduli).

We have also determined for the LVS axiverse the couplings of axions to matter fields, both at tree level and one loop. The results at one loop generalise the existing non-supersymmetric formulae to supersymmetric models. However, these rely upon the conjectured form of the K\"ahler metric for matter fields; it would be interesting to calculate these couplings in specific examples.

Moreover, we have shown how to build models that exhibit a QCD axion with a decay constant $f_a\sim M_P / \left(4\pi\sqrt{\vo}\right)\sim 10^{10}\div 10^{12}$~GeV, and an additional axion-like particle (or even more than one) having the same decay constant and coupling to the photon\footnote{With potentially many much more weakly coupled axions.}
$C_{i\gamma}\sim C_{a\gamma}\sim \mc{O}(1)$ (see section~\ref{ManyLocalAxions}). At both ends of the above range 
of the decay constant there are exciting phenomenological opportunities. For $f_a\sim M_P / \left(4\pi\sqrt{\vo}\right)\sim 10^{12}$~GeV, the QCD axion can be the dominant part of dark matter and be detected in haloscopes exploiting microwave cavities, cf. appendix \ref{APP:COSMIC}. For $f_a\sim M_P / \left(4\pi\sqrt{\vo}\right)\sim 10^{10}$~GeV,
the additional ALPs could explain astrophysical anomalies (see appendix \ref{APP:CONSTRAINTS}) and
be searched for in the next generation
of helioscopes (see section~\ref{APP:STELLAR}) or in the next-to-next generation of light-shining-through-a-wall
experiments (see section~\ref{APP:LAB}).

The particular model of inflation is also relevant for the cosmological constraints.
In the LARGE Volume Scenario there are well-defined models of inflation \cite{LVScosmo},
and so these questions can be readily addressed.

If the Hubble scale during inflation is larger than the gravitino mass,
the latter can be of the order the TeV scale giving rise to an intermediate string scale
which allows couplings for the axions compatible with the astrophysical hints.
However, reheating at the end of inflation is problematic
(since in this case the inflaton is the lightest modulus, i.e. the volume mode)
and it is an interesting question as to how this can be achieved.

Alternatively, if the Hubble scale during inflation is lower than the gravitino mass,
the latter has to be much above the TeV scale. This singles out sequestered models
where the soft terms can be around the TeV scale even if the gravitino mass is
at the intermediate scale \cite{sequester}. The string scale turns out to be around the GUT-scale,
allowing again for good QCD axion candidates. Even in this case there can be
heavy particles that decay and dilute other relics, but now, due to a higher string scale,
the axion will be a component of non-thermal dark matter.
Contrary to the situation with $H_{\rm inf}>m_{3/2}$, inflation and reheating
are better understood, and so we can have more control over all the cosmological
evolution of the axions and the moduli from the early Universe till today.

Since the phenomenology of axions is so rich, with astrophysical constraints and hints, and many experiments searching for them, it is highly worthwhile to investigate what values of the couplings can be predicted from string theory. We have compiled a comprehensive analysis of these in the LVS context
where both local and global questions can be asked about the models in a consistent way. We hope that this will be of use to experimentalists and model builders alike.

\section*{Acknowledgments}

MDG was supported by ERC advanced grant 226371.
We would like to thank Bobby Acharya, Andr\`es Collinucci, Joe Conlon, I$\tilde{{\rm n}}$aki Garc\'ia-Etxebarria, J\"org J\"ackel,
Hyun-Min Lee, Fernando Quevedo and Javier Redondo for helpful discussions.

\appendix

\section{Phenomenology of theories with many axions}
\label{sec:pheno}

\subsection{Effective Lagrangian at the weak scale}

For the low energy phenomenology of axions, their couplings to gauge bosons other than gluons, specifically to
photons, and to SM fermions, specifically to light quarks and electrons, are of uttermost importance.
It is known that for the QCD axion a great deal about its low energy couplings can be learned from effective field theory techniques,
which allow to take electro-weak symmetry breaking and QCD effects into account. In this spirit, generalising the considerations in~\cite{Georgi:1986df} to the case of several axion-like fields, $a_i$, we write down the most general effective Lagrangian at the weak scale, to first order in the fields $a_i$, which consists of anomalous couplings to gauge bosons and derivative couplings to matter:
\begin{align}
\nonumber
\mc{L} &\supset \frac{1}{2}\, \partial_\mu a_i\, \partial^\mu a_i
- \frac{g_{3}^2}{32\pi^2} \left( \theta_0 + \aC[i, 3] \frac{a_i}{f_{a_i}}\right) F_{3,\mu\nu}^b \tilde{F}_{3}^{b,\mu\nu} \\
& - \frac{g_2^2}{32\pi^2}\,\aC[i, W] \frac{a_i}{f_{a_i}}\, F_{W,\mu\nu}^b\tilde{F}_W^{b,\mu\nu}
- \frac{g_Y^2}{32\pi^2}\, \aC[i,Y] \frac{a_i}{f_{a_i}}\, F_{Y,\mu\nu} \tilde{F}_Y^{\mu\nu} \nn \\
&  -  \sum_{\psi_L, \psi_R}  (\psi_R M \psi_L + \ov{\psi}_L M^\dagger \ov{\psi}_R)+ \sum_{\psi_L}\frac{X_{\psi_L}^i}{f_{a_i}} \psi_L  \sigma^\mu \ov{\psi}_L \partial_\mu a_i + \sum_{\psi_R} \frac{X_{\psi_R}^i}{f_{a_i}} \psi_R \sigma^\mu \ov{\psi}_R \partial_\mu a_i  \nn\\
& - \sum_{\psi_L, \psi_R} (i \psi_R g^i \psi_L \frac{a_i}{f_{a_i}} + h.c.) - V(a_i),
\label{leff}
\end{align}
where $\theta_0$ is an initial theta-angle, $F_W$ and $F_Y$ are the $SU(2)$ and hypercharge field strengths, respectively,
$g_2$ and $g_Y$ the corresponding gauge couplings,
$M$ is the mass matrix of the fermions, written as Weyl fermions $\psi_L$ and $\psi_R$.
Note that in general the couplings to fermions need not conserve parity (see \cite{Berenstein:2012eg} for a related discussion on this issue). Furthermore, the Yukawa couplings on the last line are the leading terms in the expansion of Yukawa couplings of the form:
\be
\mc{L} \supset - e^{- i b \phi} H \psi_R \psi_L + h.c.
\ee
since we expect the full theory to respect the axionic shift symmetries, only broken by non-perturbative effects. Those contributed other than from QCD are encapsulated in the potential $V(a_i)$, which we expect to give a mass to all ALPs at some scale - except for those that are gauged, and obtain their mass through being eaten (these shall be discussed in section \ref{SEC:MIXINGCPDM}). When an ALP couples to a gaugino condensate or D-brane instanton, we expect the mass to be large, and we can simply exclude it from the analysis to follow. These are leading contributions in the superpotential. On the other hand, for ALPs that do not obtain masses in that way, the masses induced are many orders of magnitude smaller, coming from multi-instanton effects or K\"ahler potential corrections, and have been argued to be many orders of magnitude below the
QCD axion mass \cite{Joe}. Therefore in the following we shall neglect ALP masses and comment on their inclusion at the end.

In the text of the paper have derived the predictions for the dimensionless couplings $\aC[i,3]$, $\aC[i,W]$, $\aC[i,Y]$, $X_{\psi_L}^i$, $X_{\psi_R}^i$,
and $g^i$, and the decay constants $f_{a_i}$ from particular string compactifications.
Here, in the next subsection, we will further scale the theory down through electro-weak symmetry breaking, heavy quark thresholds and QCD
chiral symmetry breaking to arrive at the low energy Lagrangian involving only the QCD axion, ALPs,
photons and electrons, which are then be compared to astrophysical and cosmological observations, as well as to laboratory experiments, in section \ref{APP:CONSTRAINTS}.

\subsection{Effective Lagrangian below chiral symmetry breaking}

After electro-weak symmetry breaking, we diagonalise the fermion mass matrix, transforming to the physical basis with real masses $m_\Psi$,  $M \rightarrow U_R^\dagger M U_L = m_\Psi $, $\tilde{X}_{\Psi_L}^i \equiv U_L^T X_{\Psi_L}^i U_L^*, \tilde{X}_{\Psi_R}^i \equiv U_R^\dagger X_{\Psi_R}^i U_R$ (noting that this can lead to flavour-changing axionic couplings) and write $ \Psi = \left( \begin{array}{c} \psi_L \\ \ov{\psi}_R \end{array}\right)$:
\begin{align}
\mc{L} &\supset \frac{1}{2}\, \partial_\mu a_i\, \partial^\mu a_i
- \frac{g_{3}^2}{32\pi^2} (\theta_0 + \mathrm{arg \ det} (M) + \aC[i,3] \frac{a_i}{f_{a_i}} ) F_{3,\mu\nu}^b \tilde{F}_{3}^{b,\mu\nu} \nn\\
& - \frac{e^2}{32\pi^2} (\aC[i, W] + \aC[i,Y]) \frac{a_i}{f_{a_i}} \, F_{em,\mu\nu} \tilde{F}_{em}^{\mu\nu} \\
& + \sum_{\Psi} m_{\Psi} \ov{\Psi} \Psi + \sum_\Psi \bigg[\frac{1}{2} (\tilde{X}_{\psi_R}^i + \tilde{X}_{\psi_L}^i) \ov{\Psi} \gamma^\mu\gamma_5   \Psi  + \frac{1}{2} (\tilde{X}_{\psi_R}^i - \tilde{X}_{\psi_L}^i) \ov{\Psi} \gamma^\mu \Psi \bigg] \frac{\partial_\mu a_i}{f_{a_i}}\, .
\nn
\end{align}
The mass of the QCD axion is generated by the coupling to QCD, but is complicated by the presence of the light quarks - if there were none then its mass would be that the $\eta^\prime$ acquires. One can however use chiral perturbation theory to calculate it: with chiral rotations of the quarks one may eliminate the $F_3\tilde{F}_3$ term and consider the effective pion Lagrangian. In the approximation $m_s \gg m_u, m_d$, one can consider only the $\pi^0$ and its mixing with the axions (one combination of $\eta$ and $\eta^\prime$ is made massive by the strange quark mass, the other by the anomaly). Making the rotations $\Psi \rightarrow \exp (-i \alpha_\Psi \gamma_5) \Psi $, with:
\begin{align}
 \alpha_u + \alpha_d = - \frac{c_u + c_d}{2} \bigg(\theta + \aC[i,3] \frac{a_i}{f_{a_i}} \bigg),\hspace{2ex}
c_u + c_d =1,\hspace{2ex}
\theta \equiv \theta_0 +  {\rm arg \ det} (M),
\end{align}
will eliminate the $F_3 \tilde{F}_3$ term and couple the axions to the quarks.

We are now ready to define the QCD axion $a$, with decay constant $f_a$ and setting $C_{a3}=1$, cf.
eq.~\eqref{axion}, and the remaining ALP fields $a_i^\prime$,
in terms of linear combinations of the original axion-like fields $a_i$:
\begin{eqnarray}
&&\frac{a}{f_a} \equiv \theta + \aC[i,3] \frac{a_i}{f_{a_i}}, \qquad
a_i = R_{ij} a_j^\prime, \qquad
\sum_{i}\frac{\aC[i,3]}{f_{a_i}} R_{ij} = \frac{\delta_{aj}}{f_a}, \qquad \alpha_{u,d} = - \frac{c_{u,d}}{2} \, a \nn\\
&& a = \hat{a} + f_a \,{\rm arg \ det} (M),\quad
(\tilde{X}_{\psi_R}^i + \tilde{X}_{\psi_L}^i) \rightarrow
(\tilde{X}_{\psi_R}^j + \tilde{X}_{\psi_L}^j)R_{ji} - c_\psi \delta_{ia} \equiv \hat{X}_\psi^i  - c_\psi \delta_{ia}, \nn
\end{eqnarray}
resulting in the effective Lagrangian:
\begin{align}
\mc{L} \supset &  - \frac{1}{2} \partial_\mu \pi^0 \partial^\mu \pi^0  - \frac{1}{2} \partial_\mu a_i \partial^\mu a_i \nn\\
& -\frac{e^2}{32\pi^2}\bigg[ \frac{4}{3}  (4\alpha_u  + \alpha_d)+ (\aC[i, W] + \aC[i,Y]) \frac{a_i}{f_{a_i}} - 2\frac{\pi^0}{f_\pi} \bigg]  F_{{\rm em},\mu\nu} \tilde{F}_{\rm em}^{\mu\nu} \nn\\
&  -m_u \,v^2  \cos\left( \frac{\pi^0}{f_\pi} + c_u \frac{a}{f_a}\right)  -m_d \,v^2  \cos\left(-\frac{\pi^0}{f_\pi} + c_d \frac{a}{f_a} \right)  \nn\\
& + \frac{1}{4} (\hat{X}_d^i  - \hat{X}_u^i + (c_u - c_d) \delta_{ia})\frac{f_\pi}{f_{a_i}} \,\partial^\mu \pi^0 \partial_\mu a_i^\prime.
\end{align}
The kinetic mixing term on the last line between the QCD axion, ALPs and pion can be ignored provided that $\hat{X}_u - \hat{X}_d \ll f_{a_i}/f_\pi$; this is because the non-unitary transformation that diagonalises the kinetic terms can be chosen so that the pion merely rescales $\pi^0 \rightarrow \pi^0 (1 - \mc{O} (\frac{f_\pi^2}{f_{a_i}^2}))$ while the other fields shift by factors $a_i^\prime \rightarrow a_i^\prime + R^\prime_{ij} a^\prime_j \mc{O} (\frac{f_\pi}{f_{a_i}})$; these lead to subleading (and therefore negligible, for the expected sizes of $f_{a_i}$ compared to $f_\pi$) corrections to the physical masses and couplings. In fact, the all-orders diagonalisation is performed by:
\begin{eqnarray}
&&\chi_i \equiv \frac{1}{4} (\hat{X}_d^i  - \hat{X}_u^i + (c_u - c_d) \delta_{ia})\frac{f_\pi}{f_a}, \qquad
F \equiv \sqrt{1 - \sum_i \chi_i^2}, \nn\\
&&\pi^0 \rightarrow \frac{\pi^0}{F}, \qquad
a^\prime_i \rightarrow a^\prime_i - \frac{\chi_i}{F} \,\pi^0, \nn
\end{eqnarray}
and thus there are no new couplings induced between the ALPs and the photon at any order.

\subsubsection{Coupling to photons}

Neglecting the mixing terms $\chi_i$, the mass of the pion, the mass of the QCD axion and
its coupling to photons are then found as usual by choosing the coefficients $c_u = \frac{1}{1 + z}, c_d = 1 - c_u$, where $z \equiv m_u/m_d$, to eliminate the mass mixing term above; this gives:
$m_a = m_\pi \frac{f_\pi}{f_a} \frac{\sqrt{z}}{1+z}$ and the QCD axion two-photon coupling:
\begin{align}
\mc{L} \supset& -\frac{1}{4} g_{a\gamma}a  F_{{\rm em},\mu\nu} \tilde{F}_{\rm em}^{\mu\nu}, \nn\\
g_{a\gamma} \equiv& \,\frac{\alpha}{2\pi f_a} \bigg(\aC[a,W] + \aC[a,Y] + \Delta \aC[a,\gamma] \bigg),\nn\\
\Delta \aC[a,\gamma] \equiv& -\frac{2}{3} \frac{4+z}{1+z} \approx -1.95,
\end{align}
while the couplings of the ALPs $a^\prime_i$ to photons are just:
\begin{align}
\mc{L} \supset& -\frac{1}{4}g_{i\gamma} a_i^\prime F_{{\rm em},\mu\nu} \tilde{F}_{\rm em}^{\mu\nu}, \qquad
g_{i\gamma} \equiv \frac{\alpha}{2\pi} \sum_j \left(\frac{\aC[j,W] + \aC[j,Y]}{f_{a_j}}\right)R_{ji}.
\end{align}
The upshot is that the QCD axion has a `universal' coupling to photons
$\propto \Delta \aC[a,\gamma]$, but the ALPs do not; their coupling to photons and electrons comes entirely from the high-energy theory.

Up to this stage, we did not consider mass terms for ALPs. These will have effects on the `universal' couplings to photons. In fact, we shall treat any ALP heavier than the pion as being already integrated out and ignore it for these purposes, and for ALPs of comparable mass to the QCD axion or heavier the analysis is necessarily involved. However, in the case that additional ALPs are \emph{lighter} than the
QCD axion, we can neglect the kinetic mixing terms as before, and, introducing ALP masses $m^2_{i}$, the QCD axion is approximately shifted by:
\be
a \rightarrow a - \sum_{i,j \ne a} \frac{m_j^2}{m_a^2} R_{ja} R_{ji} a_i,
\ee
and so the ALPs $a_j$ acquire universal couplings, too, which are, however, suppressed in comparison
to the universal coupling of the axion, by the ratio $m^2_i/m_a^2$:
\be
g_{i\gamma} \approx \frac{\alpha}{2\pi f_a} \bigg(\aC[i,W] + \aC[i,Y] - \Delta \aC[a,\gamma]\sum_{j \ne a} \frac{m_j^2}{m_a^2} R_{ja} R_{ji}  \bigg).
\ee
Conversely, since the ALPs are shifted by a similar amount, we expect the QCD axion to photon coupling to be shifted by an amount:
\be
\Delta g_{a\gamma} \sim  \frac{\alpha}{2\pi} \sum_i \frac{\aC[j,W] + \aC[j,Y]}{f_{a_i}}\sum_{j \ne a} \frac{m_j^2}{m_a^2} R_{ja} R_{ji}.
\ee
This is relevant in the case that an ALP couples much more strongly than the QCD axion to the photon \emph{and} has some coupling to QCD.

\subsubsection{Couplings to electrons}

Finally, we consider the coupling to electrons which is also crucial for low energy phenomenology. Other than the direct coupling,
there will also be one generated through the two-photon coupling at one loop. In supersymmetric theories, the two-photon coupling is accompanied by a coupling to gauginos:
\be
\mc{L} \supset - \int d^2 \theta\ (i \,a_i) \frac{g_{i\gamma}}{4} \,W^\alpha W_\alpha
\supset  \frac{1}{4}\,g_{i\gamma} a_i F_{{\rm em},\mu\nu} \tilde{F}_{\rm em}^{\mu\nu}
+ \frac{1}{2}\,g_{i\gamma} \partial_\mu a_i \lambda^\alpha \sigma^\mu \ov{\lambda},
\ee
and so we have a quantitatively different result. We also have a qualitatively different result, in that we find parity violating vector-like terms arising from the differences in squark masses. We can parameterise the coupling as:
\begin{align}
\mc{L} \supset& \frac{C_{ie}^A}{2 f_{a_i}} \,\bar{e} \gamma^\mu \gamma_5 e \partial_\mu a_i
+  \frac{C_{ie}^V}{2 f_{a_i}} \,\bar{e} \gamma^\mu e \partial_\mu a_i, \nn\\
C_{ie}^{A,V} =& \,\hat{X}_{e}^{A,V\, j} R_{ji} + \Delta_{i \gamma\gamma}[ C_{ie}^{A,V}] +  \delta_{ai} \Delta_{\rm QCD} [C_{ae}^A ]\,,
\label{ElectronCoupling}
\end{align}
where $\hat{X}_{e}^{A,V\, j} \equiv \frac{1}{2}( \tilde{X}_{e_R}^j \pm \tilde{X}_{e_L}^j)$ and $\Delta_{\rm QCD} [C_{ae}^A] = \frac{3 \alpha^2}{4\pi} \Delta \aC[a,\gamma] \ln (\Lambda_{\rm QCD}/m_a)  $ is a correction from pion loops to the QCD axion to electron coupling, unchanged in SUSY theories. We shall calculate the full contribution $\Delta_{i \gamma\gamma} C_{ie}^{A,V}$ in section \ref{SEC:LOOPS}, but we note that to a rough approximation we can take:
\begin{align}
\Delta_{i \gamma\gamma}[ C_{ie}^{A}] \approx& \,\frac{3 \alpha^2}{4\pi^2}  \aC[i,\gamma] \ln \left(\frac{M_{\rm soft}}{m_e}\right) +  \frac{2\alpha^2}{4\pi^2}  \aC[i,\gamma] \ln \left(\frac{\Lambda}{M_{\rm soft}}\right),  \nn\\
\Delta_{i \gamma\gamma}[ C_{ie}^{V}] \approx& \,\frac{2\alpha^2}{4\pi^2}  \aC[i,\gamma] \ln \left(\frac{\Lambda}{M_{\rm soft}}\right),
\label{EQ:APPROXELECTRONCOUPLING}\end{align}
where $M_{\rm soft}$ is the scale of superpartner masses, and $\Lambda$ the cutoff of the theory, of the order of the string scale.

\section{Axion searches in astrophysics, cosmology, and laboratory}
\label{APP:CONSTRAINTS}

The couplings of QCD axions and ALPs to photons and electrons give rise to a number of
effects which can be searched for in astrophysics, cosmology, and in laboratory experiments.
In the remainder of this section, we will summarise the present knowledge about the
dimensionless couplings and the decay constants.
Some astrophysical observations seem to point to a preferred scale of order $\sim 10^9$~GeV.

\subsection{Axions from astrophysical sources}

\subsubsection{Stellar emission}
\label{APP:STELLAR}

Light and weakly interacting particles such as the QCD axion or ALPs are produced in hot
astrophysical plasmas via their coupling to photons or via their coupling to electrons
or nucleons. One of the strictest bounds on the
coupling of axions to photons\footnote{For a recent update on the bounds on this coupling, see~\cite{Arias:2012mb}.} arise from either the non-observation of an anomalous energy loss of Horizontal Branch (HB) stars due to axion emission~\cite{Raffelt:1996wa} or (slightly stronger) from the non-observation of solar axions in the CAST experiment~\cite{Andriamonje:2007ew}:
\begin{align}
\mc{L} \supset -\frac{1}{4} g_{a \gamma} a F_{\mu\nu} \tilde{F}^{\mu\nu}, \
{\rm with}\ \ g_{a \gamma}\equiv   \frac{\alpha}{2\pi} \frac{\aC[i, \gamma]}{f_{a_i}}
< 10^{-10}\ {\rm GeV}^{-1}.
\label{EQ:AXIONPHOTONPHOTON}
\end{align}
This translates into a bound
\begin{align}
\frac{f_{a_i}}{\aC[i,\gamma]} \gtrsim \frac{\alpha}{2\pi} \ 10^{10}\ {\rm GeV} \gtrsim  10^7\ {\rm GeV}.
\label{CASTbound}\end{align}
This is usually used to provide a lower bound for the decay constant of the QCD axion; due to the `universal' contribution to its two-photon coupling, for a generic field-theoretical model, $\aC[a,\gamma]\sim 1$. On the other hand, there are some models with $\aC[a,\gamma]\ll 1$ where the axion decay constant can be lowered. The CAST bounds (\ref{EQ:AXIONPHOTONPHOTON}) and (\ref{CASTbound}) can be improved in the next generation of helioscopes to $g_{i\gamma}\sim  10^{-12}$~GeV$^{-1}\rightarrow f_{a_i}/\aC[i,\gamma]\sim 10^9$~GeV~\cite{Irastorza:2011gs}.

Turning now to the axial-vector couplings to electrons, let us first note that,
in many tree-level interactions, an axial-vector vertex is equivalent to a scalar one:
\be
\mc{L} \supset \frac{g_{i\psi}^A}{2m_\psi} \, \bar{\psi} \gamma^\mu \gamma_5\psi \partial_\mu a_i
\rightarrow - \frac{g_{i\psi}^A}{2m_\psi} \, a_i \partial_\mu ( \bar{\psi} \gamma^\mu \gamma_5 \psi )
= -i g_{i\psi}^A  a_i  \bar{\psi}  \psi\,,
\label{EQ:AXIALELECTRON}
\ee
where we have used the equations of motion. Of course, this only applies for tree-level processes involving the fermions as on-shell external states.

Excessive axion emission from red giants places the following constraint on axion couplings to electrons \cite{Raffelt:1996wa}:
\begin{align}
\mc{L} \supset \frac{\mC[e,i]}{2 f_{a_i}} \,\bar{e} \gamma^\mu\gamma_5 e \partial_\mu a_i
\rightarrow \mC[i,e] = \hat{X}^i_e, \nn\\
g_{ie}^A \equiv  \frac{\mC[i,e] m_e}{f_{a_i}}
\lesssim 2.5 \times 10^{-13}
\rightarrow \frac{f_{a_i}}{\mC[i,e]} \gtrsim \,& 2.0 \times 10^{9}\ {\rm GeV}.
\end{align}
A slightly better limit arises from pulsationally
unstable white dwarfs (ZZ Ceti stars), where the period increase
provides a measure of the cooling speed. From the measured period increase of the star
G117-B15A, it was found~\cite{Corsico:2001be,Isern:2003xj,Nakamura:2010zzi,Corsico:2012ki}:
\be
g_{ie}^A < 1.3\times 10^{-13} \rightarrow \frac{f_{a_i}}{\mC[i,e]} > 3.9 \times 10^9\ {\rm GeV}.
\ee
Hence, for $\mC[i,e] \gtrsim 10^{-2}\, \aC[i,\gamma]$, the limits from electron couplings are more constraining than the direct two-photon coupling. Noting that the `universal' coupling of the QCD axion to electrons is suppressed by $\mc{O} (10^{-4})$, we see that the direct couplings could be important, as we stressed in their discussion in the context of the LARGE Volume Scenario.

However -- and intriguingly -- reanalysis of the same white dwarf with improvements in the accuracy and precision of the pulsational data identifies a non-standard energy loss~\cite{Isern:2008nt,Isern:2008fs,Isern:2012ef}. Improvements in the luminosity distribution of white dwarf stars were also found to require additional energy loss, an absence thereof predicting an excess of white dwarfs at luminosities $\log L/L_\odot \simeq -2$. The observed excess in cooling speed is consistent with the existence of an
axion with mass $m_i\lesssim\ {\rm keV}$
and a coupling to electrons of strength:
\be
g_{ie}^A = (2.0\div 7.0)\times 10^{-13}
\rightarrow \frac{f_{a_i}}{\mC[i,e]}\simeq
(0.7\div 2.6)\times 10^9\ {\rm GeV}\,
\label{dec_const_wd}
\ee
which is compatible, within the uncertainties, with the limits from red giants.

On parity violating couplings to the electron, there
is a limit \cite{Grifols:1986fc,Raffelt:1996wa} on a scalar Yukawa coupling $g_{Yie}^V \ov{e}  e a_i$ of the order:
\be
g_{Y ie}^V \lesssim 10^{-14}.
\ee
These would arise only from non-perturbative corrections to the Yukawa couplings of the form $W \supset e^{-b \frac{a_i}{f_{a_i}}} h_d e_L e_R$; they are difficult to determine in a string model without knowing the flavour structure of the Yukawa couplings but generically we could estimate $g_{Yie}^V \sim \frac{m_e}{f_{a_i}} $ - if they are present at all - which would translate into a bound of the order $f_{a_i} \gtrsim 5\times 10^{10}$ GeV.

On the other hand, we will generically find in string models parity-violating vector-like couplings
\be
\mc{L} \supset g_{i\psi}^V  \bar{\psi} \gamma^\mu \psi \partial_\mu a_i.
\ee
Here however the integration by parts yields a coupling to the divergence of the electron number current; hence we expect the simplest amplitudes involving these couplings to vanish, and the constraints upon them to be very weak.

The limits on couplings to nucleons $\aC[a,N]$ are most strongly constrained by (lack of) energy loss from supernova SN1987A, and applies for the range $0.002\ {\rm eV} \lesssim \aC[a,N] m_a \lesssim 2\ {\rm eV}$ \cite{Raffelt:1996wa}:
\begin{align}
3\times 10^9 \ {\rm GeV} \lesssim f_a/\aC[a,N] .
\end{align}

\subsubsection{Conversion of astrophysical photon fluxes}
\label{APP:FLUXES}

Due to the $F_{\rm em}\tilde F_{\rm em}\propto \mathbf E\cdot \mathbf B$ coupling, photons
can oscillate into axions in large-scale magnetic fields. These oscillations can
lead to observable effects in astrophysical photon spectra, but with high sensitivity
only for very small masses. Correspondingly, these observations typically constrain only ALPs,
but not the QCD axion.

The absence of a $\gamma$-ray burst in coincidence with SN 1987A neutrinos provides, for
$m_i\lesssim 10^{-9}$~eV,  the
most restrictive limit on the two-photon coupling of ALPs~\cite{Brockway:1996yr,Grifols:1996id}:
\begin{align}
g_{a \gamma}\equiv   \frac{\alpha}{2\pi} \frac{\aC[i, \gamma]}{f_{a_i}}
\lesssim 10^{-11}\, {\rm GeV}^{-1} \rightarrow
\frac{f_{a_i}}{\aC[i,\gamma]} \gtrsim \frac{\alpha}{2\pi} \ 10^{10}\ {\rm GeV}
\gtrsim  10^8\,{\rm GeV},
\end{align}
an order of magnitude better than the one inferred from the lifetime of HB stars and from CAST.

This constraint challenges the ALP interpretation of recent measurements of $\gamma$-ray spectra from
distant active galactic nuclei (AGN), extending to energies $\gtrsim$~TeV -- a puzzling observation,
since the $\gamma$-ray absorption due to $e^+/e^-$ pair production off the extragalactic background light is expected to cut-off the spectra at very high energies\footnote{For a recent compilation of the available data and comparison with the expectation, see~\cite{Horns:2012fx}.}.
Intriguingly, apart from conventional explanations, such a high transparency
of the Universe may be explained by photon $\leftrightarrow$ ALP oscillations:
the conversion of gamma rays into ALPs in the
magnetic fields around the AGNs or in the intergalactic medium, followed by their unimpeded
travel towards our galaxy and the consequent reconversion into photons in the galactic/intergalactic magnetic
fields, requiring a very light, $m_i\lesssim 10^{-9}~ {\rm eV}$, ALP coupling to two photons with strength~\cite{DeAngelis:2007dy,Simet:2007sa,SanchezConde:2009wu}:
\begin{equation}
g_{i\gamma}\sim  10^{-11} \ {\rm GeV}^{-1} \rightarrow
f_{a_i}/\aC[i,\gamma] \sim 10^8\ {\rm GeV},
\label{dec_const_transp}
\end{equation}
in the same ball-park as the constraint from the non-observation of $\gamma$-rays from
SN~1987A.

A similar effect can occur when gamma ray-ALP conversions are induced by intergalactic magnetic fields $B$, which have a strength of $0.1 \div 1$ nG. In this case, steps in the power spectrum can be observed at a critical energy \cite{Hooper:2007bq}:
\be
\frac{E_{\rm crit}}{{\rm GeV}} = \left( \frac{m_{\rm ALP}}{ \mu \rm eV}\right)^2
\frac{10^{-11}\ {\rm GeV}^{-1}}{0.4 g_{a\gamma}} \frac{\mathrm{ Gauss}}{ B} .
\ee
It has been reported in \cite{Dominguez:2011xy} that such steps are indeed seen at a critical energy of $\sim 100$ GeV, with the spectrum consistent with $g_{i\gamma} \sim 10^{-11}\ {\rm GeV}^{-1}$, giving:
\be
m_{\rm ALP} \sim 10^{-9} \div 10^{-10} \ {\rm eV} .
\ee

Very recently, it was pointed out~\cite{Tavecchio:2012um} that a similar value of the photon coupling is required for
a possible explanation of the rapidly varying very high energy ($E>50$ GeV) emission from the flat spectrum radio quasar PKS 1222+216, which represents a challenge for standard blazar scenarios: in the latter one
is forced to invoke the existence of a very compact ($r\sim 10^{14}$ cm) emitting region at a large distance ($R>10^{18}$ cm) from the jet base, in order to avoid absorption of gamma rays in the dense ultraviolet radiation field of the broad line region (BLR). In ref.~\cite{Tavecchio:2012um} it was shown that one can also use a
standard blazar model for PKS 1222+216 where gamma rays are produced close to the central engine,
if one assumes that inside the source photons can oscillate into ALPs, with a similar coupling
corresponding to $f_{a_i}/\aC[i,\gamma] \sim 10^8\ {\rm GeV}$.

Remarkably, the values of $f_{a_i}/\mC[i,e]$ and $f_{a_i}/\aC[i,\gamma]$, inferred from very
different astronomical observations, turn out to be in the same region\footnote{There are also other hints of signals for ALPs in other parameter regions, for example in \cite{DiLella:2002ea,Zioutas:2010ya}.}.
However, there is one
crucial difference: while the WD value~(\ref{dec_const_wd}) can be realised both with the QCD axion
or with a generic ALP,
the anomalous transparency value~(\ref{dec_const_transp}) can not be realised with the
QCD axion, since the mass of the latter,
$m_a \sim m_\pi f_\pi/f_a\sim 10\ {\rm meV} \left( 10^9\ {\rm GeV}/f_a\right)$,
is much higher than the phenomenologically required values, which are in the sub-neV range.

\subsubsection{Black hole super-radiance}

Any axion that has a Compton wavelength comparable to the radius of black hole event horizons
can cause the spin-down of the black hole. Since black holes with masses between $2M_{\odot}$ and $10^{10} M_{\odot} $ have been observed,
this corresponds to masses of the order $10^{-10}$ to $10^{-21}$ eV.
The more precise upper bound, independent of the cosmological abundance of the axions, is then \cite{Arvanitaki:2009fg,Arvanitaki:2010sy}:
\be
m_a > 3 \times 10^{-11} {\rm eV}.
\ee
Interestingly, this is just below the range required for the astrophysical hints.

\subsection{Cosmic axions}
\label{APP:COSMIC}

\subsubsection{Dark matter}

The QCD axion~\cite{Preskill:1982cy,Abbott:1982af,Dine:1982ah} and
more generally ALPs~\cite{Arvanitaki:2009fg,Acharya:2010zx,Marsh:2011gr,Higaki:2011me,Arias:2012mb}
may constitute populations of cold dark matter (CDM), produced non-thermally via the misalignment
mechanism. The latter relies on assuming that fields in the early Universe have a random initial state
(arising from quantum fluctuations during inflation) which is fixed by the expansion of the Universe; fields with mass $m_i$ evolve on timescales $t \sim m_i^{-1}$. After such a timescale, the fields respond by attempting to minimise their potential, and consequently oscillate around the minimum. If there is no significant damping via decays, these oscillations can behave as a cold dark matter fluid since their energy density is diluted by the expansion of the Universe as $\rho_i\propto a^{-3}$, where $a$ is the Universe scale factor. Without fine-tuning of the
initial conditions and without further dilution by e.g. late entropy production,
the expected cosmic mass fraction in QCD axion CDM is then~\cite{Sikivie:2006ni}:
\be
\frac{\Omega_a h^2}{0.112}\approx  6.3 \times \left( \frac{f_a}{10^{12}\ \rm GeV} \right)^{7/6} \left( \frac{\Theta_a}{\pi} \right)^2
\ee
where $\Theta_a$ is the initial misalignment angle, while the one in ALP CDM is~\cite{Arias:2012mb}:
\be
\frac{\Omega_{a_i} h^2}{0.112} \approx 1.4 \times
\left( \frac{m_i}{\rm eV} \right)^{1/2} \left( \frac{f_{a_i}}{10^{11}\ \rm GeV} \right)^{2}  \left( \frac{\Theta_i}{\pi} \right)^2
\ee
where $m_i$ is the mass of the ALP.

Galactic halo QCD axions, as well as ALPs -- if they couple to photons,
$\aC[i,\gamma]\neq 0$ -- can be directly searched for via their resonant conversion into
a quasi-monochromatic microwave signal with frequency $\nu=m_a/(2\pi )=0.24\ {\rm GHz}\times (m_a/\mu{\rm eV})$,
in a high-Q electromagnetic cavity permeated by a strong static magnetic field~\cite{Sikivie:1983ip}.
A number of experiments of this type have already been done~\cite{DePanfilis:1987dk,Wuensch:1989sa,Hagmann:1990tj,Asztalos:2001tf,Asztalos:2009yp} and further improvements are
underway~\cite{Baker:2011na,Asztalos:2011bm,Heilman:2010zz}.
These haloscope searches reach currently sensitivities of order
\begin{equation}
g_{a\gamma}\sim g_{i\gamma}=\frac{\alpha}{2\pi} \frac{\aC[i,\gamma]}{f_{a_i}}\lesssim 10^{-(15\,\div 13)}\ {\rm GeV}^{-1}
\rightarrow
\frac{f_{a_i}}{\aC[i,\gamma]}\gtrsim 10^{10\,\div 12}\ {\rm GeV},
\end{equation}
in the $(1\div 10)\ \mu$eV mass range.

Recently, \cite{Graham:2011qk} reported the possibility of probing for QCD axion dark matter for decay constants between $10^{16}$ and $10^{19}$ GeV via molecular interferometry experiments. While it would not apply to ALPs - relying on the coupling to QCD - it would be a very interesting project to probe such weak couplings.

\subsubsection{Isocurvature and tensor modes}
\label{Iso}

If the axions constitute a significant proportion of dark matter, then they produce isocurvature and tensor fluctuations, which may be used to place bounds on the inflationary scale. These have been discussed for the case of many axions, in, for example \cite{Arvanitaki:2009fg}. The proportion of isocurvature fluctuations $\bra |S^2| \ket$ to the total $A_S \equiv \bra |S^2 (k_0)| \ket + \bra |R^2 (k_0)| \ket \simeq \bra |R(k_0)|^2\ket$ (where $R$ are the adiabatic fluctuations and $k_0$ is the pivot scale) is constrained by \cite{arXiv:1001.4538} to be less than $0.077$; we have:
\begin{align}
A_S =& (2.430 \pm 0.091) \times 10^{-9}\nn\\
\alpha \equiv& \frac{\bra |S^2| \ket}{\bra |S^2| \ket + \bra |R^2| \ket} = \sum^{n_{\rm ax}}_{i=1} \frac{1}{A_S}\left(\frac{\Omega_{a_i}}{\Omega_m} \right)^2 \frac{2\sigma_{\Theta_i}^2 (2 \Theta_i^2 + \sigma_{\Theta_i}^2)}{(\Theta_i^2 + \sigma_{\Theta_i}^2)^2}  < 0.077
\end{align}
where $\sigma_{\Theta_i} = \frac{H_{\rm inf}}{2\pi f_{a_i}}$ is the standard deviation of axion angle fluctuations and $\Omega_m$ is total matter fraction of the critical energy density.  Assuming that these fluctuations are dominated by one axion with $\Theta_i \gg \sigma_{\Theta_i}$ (i.e. no tuning of the angle) we have
\be
H_{\rm inf} < 4.3 \times 10^{-5} \left( \frac{\Omega_m}{\Omega_{a_i}} \right) \Theta_i f_{a_i}.
\ee
For the single-ALP model to explain the astrophysical anomalies (section \ref{APP:CONSTRAINTS}), this gives $H_{\rm inf} \lesssim 10^4$ GeV.
Since local axions will dominate over non-local ones in terms of energy density (since they couple much more strongly to the non-perturbative effects) we can use $f_{a_i} \sim M_P/\sqrt{\vo}$. Assuming that the dark matter is entirely one axion gives us the restriction $H_{\rm inf} < m_{3/2} \sim \frac{M_P}{\vo}$ for $\vo \gtrsim 10^9$, which restricts the possibility of entropy injection by large numbers of moduli for large volumes. Thus, if we insist upon axion dark matter and entropy injection, we are restricted to GUT-scale strings; but on the other hand, intermediate-scale strings produce axions only within the cosmologically allowed window, eliminating the need for entropy injection.

Tensor fluctuations of axions are typically negligible compared to those induced by metric fluctuations; the constraint upon the ratio of tensor to scalar fluctuations is $r\equiv A_T/A_S < 0.36$, where $A_T = \frac{16H_{\rm inf}^2}{\pi (8\pi M_P^2)}$, giving:
\be
H_{\rm inf} < 1.6 \times 10^{14} \ {\rm GeV}\,,
\ee
which is typically much weaker than the isocurvature bound.

\subsubsection{Rotation of CMB polarisation}

For very light axions/ALPs that begin to oscillate between the CMB era and today (corresponding to masses between $4\times 10^{-28} $ and $10^{-33} $ eV), that in addition have a significant coupling to photons, polarisation of the CMB may occur \cite{Arvanitaki:2009fg}. The limit is:
\be
\frac{\aC[i, \gamma] \alpha}{2\sqrt{3}} \lesssim 3.5 \times 10^{-2}.
\ee
This is only potentially constraining for \emph{local} axions, since non-local ones will have a parametrically suppressed $\aC[i, \gamma]$. With a combination of different cosmological observations, including galaxy redshift and weak lensing surveys, it should be possible to detect a fraction of ultralight axions to dark matter of a few percent~\cite{Marsh:2011bf}.

\subsection{Laboratory searches for axions}
\label{APP:LAB}

Currently, pure laboratory searches for very light (mass $<$~eV) axions can not compete with the
constraints from astrophysics and cosmology. The best sensitivity is obtained in light-shining-through-a-wall
experiments, which are searching for $\gamma \to a_i \to \gamma$ conversions in an external magnetic field
and thus probing the axion to two-photon coupling. Recently, a number of such experiments were performed; among
them, ALPS was the most sensitive one, improving the limit to~\cite{Ehret:2010mh}:
\be
g_{i\gamma}
< 6.5\times 10^{-8}\ {\rm GeV}^{-1}
\rightarrow
\frac{f_{a_i}}{\aC[i,\gamma]}\gtrsim 1.7\times 10^{5}\ {\rm GeV},
\ee
for masses below $m_i\ll$ meV. The next generation of these experiments is designed to reach a
sensitivity of $g_{i\gamma}\sim  10^{-11}$~GeV$^{-1}\rightarrow f_{a_i}/\aC[i,\gamma]\sim 10^8$~GeV~\cite{Redondo:2010dp},
thus starting to probe the region of interest for the possible explanation of the anomalous features
in AGN spectra reported above.

\section{Axionic couplings to matter at one-loop}
\label{SEC:LOOPS}

In addition to the tree level axionic couplings computed in section \ref{AxionicCouplings}, there will also be corrections to the K\"ahler metric at one loop. These have not yet been calculated for all cases of interest: the result is only known for the coupling of non-local axions to matter at a singularity \cite{Bain:2000fb,Berg:2005ja} and to adjoint matter \cite{Berg:2011ij}. It would be very interesting to calculate the corrections involving local axions (i.e. blow-up moduli) for matter at a singularity along the lines of \cite{Conlon:2011jq} but we leave this for future work. Here we shall analyse the contributions from field theory, and how they relate to the existing and future results.

The couplings to matter fields at one loop are generated via the coupling to photons or other gauge groups. This is to be distinguished from renormalisation of the axion-photon coupling, since we do not perform a chiral rotation - which would introduce axion couplings to mass terms. In non-supersymmetric theories, the correction from a loop involving photons and electrons is
$\Delta_{i \gamma\gamma}[ C_{ie}^{A,V}] = \frac{3\alpha^2}{4\pi^2} \,\aC[i,\gamma]  \ln \left(\frac{f_{a_i}}{m_e}\right) $ -
which depends upon the axion decay constant as a cutoff in the loop, or can be interpreted as a renormalisation of the axion-matter coupling.
However, in supersymmetric theories,
the two-photon coupling is related to a coupling to gauginos:
\be
\mc{L} \supset \int d^2 \theta\ (i a_i) \frac{g_{i \gamma}}{4} W^\alpha W_\alpha
\supset \frac{1}{4}g_{i\gamma} a_i F_{em,\mu\nu} \tilde{F}_{em}^{\mu\nu}
- \frac{1}{2}g_{i\gamma} \partial_\mu a_i \lambda^\alpha \sigma^\mu \ov{\lambda}, \nn
\ee
and so we also have loops involving gauginos and selectrons which will change the result. The cutoff-dependent contributions arise even before supersymmetry is broken, and are related to the corrections to the K\"ahler metric at one loop, whereas there are also more complicated finite contributions that arise when supersymmetry is broken. Let us recall that the K\"ahler potential at one loop is given by \cite{Brignole:2000kg}:
\begin{align}
\Delta K &= \frac{\Lambda^2}{16\pi^2} \left[ \ln \det \hat{K} - \ln \det \hat{H}\right] \nn\\
&- \frac{1}{32\pi^2} \left[ \tr\left( \mc{M}_\Phi^2 \left( \ln \frac{\mc{M}_\Phi^2}{\Lambda^2} - 1\right)\right)
- 2 \tr\left( \mc{M}_V^2 \left( \ln \frac{\mc{M}_V^2}{\Lambda^2} - 1\right) \right) \right],
\end{align}
where $\hat{K}$ is the K\"ahler metric, $\hat{K}^{-1/2}$ is the matrix that diagonalises $\hat{K}$: $\hat{K}^{-1/2\ T} \hat{K} \hat{K}^{-1/2} = \delta_{\bar{i}j}$, $\hat{w} \equiv \partial^2_{ij} W$ and:
\begin{align}
\hat{H}_{gh} \equiv& \,{\rm Re} (f_{gh}), \qquad
\mc{M}^2 \equiv \left(\hat{K}^{-1/2} \ov{w} \,\hat{K}^{-1\ T} w \,\hat{K}^{-1/2} \right)\nn\\
(\mc{M}_V^2)_{gh} \equiv& \,\hat{H}^{-1/2}_{gk} \left[ (\ov{\Phi} T_k)^{\ov{i}} K_{\ov{i} j}  (T_l \Phi)^j
+ (\ov{\Phi} T_l)^{\ov{i}} K_{\ov{i} j}  (T_k \Phi)^j \right] \hat{H}^{-1/2}_{kh}.
\end{align}
This is a global supersymmetry calculation, so we must be careful how we interpret the result.
Considering a model where $f_{gh} = \frac{\delta_{gh}}{g_g^2} ( 1 + g_{ig} T_i)$
for superfields $T_i \supset i c_i $ containing the axions $c_i$,
we can compute the gauge corrections to the K\"ahler potential
involving quadratic terms in matter fields $\Phi_i$ at first order in $T (\theta, \ov{\theta})$:
\begin{align}
\Delta K_{\rm{gauge}}\bigg|_{T} =& \frac{4}{32\pi^2}  \,g_g^2 (T_\alpha + \ov{T}_\alpha) \hat{K} \Phi_j^\dagger \Phi_j C_{2} ( \Phi_j )
\left[ \partial_{T_\alpha} \left(\ln \hat{K}\right) - \frac{g_{\alpha g}}{2}  \right] \ln\left( \frac{\Lambda^2}{m_{V}^2}\right), \nn
\end{align}
where we have inserted a mass $m_V$ for the gauge boson as an infra-red regulator. This is necessary in a theory without a superpotential and thus massless particles, but when the matter fields obtain masses these should appear instead. We thus find the physical coupling:
\begin{align}
\Delta \mc{L}_{\rm{gauge}} =& \frac{4}{32\pi^2}  \,g_g^2  K_{j \bar{j}} \psi_j \sigma^\mu \ov{\psi}_j \partial_\mu c_\alpha C_{2} ( \Phi_j )
\left[ \partial_{T_\alpha} \left(\ln \hat{K}\right) - \frac{g_{\alpha g}}{2}  \right] \ln\left( \frac{\Lambda^2}{m_{V}^2}\right). \nn
\end{align}
The first term in square brackets is generated by loops involving the tree-level coupling and
thus we expect should be corrected to $\ln \left(e^{-\frac{K_0}{2}} \hat{K}\right) $ in a supergravity calculation,
whereas the second term is the one truly generated by the axionic coupling.
The above is related to the dependence on the moduli of the anomalous dimension of the fields.
If we add a superpotential $W \supset Y \Phi_1 \Phi_2 \Phi_3$, we have:
\begin{align}
\Delta K_{\rm{Yukawa}}\bigg|_{T} =& - \frac{2}{32\pi^2}  |Y|^2 (T_\alpha + \ov{T}_\alpha)  \frac{\Phi_j^\dagger \Phi_j}{\hat{K}^2}
\,\partial_{T_\alpha} \left(\ln \hat{K}\right) \ln \left( \frac{\Lambda^2}{m_\Phi^2} \right). \nn
\end{align}
These two terms are apparently quite different. However, we list it for completeness to compare with the results for the behaviour of models of branes at orbifold singularities, where the anomalous dimension vanishes. There the gauge coupling is given by the dilaton, we have $\hat{K}= e^{\frac{K_0}{3}}$ and we should take $Y = \sqrt{2} \,g\, e^{\frac{K_0}{2}}$ with $g^2 = \left(S+\bar{S}\right)^{-1}$. Hence the dilaton-axion coupling cancels between the two terms. However, it does indicate the very interesting result that at one loop the corrections to the K\"ahler potential involving twisted (blow-up) moduli should not only not vanish but involve a large logarithmic factor. It would be interesting to confirm this with a calculation along the lines of \cite{Conlon:2011jq} where we could learn about the cutoff scale $\Lambda$ introduced in string theory but we leave this to future work.

Since the corrections at one loop due to the $\partial_{T_i} \left(\ln \hat{K}\right)$ term should be smaller than the tree-level term,
we can take the dominant loop contribution to be:
\begin{align}
\mc{L}\supset& - \frac{g_{ig}  \alpha_g}{2\pi}  \partial_\mu a_i   \psi_j \sigma^\mu \ov{\psi}_j C_{2} ( \psi_j )
\ln\left( \frac{\Lambda}{m_{V}}\right) \nn\\
\supset& - \frac{  \alpha_g^2}{4\pi^2 f_{a_i}}\mC[i,g]   \partial_\mu a_i   \psi_j \sigma^\mu \ov{\psi}_j C_{2} ( \psi_j )
\ln\left( \frac{\Lambda}{m_{V}}\right), \nn\\
\Delta \mC[i,g] =& - \frac{ 2 \alpha_g^2}{4\pi^2}\aC[i,g] C_{2} ( \psi_j ) \ln\left( \frac{\Lambda}{m_{V}}\right),
\end{align}
in the physical basis where the fields $\psi_j$ are canonically normalised. Note that this is for a left-handed fermion, so to obtain the electron couplings we should add together the couplings for left- and right-handed components. If we take the approximation that the electron couples only non-chirally to the photon, then no parity-violating vector-like coupling is generated. However, when we take the chiral nature of the electro-weak interactions into account we obtain the approximation (\ref{EQ:APPROXELECTRONCOUPLING}).

Once the supersymmetry breaking terms are included, there are also finite pieces that should in principle be included, even if they are subdominant to the above. Then the effect of electro-weak symmetry breaking must be taken into account. This leads to non-diagonal gaugino and gauge boson couplings, and the final result is rather complicated. We simply note for reference that the contribution to the coupling of a Weyl fermion (i.e. the left- or right-handed component of a matter fermion) will be derived from coupling terms of the form:
\begin{align}
\mc{L} \supset& \frac{g_{gh}}{4} \, a F^g_{\mu\nu} \tilde{F}^{h\,\mu \nu}
- \frac{\tilde{g}_{12}}{2} \,\partial_\mu a \lambda_1 \sigma^\mu \ov{\lambda}_2
- \sqrt{2} g_1 [\Phi^* (\psi \lambda_1) + h.c.] - \sqrt{2} g_2 [ \Phi^* (\psi\lambda_2) + h.c.], \nn
\end{align}
where $g_1, g_2, \tilde{g}_{12}$ involve gauge couplings and mixing matrices of the gaugino mass eigenstates. The result is:
\begin{eqnarray}
\Delta \mc{L} &=& \Delta_{\rm{boson}} \mc{L} + \Delta_{\rm{fermion}}\mc{L} \nn\\
\Delta_{\rm{boson}} \mc{L} &=& - (\partial_\mu a) \psi \sigma^\mu \ov{\psi} \frac{3g_g g_h g_{gh}}{2(4\pi)^2}
\left[ f( m_\psi^2, m_g^2) + \frac{m_h^2}{m_g^2 - m_h^2} \left( f(m_\psi^2, m_g^2) - f( m_\psi^2, m_h^2) \right) \right] \nn\\
\Delta_{\rm{fermion}}\mc{L} &=& (\partial_\mu a) \psi \sigma^\mu \ov{\psi} g_1 g_2 \tilde{g}_{12}
\left[ -i \int \frac{d^4q}{(2\pi)^4} \frac{\frac{1}{2} q^2 + M_1 M_2}{(q^2 - M_1^2) (q^2 - M_2^2)  (q^2 - M_\Phi^2) }\right] \nn\\
&=& (\partial_\mu a) \psi \sigma^\mu \ov{\psi} \frac{g_1 g_2 \tilde{g}_{12}}{(4\pi)^2}
\left[ \frac{1}{2} f( M_1^2, M_2^2) + \frac{ M_1 M_2 + \frac{1}{2} M_\Phi^2}{M_2^2 - M_\Phi^2}
\left( f(M_1^2, M_2^2) - f(M_1^2, M_\Phi^2)\right) \right] \nn\\
f( m_1^2, m_2^2) &\equiv& 1 + \left[  \frac{ m_2^2 \ln \left(\frac{\Lambda^2}{m_2^2}\right)
-  m_1^2 \ln \left(\frac{\Lambda^2}{m_{1}^2}\right) }{m_2^2 - m_1^2} \right]. \nn
\end{eqnarray}
Here we have given expressions in terms of gauge boson masses $m_g, m_h$,
fermion masses $m_\psi$, sfermion masses $M_\Phi$ and gaugino masses $M_1, M_2$.
In the case that $m_g = m_h = m_V$, $g_1 = g_2 = g_g = g_h$,
the leading cutoff-dependent part recovers the diagonal supersymmetric result above.
Note that in general we have parity-violating terms.

\section{Axion decay constants for general fibred Calabi-Yau manifolds}
\label{GenfibCY}

The analysis of the axion decay constants performed in section \ref{fibCY} is simple
since the volume form can be written explicitly in terms of the $\tau$ moduli. However, when we have a more complicated model, this may not be the case: we may only have the expression in terms of the K\"ahler parameters $t_\alpha$ and the $\tau_\alpha$ as quadratic expressions of these. In this case, we must use these to examine the metrics $\mc{K}_0$ and $\mc{K}_0^{-1}$. An interesting example of this is a general K3 or $T^4$ fibration in the anisotropic limit (which may in principle involve many moduli), where some $t_\alpha \sim x t_\alpha^0$ become large; we can expand all of the two-cycles, four-cycles and the volume as:
\be
t_A \equiv \,x\, t_A^0 +  t_A^1, \qquad
\tau_A \equiv \,x\, \tau_A^0 + \tau_A^1, \qquad
\vo = \,x \,\vo_0 + ...\,,\nn
\ee
and thus:
\begin{align}
(\mc{K}_0)^{AB} \sim& \,\vo \left( - x  K_{AB}^0 + x \frac{\tau_A^0 \tau_B^0}{\vo_0} - K_{AB}^1    + \frac{\tau_A^0 \tau_B^1 + \tau_A^1 \tau_B^0 - \frac{\vo_1}{\vo_0}\tau_A^0 \tau_B^0 }{\vo_0} + ...... \right) \nn\\
(\mc{K}_0)_{AB} \sim& \,\frac{1}{\vo} \left( \frac{x}{2} \frac{t_A^0 t_B^0}{\vo_0} + \frac{t_A^0 t_B^1 + t_A^1 t_B^0 - \frac{\vo_1}{\vo_0}t_A^0 t_B^0 }{2\vo_0} - (K^{AB})^0 - \frac{1}{x} (K^{AB})^1 +  ... \right). \nn
\label{EQ:OVERCOMPLICATED}
\end{align}
The matrix $\CC_{AB}$ diagonalises both of these, and we can determine it order by order by considering the two matrices separately. Since the leading matrix in the expansion of $(\mc{K}_0)_{AB} $ is of order one and rank one (it is proportional to $t_A^0 t_B^0$)  clearly there is exactly \emph{one} axion that has a decay constant of the order of the Planck mass. This is the axion corresponding to the fibre: it is proportional to $t_A^0 \rhoo_A $ so the associated modulus is proportional to $t_A^0 \tau_A $; since $t_A^0 \tau_A^0 = 0$ this must be a small cycle, and the coupling to matter is therefore Planck suppressed.

There are also several axions with string scale decay constants, but on the other hand, by examining $(\mc{K}_0)^{AB}$ we see that there are axions that have decay constants of order $M_P \vo^{-1}$! These would be very interesting phenomenologically if they coupled strongly to small cycles; however, (\ref{EQ:LOCALITY}) tells us that they must correspond to cycles with very small gauge couplings, and since in this limit there are only cycles $\tau_A^0 \sim \vo$ or $\tau_A \sim 1$ the former case must apply: hence they have Planck-suppressed couplings to gauge groups. They are given by the eigenvectors $c_{B}^i \rhoo_B$ of the matrix $- \vo_0  \mc{K}_{AB}^0 + \tau_A^0 \tau_B^0$; note that $\mc{K}_{AB}$ has signature $(+,-,-,-,...)$ and $\mc{K}_{AB}^0$ is the matrix of intersections that have large sizes. Hence there may be more than one large cycle: in this case, they must all be fixed relative to each other via D-terms to give an effective canonical fibration volume form for moduli stabilisation \cite{HiddenPhotons}.

\end{document}